\newif\ifuseprd
\newif\iftoomuchdetail
\newif\ifeprint

\useprdtrue
\eprinttrue
\toomuchdetailfalse

\newif\ifrr
\rrfalse
\newcommand\RR{{\ifrr{RR}\else{Ramond-Ramond (RR)}\fi\global\rrtrue}}

\ifuseprd 
\documentclass[\ifeprint preprint,\fi
tightenlines,%
floats,prd,eqsecnum,nobibnotes%
,nofootinbib,aps,12pt]{revtex4}
\else
\documentclass[final,11pt,letterpaper]{JHEP3}
\fi 
\usepackage{amsmath,amssymb}
\allowdisplaybreaks[3]

\makeatletter\@addtoreset{equation}{section}\makeatother

\newcounter{saveequation}
\newcounter{detailnum}
\newcommand\savetheequation{\theequation}
\newcommand\detailtheequation{%
	  $\delta$\Roman{detailnum}:\roman{equation}}

\newenvironment{detail}{\iftoomuchdetail\sf
         \setcounter{saveequation}{\value{equation}}%
         \setcounter{equation}{0}\addtocounter{detailnum}{1}%
         \renewcommand\theequation\detailtheequation%
         \fi}{
     \iftoomuchdetail%
     \ifnum\value{equation}=0\addtocounter{detailnum}{-1}\fi%
     \setcounter{equation}{\value{saveequation}}%
     \renewcommand\theequation\savetheequation%
     \fi%
     }

\DeclareMathOperator{\real}{Re}

\DeclareMathOperator{\diag}{diag}

\newcommand\p{\ensuremath{\partial}}

\newcommand\abs[1]{\ensuremath{\left\lvert{#1}\right\rvert}}

\newcommand\transpose{{\ensuremath{\text{\sf T}}}}
\newcommand\field[1]{{\ensuremath{\mathbb{{#1}}}}}
\newcommand\ZZ{{\field{Z}}}

\newcommand\ZR{{\field{R}}}
\newcommand\order[1]{{\ensuremath{{\mathcal O}\left({#1}\right)}}}
\newcommand\vev[1]{{\ensuremath{\left\langle{#1}\right\rangle}}}
\newcommand\anti[2]{\ensuremath{\left\{{#1},{#2}\right\}}}
\newcommand\com[2]{\ensuremath{\left[{#1},{#2}\right]}}
\DeclareMathOperator{\Tr}{Tr}

\newcommand\mathone{{\rlap{\kern .25em l}1}}
\newcommand\one{{\ifmmode{\text{\mathone}}\else{\mathone}\fi}}
\newcommand\proj{{\ensuremath{{{\mathbb P}}}}}
\newcommand\apr{{\ensuremath{{\alpha'}}}}
\newcommand\su{{\ensuremath{{\mathfrak su}}}}

\newcommand\lie[2]{\ensuremath{\pounds_{{#1}} {#2}}}
\newcommand\cg[6]{\ensuremath{{\left(\begin{smallmatrix} 
           \vphantom{#5} {#1} & {#3}%
	\\ \vphantom{#6} {#2} & {#4} \end{smallmatrix}%
     \right|\left. \begin{smallmatrix} 
           \vphantom{{#1}{#3}} {#5} \\ \vphantom{{#2}{#4}}{#6} 
     \end{smallmatrix} 
     \right)}}}
\newcommand\thj[6]{\ensuremath{{\left(\begin{smallmatrix} 
           {#1} & {#3} & {#5} \\ {#2} & {#4} & {#6} %
     \end{smallmatrix}\right)}}}
\newcommand\sixj[6]{\ensuremath{{\left\{\begin{smallmatrix} 
           {#1} & {#2} & {#3} \\ {#4} & {#5} & {#6} %
     \end{smallmatrix}\right\}}}}

\providecommand\FIGURE[2][]{\begin{figure}[#1]\begin{center}{#2}\end{center}
                       \end{figure}}
\providecommand\TABLE[2][tbh]{\begin{table}[#1]\begin{center}{#2}\end{center}
                       \end{table}}
\providecommand\putabstract[1]{\ifuseprd\begin{abstract} {#1} \end{abstract}%
                           \else \abstract{{#1}} \fi}
\providecommand\ajp[3]{{Am.\ J.\ Phys. {\bf {#1}}, {#3} ({#2})}}
\providecommand\plb[3]{{Phys.\ Lett.\ B {\bf {#1}}, {#3} ({#2})}}
\providecommand\npb[3]{{Nucl.\ Phys.\ {\bf B{#1}}, {#3} ({#2})}}
\providecommand\jhep[3]{{J.\ High Energy Phys.\ {\bf #1}, {#3} ({#2})}}
\providecommand\npps[3]{{Nucl.\ Phys.\ {\bf {#1}} Proc.\ Suppl.\ {#3} ({#2})}}
\providecommand\forp[3]{{Fortschr.\ Phys.\ {\bf {#1}} {#3} ({#2})}}

\providecommand\cqg[3]{{\ifuseprd\else\begingroup\em\fi Class.\ Quant.\ %
     Grav.\ %
     \ifuseprd\else\endgroup\fi {\bf {#1}}\ifuseprd, {#3} ({#2})\else%
     \ ({#2}) {#3}\fi}}

\providecommand\ijmpa[3]{{\ifuseprd\else\begingroup\em\fi Int.\ J.\ %
     Mod.\ Phys.\ \ifuseprd\else\endgroup\fi {\bf A{#1}}\ifuseprd, %
     {#3} ({#2})\else\ ({#2}) {#3}\fi}}

\providecommand\atmp[3]{{\ifuseprd\else\begingroup\em\fi Adv.\ Theor.\ %
     Math.\ Phys.\ \ifuseprd\else\endgroup\fi {\bf {#1}}%
     \ifuseprd, {#3} ({#2})\else\ ({#2}) {#3}\fi}}
\newcommand\citeap[3]{{\ifuseprd{Ann.\ Phys.\ (NY) {\bf {#1}}, {#3} ({#2})}%
                        \else{\ap{#1}{#2}{#3}}\fi}}
\newcommand\citeprd[3]{{\ifuseprd{Phys.\ Rev.\ D {\bf {#1}}, {#3} ({#2})}%
                        \else{\prd{#1}{#2}{#3}}\fi}}

\providecommand\hepth[1]{{\ifuseprd{\eprint{{\ifeprint\tt\fi hep-th/#1}}}%
                \else{\tt hep-th/{#1}}\fi}}

\newcommand\phepth[1]{{\ifuseprd\else\tt\fi [\hepth{#1}]}}
\newcommand\ct[1]{{\ifeprint\ifuseprd{\em{#1}},\else{\sf {#1}},\fi\fi}}
\newcommand\bt[1]{{\em {#1}},}

\newcommand\skipthis[1]{{}}

\newcommand\mass{M}
\newcommand\ang{{\mathcal J}}
\newcommand\wastheta[1]{X^{{#1}8}_0}
\newcommand\ppwave{{\em pp\/}~wave}
\newcommand\ppWave{{\em pp\/}~Wave}

\ifuseprd
\begin{document} 
\fi 

\title{Fuzzy Spheres in \ppWave\ Matrix String Theory}
\ifuseprd
\author{Sumit R. Das}\email{das@pa.uky.edu}
\author{Jeremy Michelson}\email{jeremy@pa.uky.edu}
\author{Alfred D. Shapere}\email{shapere@pa.uky.edu}
\affiliation{Department of Physics and Astronomy,
       University of Kentucky;
       600 Rose Street;
       Lexington, KY \ 40506}
\else 
\author{Sumit R. Das\thanks{\tt das@pa.uky.edu},
Jeremy Michelson\thanks{\tt jeremy@pa.uky.edu}
and
Alfred D. Shapere\thanks{\tt shapere@pa.uky.edu} \\
Department of Physics and Astronomy,
       University of Kentucky;
       600 Rose Street;
       Lexington, KY \ 40506
} 
\fi 

\putabstract{The behaviour of matrix string theory in the background of 
a type IIA \ppwave\ at small string coupling, $g_s \ll 1$,
is determined by the combination $M g_s$ where $M$ is a
dimensionless parameter proportional to the strength of the Ramond-Ramond
background.  For $M g_s \ll 1$, the
matrix string theory is conventional; only the degrees of freedom in
the Cartan subalgebra contribute, and the theory reduces to copies of
the perturbative string.  For $M g_s \gg 1$, the theory
admits degenerate vacua representing fundamental strings 
blown up into fuzzy spheres with nonzero lightcone momenta.
We determine the spectrum of small fluctuations around these
vacua. Around such a vacuum all $N^2$ degrees of freedom are
excited with comparable energies. The spectrum of masses has a
spacing which is independent of the radius of the fuzzy sphere, in
agreement with expected behaviour of continuum giant
gravitons. Furthermore, for fuzzy spheres characterized by reducible
representations of SU$(2)$ and vanishing Wilson lines, 
the boundary conditions on the field are
characterized by a set of continuous 
angles which shows that generically the blown up strings do not
``close''.}
\preprint{UK/03-09\ifuseprd,~\else\\\fi 
   {\tt hep-th/0306270}}

\ifuseprd
\maketitle
\else
\begin{document}
\fi 

\ifuseprd
\tableofcontents
\pagebreak
\fi

\newcommand\startylmprops{\eqref{startylmprops}}
\newcommand\startyvecprops{\eqref{startyvecprops}}
\newcommand\enylmprops{\eqref{endylmprops}}
\newcommand\enyvecprops{\eqref{vnorm}}
\newcommand\startSjlmprops{\eqref{Sjlmmajorana}}
\newcommand\enSjlmprops{\eqref{endSjlmprops}}

\newcommand\ben{\begin{equation}}
\newcommand\een{\end{equation}}
\newcommand\nn{\nonumber}
\newcommand\hx{{\hat x}}
\newcommand\hR{{\hat R}}
\newcommand\hg{{\hat g}}
\newcommand\tg{{\tilde g}}
\newcommand\tl{{\tilde \ell}}
\newcommand\hl{{\hat \ell}}
\newcommand\hF{{\hat F}}

\section{Introduction} \label{sec:intro}

The \ppwave\ is one of the few nontrivial backgrounds in which
perturbative string theory is exactly solvable~\cite{met}. Its
realization as a Penrose limit of $AdS \times S$ spacetimes~\cite{pen}
provides a candidate for a nonperturbative definition of string
theory, and has recently led to progress in understanding the
correspondence between gauge theories and IIB string theories in such
backgrounds~%
\cite{bmn}. In a slightly different direction, BMN~%
\cite{bmn}\ also proposed a definition of a BFSS
type Matrix theory~\cite{bfss}\ in \ppwave\ backgrounds, thus giving a
nonperturbative definition of M-theory in a nontrivial
background. The \ppwave\ Matrix theory has been studied quite
extensively~\cite{dsv1,dsv2,kp,kp2} from various points of view.

In this paper we begin an investigation of several nonperturbative
aspects of IIA strings in \ppwave\ backgrounds using matrix string theory.
The BMN matrix model represents DLCQ M-theory in the \ppwave\ 
background. To derive a matrix string theory we first need to compactify an
additional spacelike compact direction.  Circle compactifications of
\ppwave\ backgrounds were found in~\cite{jm1}, by identifying 
an appropriate isometry direction. Following a standard 
procedure~\cite{lm,bs,dvv}, we may then
obtain a matrix string theory on a IIA \ppwave\ background.
(This matrix string theory
was constructed in~\cite{gb,sy} and
the corresponding perturbative string has been studied in~\cite{hs,sy,hs2}.
Some properties of the
matrix string theory for the maximally supersymmetric type IIB
\ppwave~\cite{bfhp}
have been discussed in~\cite{gopakumar}.)
Applying standard arguments involving a ``9-11'' flip, the corresponding IIA
string theory has no zerobrane charge, but has nonzero light cone
momentum.

The IIA \ppwave\ matrix string theory is
a deformed $1{+}1$ dimensional SU$(N)$ Yang-Mills theory which should
describe IIA strings with a compact lightcone direction of radius
$R$ and a momentum $p_- = N/R$ along it.
The dynamics is
characterized by two dimensionless parameters, the string coupling
$g_s$ and the quantity
\begin{equation}
M = \frac{\mu \ell_s^2}{R},
\label{eq:one}
\end{equation}
where $\mu$ is the strength of the \RR\ field strength, $\ell_s$ is the
string length and $R$ is the radius of a null direction. 
By using a chain of dualities the corresponding matrix string theory
is essentially two-dimensional Yang-Mills theory with additional terms
that represent the nontrivial background. The dimensional coupling
constant of the gauge theory $g_{YM}$ is related to $g_s$ by 
\ben
g_{YM} = \frac{R}{g_s \ell_s^2},
\label{eq:two}
\een
in units for which the circumference of the spatial circle on which
the Yang-Mills
theory lives is
\ben
L = \frac{2\pi \ell_s^2}{R}.
\label{eq:three}
\een 

The various regimes of the theory are characterized by the combination
$Mg_s$. Observe that this quantity is proportional to the {\em effective
string coupling} in this background 
\hbox{$g_{\text{eff}} \sim g_s \mu p_- l_s^2$}
\cite{constable,gopakumar}.  When $g_s \rightarrow 0$ with $M$ finite,
so that $M g_s \ll 1$, the vacua of matrix string theory are
essentially the same as in flat space, {\em i.e.\/}\ the matrices $X^i$ can be
all chosen to be diagonal. Well known arguments then lead to light
cone strings in the IIA \ppwave\ with various lengths whose sum is the
total lightcone momentum.

However, matrix string theory is in principle a nonperturbative definition
of string theory. The \ppwave\ is in fact the only known nontrivial
background in which a matrix string theory can be formulated unambiguously.
We therefore have a unique opportunity to probe nonperturbative aspects
of string theory in a controlled setting.
Indeed, when $g_s \rightarrow 0$, but $M \rightarrow \infty$ keeping
$Mg_s$ finite, there are degenerate vacua corresponding
to fuzzy spheres along the directions of the \RR\ field. In the IIB
language these are actually ``fuzzy cylinders'' since there is an
additional compact direction along which the ground state solution is
constant. By the chain of dualities reviewed below this means that 
IIA fundamental strings with  momenta along a 
compact null direction have blown up into
fuzzy spheres. In the following we will, therefore, refer to these
simply as fuzzy spheres. In the ground state the matrices which
correspond to the other transverse directions are constrained to
vanish, {\em except\/} for one of the directions, for which the matrix can
have a set of ``$\theta$-parameters'', the number of which is determined
by the SU(2) representation content of the fuzzy sphere. Such fuzzy
sphere configurations retain all eight linearly realized supersymmetries of the
theory.  The mechanism is in fact the Myers' effect~\cite{myers,tv} with
two crucial differences. First the various fuzzy sphere configurations
have zero lightcone energy and are, therefore, degenerate with the trivial
vacuum for which all the matrices vanish. This is what happens in the
giant graviton effect~\cite{giant}. Indeed we have an {\em exact} microscopic
description of giant gravitons in a nontrivial background. In earlier works
giant gravitons have been described in certain D-brane backgrounds in
an approximate fashion \cite{microgiant}.%
\footnote{An exact description, based in part on the extension~\cite{bjy,jy1}
of matrix theory in weak backgrounds~\cite{tv}
to matrix string theory, and on the nonabelian Born-Infeld action
defined using the symmetric trace prescription, was proposed
in~\cite{jy2}.
However, as the symmetric trace prescription is known to break
down by order $F^6$~\cite{brs}, this proposal should be viewed cautiously.}
Secondly,
the gauge field, though non-dynamical, plays an important role,
particularly in the emergence of the $\theta$-parameters.
 
We then investigate the spectra of small fluctuations around these
degenerate fuzzy vacua. This is done in the limit of large $Mg_s$
where a perturbative analysis is valid. The matrix theory fuzzy sphere
fluctuation spectrum has been investigated earlier in~\cite{dsv1}.
For matrix strings, there are several important modifications since the
gauge field couples nontrivially and renders
the analysis rather involved. Nevertheless the spectrum is rather
simple. For the fuzzy sphere in the irreducible representation, the
fluctuations are described by a set of $1+1$ dimensional fields
labeled by angular momentum quantum numbers $(\ell,m)$ with%
\footnote{This bound on $\ell$ holds only for irreducible vacua;
the more general statement is given in Sec.~\ref{sec:redfluc}.}
$ 0 \leq \ell
\leq N-1$ and $-\ell \leq m \leq \ell$, with masses of the form $M
f(\ell)$, where
$f(\ell)$ is a linear function of $\ell$ that depends on the specific
field. Significantly, the mass does not depend directly on $N$. This feature is
in fact quite characteristic of fluctuations of giant gravitons.

When
$N \gg 1$ the fuzzy spheres may be approximated by continuous
spheres. In this limit we show that the radius of the giant graviton
agrees with that of the fuzzy sphere, which is
\ben
r_0 = \frac{\pi \mu g_s \ell_s^3 N}{3R} = (\frac{Mg_s}{3}) \pi N \ell_s.
\een
A simple argument then shows that the fluctuation spectrum of these
giant gravitons is governed by a level spacing which is given by $\mu$
and {\em independent of the radius\/} $r_0$. This kind of spectrum has
been noticed for giant gravitons in $AdS \times S$ backgrounds earlier~%
\cite{djm}. Their description in terms of the holographic
Yang-Mills theory~\cite{giantholo1,giantholo2} when applied to the \ppwave\ 
background also reveals a similar
spectrum~\cite{ppgiantspectrum1,ppgiantspectrum2}.

For matrix strings in flat space, the length of the string is
determined by the boundary conditions of the fields in a description
in which there is no background Wilson line. As is well known these
boundary conditions correspond to conjugation by U$(N)$ group elements
which belong to the subgroup $S_N$. They represent a collection strings whose
lengths add up to the total light cone momentum. In the
\ppwave\ background there is a similar story for $Mg_s \ll 1$. However
for $Mg_s \gg 1$,
the fuzzy spheres generically  form $N {\times} N$ dimensional 
reducible representations of SU$(2)$. Sticking to a gauge in
which the background Wilson lines vanish, we will find that for $Mg_s$ large,
the fluctuations in each of the irreducible
blocks are necessarily periodic as we go around the compact
circle. However the boundary conditions on the fluctuations in the
off-diagonal blocks are characterized by a set of continuous angles.
This implies that the momenta along the circle may be written in the
form $\frac{1}{R}(n + \frac{1}{\chi})$, with $\chi$ generically
irrational. In other words the blown up string never
closes. Alternatively, in a gauge in which the fields
are strictly periodic, allowed Wilson lines are characterized by a 
set of U(1) angles.

The presence of fuzzy sphere vacua suggests that the
density of states
at intermediate energies has a rather different behaviour than the
Hagedorn density of states appropriate for strings.
In fact, as discussed in Sec.~\ref{sec:thermo}, this
appears to be an overly na\"{\i}ve expectation, and the Hagedorn
transition persists.
Certainly, though, there are
nontrivial consequences for the thermodynamics of strings in this regime.

This paper is organized as follows. In Sec.~\ref{sec:setup} we describe the
basic setup which relates Type IIA strings in \ppwave\ backgrounds to a
matrix string theory. In Sec.~\ref{sec:giant} we show the existence of stable
giant gravitons in these \ppwave\ backgrounds and discuss their
fluctuation spectra. Sec.~\ref{sec:action} contains our derivation of
the \ppwave\
matrix string action. In Sec.~\ref{sec:prop} we discuss supersymmetry
properties
of the action and its vacua in various regimes, with particular attention to 
fuzzy sphere vacua for large $Mg_s$. Sec.~\ref{sec:spectrum}
contains the bulk of our
results for fluctuations around fuzzy sphere vacua and allowed
boundary conditions. In Sec.~\ref{sec:thermo} we make brief comments about
implications to thermodynamics and Sec.~\ref{sec:conc} contains conclusions
and comments.  The appendix presents a detailed and self-contained
discussion of fuzzy
spherical harmonics.

\section{The Setup} \label{sec:setup}

The \ppwave\ background in $M$-theory is given by the metric
\begin{equation} \label{eq:aone}
\begin{aligned}
ds^2 &= 2dx^+dx^- - 
\Bigr[(\frac{\mu}{3})^2 (x^a)^2 + (\frac{\mu}{6})^2 (x^{a'})^2\Bigr]
(dx^+)^2 +(dx^a)^2 + (dx^{a'})^2, \\ 
F_{+123} &= \mu,
\end{aligned}
\end{equation}
where $a = 1\cdots 3$ and $a' = 4 \cdots 9$.  To make a spacelike
isometry explicit it is necessary to perform a coordinate
transformation
\begin{equation} \label{eq:atwo}
\begin{gathered}
x^+ = \hx^+, \qquad x^a = \hx^a, \qquad x^{a''} = \hx^{a''} (a''=4 \cdots 7) \\
\begin{aligned}
x^- &= \hx^- - \frac{\mu}{6} \hx^8 \hx^9, \\
x^8 &= \hx^8 \cos (\frac{\mu}{6}\hx^+) + \hx^9 \sin (\frac{\mu}{6}\hx^+), 
\\
x^9 &= -\hx^8 \sin (\frac{\mu}{6}\hx^+) + \hx^9 \cos (\frac{\mu}{6}\hx^+),
\end{aligned}
\end{gathered}
\end{equation}
so that the background becomes
\begin{equation} \label{eq:athree}
\begin{aligned}
ds^2 & =  2 d\hx^+d\hx^- - 2 \frac{\mu}{3} \hx^8 d\hx^9 d\hx^+
-\Bigl[(\frac{\mu}{3})^2 (\hx^a)^2 + (\frac{\mu}{6})^2
(\hx^{a''})^2\Bigr] (d\hx^+)^2 + (d\hx^a)^2 + (d\hx^{a'})^2, \\
\hF_{+123} & = \mu.
\end{aligned}
\end{equation}

\FIGURE[tb]{
\setlength{\unitlength}{\baselineskip} 
\begin{picture}(22,20)(-4,-20) 
\put(4.75,-4){\frame{
   \begin{picture}(7.5,4)
      \put(1,1){\shortstack{
         {\bf M-Theory} \\ \\
         $x^9{:}\; \hat{R} \qquad x^-{:}\; R$ \\
         $\ell_P$}}
   \end{picture}}}
\put(10,-12){\frame{
   \begin{picture}(8,6)
      \put(1,1){\shortstack{
         {\bf String Theory II} \\ \\
         $x^9{:}\; \hat{R}$ \\
         $\hat{g}_s = \left( \frac{R}{\ell_P} \right)^{3/2}$ \\
         $\hat{\ell}_s = \left( \frac{\ell_P^3}{R} \right)^{1/2}$}}
   \end{picture}}}
\put(11,-4){\vector(3,-2){3}} 
   \put(12,-5){\makebox(0,0)[r]{\scriptsize on $-$}}
\put(10,-20){\frame{
   \begin{picture}(8,6)
      \put(1,1){\shortstack{
         {\bf String Theory III} \\ \\
         $x^9{:}\; \hat{R}'$ \\
         $\tilde{g}_s = \frac{R}{\hat{R}}$ \\
         $\tilde{\ell}_s = \hat{\ell}_s$}}
   \end{picture}}}
\put(14,-12){\vector(0,-1){2}} \put(14,-14){\vector(0,1){2}}
   \put(11,-13.5){\shortstack{\scriptsize T-duality \\ \scriptsize on 9}}
\put(0,-12){\frame{
   \begin{picture}(8,6)
      \put(1,1){\shortstack{
         {\bf String Theory I} \\ \\
         $x^-{:}\; R$ \\
         $g_s = \left( \frac{\hat{R}}{\ell_P} \right)^{3/2}$ \\
         $\ell_s = \left( \frac{\ell_P^3}{\hat{R}} \right)^{1/2}$}}
   \end{picture}}}
\put(6,-4){\vector(-3,-2){3}}
   \put(5.5,-5){\makebox(0,0){\scriptsize on $9$}}
\put(8,-9){\vector(1,0){2}} \put(10,-9){\vector(-1,0){2}}
   \put(9,-9){\makebox(0,0){\shortstack{\scriptsize 9/11 \\
                                        \scriptsize flip}}}
\put(0,-20){\frame{
   \begin{picture}(8,6)
      \put(1,1){\shortstack{
         {\bf String Theory IV} \\ \\
         $x^-{:}\; \frac{\ell_s^2}{R}$ \\
         $\Hat{\tilde{g}}_s = \frac{\hat{R}}{R}$ \\
         $\ell_s = \ell_s$}}
   \end{picture}}}
\put(4,-12){\vector(0,-1){2}} \put(4,-14){\vector(0,1){2}}
   \put(4.5,-13.5){\shortstack{\scriptsize T-duality \\ \scriptsize on $-$}}
\put(8,-17){\vector(1,0){2}} \put(10,-17){\vector(-1,0){2}}
   \put(9,-16.5){\makebox(0,0){\tiny S-duality}}
\put(-4,-9){\bf IIA:}
\put(-4,-17){\bf IIB:}
\end{picture}
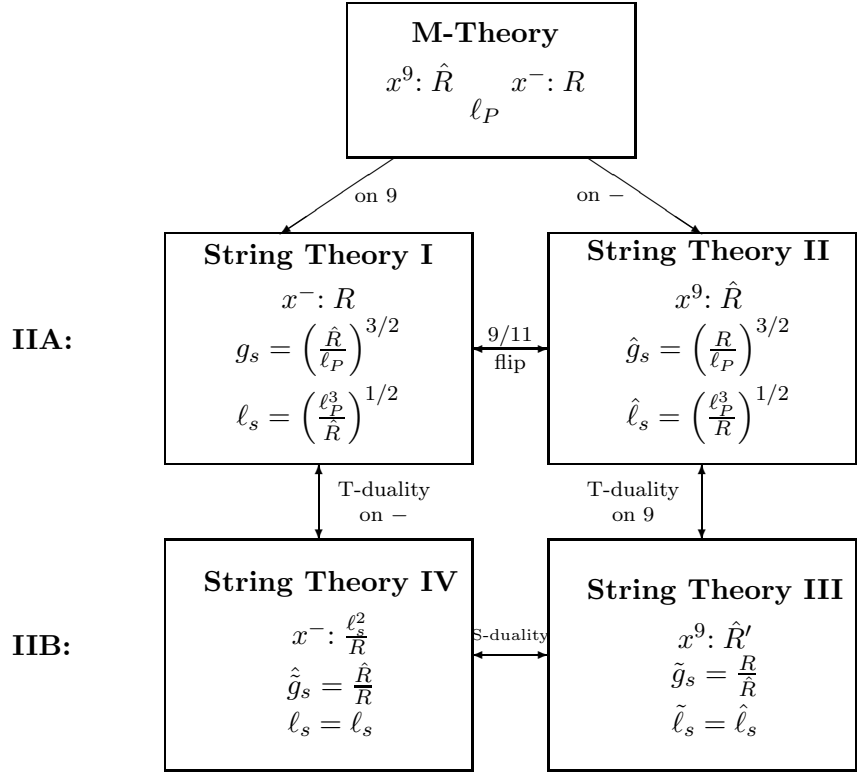
\caption{The chain of dualities that leads from matrix theory to the
matrix string.  For each theory, the length and direction 
of the compact dimensions are given. 
String Theory IV has been included for completeness, but
would be more accurately titled
``String Theory not appearing in this paper'' (with apologies to
Monty Python).
\label{fig:duality}}
}

From now on we will use these new coordinates and remove the hats from
the $x$'s. Since translations along the $\hx^9$-direction are isometries one
can now compactify along $x^9$ with a radius $\hR$ which will lead to
a IIA string theory with the string frame metric and \RR\ fields given
by 
\begin{equation} \label{eq:afour}
\begin{gathered}
ds^2 = 2dx^+dx^- -\Bigl[(\frac{\mu}{3})^2 \bigl((x^a)^2 + (x^8)^2\bigr) 
   + (\frac{\mu}{6})^2 (x^{a''})^2\Bigr](dx^+)^2 + (dx^a)^2 +(dx^{a''})^2
+(dx^8)^2, \\
\begin{aligned}
A_+ &= -\frac{\mu}{3} x^8, & C_{+ab}
& = \mu\epsilon_{abc} x^c.
\end{aligned}
\end{gathered}
\end{equation}
This IIA theory will be called String Theory No.~I in the
following (see Fig.~\ref{fig:duality}). It has a string coupling
\ben
g_s = \left(\frac{\hR}{\ell_P}\right)^{3/2},
\label{eq:afive}
\een
and a string length $\ell_s$ given by
\ben
\ell_s = \left(\frac{\ell_P^3}{\hR}\right)^{1/2}.
\label{eq:asix}
\een
We will take $x^-$ to be a compact direction with radius $R$ and consider a
state in the M-theory with a momentum
\ben
p_- = \frac{N}{R}
\label{eq:aseven}
\een
along this direction.  Then String Theory No.~I has a null compact
direction $x^-$ with a momentum $p_-$ given above.

We can construct another IIA string theory, which we will call String
Theory No.~II where the null direction $x^-$ is considered as the
extra M-theory direction. In this string theory the state under
consideration has a zerobrane charge $N$ while the string coupling
$\hg_s$ and the string length $\hl_s$ are given by
\ben
\hg_s = \left(\frac{R}{\ell_P}\right)^{3/2}, \qquad
\hl_s = \left(\frac{\ell_P^3}{R}\right)^{1/2}.
\label{eq:aeight}
\een
String Theory No.~II has a compact spacelike direction $x^9$ with
radius $\hR$.

T-dualizing String Theory No.~II 
along $x^9$ gives rise to a IIB string theory (String Theory No.~III) 
with D1-brane charge $N$ defined on a radius $\hR '$
\ben
\hR ' = \frac{\hl_s^2}{\hR} = \frac{\ell_P^3}{R \hR},
\label {eq:anine}
\een
and string coupling $\tg_s$ and string length $\tl_s$ given by
\ben
\tg_s = \hg_s \frac{\hl_s}{\hR} = \frac{R}{\hR}, \qquad
\tl_s = \hl_s.
\label{eq:aten}
\een
This IIB theory thus has a set of D1-branes wrapped on a circle with a
total D1-brane charge $N$ and is therefore described by a 1+1
dimensional Yang-Mills theory with gauge group SU$(N)$. The
dimensional coupling constant of this Yang-Mills theory is given by
\ben
g_{YM}^2 = \frac{R^2}{\hR \ell_P^3}.
\label{eq:aeleven}
\een
The dimensionless combination of the Yang-Mills coupling and the size
of the circle is easily seen to be
\ben
g_{YM} \hR ' = \frac{1}{g_s},
\label{eq:atwelve}
\een
so that String Theory No.~I at weak coupling corresponds to strongly coupled
Yang-Mills theory. This is the matrix string theory which we discuss in this
paper.

As we will see, the matrix string theory is characterized by the
combination $Mg_s$, where $M=\frac{\mu
\ell_s^2}{R}$ is also dimensionless. In view of the results of
\cite{constable} the
effective string coupling of String Theory No.~I is given by the
quantity $g_{\text{eff}} = g_s\mu p_- l_s^2$. Since $p_- = N/R$, this implies
$g_{\text{eff}} \sim
Mg_sN$. Thus one would expect that when $Mg_s \ll 1$ one has the
usual
perturbative IIA string, while for $Mg_s \sim 1$ one starts probing
nonperturbative behaviour. We will find that $1/(Mg_s)$ is essentially
the coupling constant of the deformed Yang-Mills theory. Thus for
$Mg_s \gg 1$ the Yang-Mills theory becomes weakly coupled and one can
discuss nonperturbative string theory in a controlled fashion.

\section{Giant gravitons and their fluctuations} \label{sec:giant}

In this section we will consider a single spherical M2-brane
in the background of the M-theory \ppwave~\eqref{eq:aone} and
show that there is a stable solution in the presence of a nonzero
momentum $p_-$ in the $x^-$ direction. We will then examine the nature
of the fluctuation spectrum of these branes.

\subsection{The Classical Solution} \label{sec:giantsoln}

The reparametrization invariant
action for this brane is given by
\ben
S = T_2 \int d^3\xi \left[-{\sqrt{-\det g}} +\frac{1}{6}\epsilon^{abc}
\partial_a X^\mu \partial_b X^\nu \partial_c X^\lambda
C_{\mu\nu\lambda} (X)\right],
\label{eq:bone}
\een
where $\xi^a$ denotes the worldvolume coordinates, $X^\mu (\xi)$ are the
target space coordinates, and $g_{ab}$ is the induced metric on the
worldvolume
\ben
g_{ab} = G_{\mu\nu}\partial_aX^\mu \partial_b X^\nu,
\label{eq:btwo}
\een
with $G_{\mu\nu}$ the \ppwave\ metric given in eq.~%
(\ref{eq:aone}). $C_{\mu\nu\lambda}$ is a 3-form \RR\ potential which
gives rise to the 4-form gauge field strength appearing in eq.~%
(\ref{eq:aone}).

We will use polar coordinates $(r,\theta,\phi)$ in the plane defined by
the Cartesian coordinates $x^a, a = 1\cdots 3$. In these coordinates
the field strength is 
\ben
F_{+r\theta\phi} = \mu r^2 \sin \theta,
\een
and we can choose a gauge such that the potential is
\ben
C_{+\theta\phi} = \frac{1}{3} \mu r^3 \sin\theta,
\een
with all other components zero. 

Let us first fix a gauge in which the spatial coordinates on the
worldvolume are identified with the angles $\theta$ and $\phi$. We
want to restrict our dynamics to the sector where the remaining
coordinates are independent of $\theta,\phi$. This is a consistent
truncation which respects the equations of motion. Then all
worldvolume fields depend only on the worldvolume time $\tau$ and the
action (\ref{eq:bone}) becomes
\ben
S = 4\pi T_2\int d\tau\left[-r^2(\tau) {\sqrt{-G_{AB}\partial_\tau
X^A \partial_\tau X^B}} + \frac{\mu}{3} r^3(\tau) \partial_\tau
X^+\right], 
\label{eq:bfour}
\een
where $A,B$ stand for $(r,X^{a'},X^\pm)$

Let us define the quantity
\ben
D \equiv {\sqrt{-G_{AB}\partial_\tau
X^A \partial_\tau X^B}}.
\label{eq:bfive}
\een
Then the canonical momenta are given by
\ben
P_A = 4\pi T_2 \left[ \frac{1}{D} r^2 G_{AB}\partial_\tau X^B +
\frac{\mu r^3}{3}\delta_A^+ \right].
\label{eq:bsix}
\een
The identity
\ben
G^{AB}(P_A - \frac{4\pi T_2}{3}\mu r^3 \delta_A^+)
(P_B - \frac{4\pi T_2}{3}\mu r^3 \delta_B^+) = -(4\pi T_2 r^2)^2
\label{eq:bseven}
\een
yields
\ben
P_+ = \frac{1}{2P_-}[G_{++} P_-^2 - P_{a'}^2 - P_r^2 - (4\pi T_2 r^2)^2]
+\frac{4\pi T_2}{3} \mu r^3.
\label{eq:beight}
\een

The above gauge choice still allows arbitrary reparametrization of the
worldvolume time coordinate $\tau$. We  fix this by choosing a gauge
\ben
\tau = X^+.
\een
Then the canonical Hamiltonian in this gauge is given by
\ben
H = - P_+.
\een
Using the specific form of $G_{++}$ given in (\ref{eq:afour}), 
$H$ may be written as a sum of squares 
\ben
H = \frac{1}{2P_-}\left[P_{a'}^2 + P_r^2 
  + \left(\frac{\mu}{6}\right)^2 P_-^2 (X^{a'})^2
  + r^2 (4\pi T_2 r - \frac{\mu}{3} P_-)^2\right].
\label{eq:bnine}
\een
Clearly the ground state solutions are static with $P_r = P_{a'} = 0$, 
$X^{a'}=0$, and either $r=0$ or
\ben
r = r_0 = \frac{\mu}{12\pi T_2} P_-.
\label{eq:bten}
\een
This second solution is the giant graviton. It has lightcone
energy zero, and is therefore degenerate with the lightcone vacuum.

An interesting feature of this solution is that while the momentum
$P_-$ in the direction $X^-$ is nonzero, there is no {\em velocity\/}
along $x^-$. This may be easily seen from the expressions for $P_\pm$
\begin{equation} \label{eq:beleven}
\begin{aligned}
P_+ & = \frac{(4\pi T_2 r^2)}{D}(\partial_\tau X^- + G_{++})
+ \frac{4\pi T_2}{3} \mu r^3, \\
P_- & = \frac{(4\pi T_2 r^2)}{D},
\end{aligned}
\end{equation}
by plugging in the classical solution for a giant graviton given
in eq.~\eqref{eq:bten}.

The M2-brane giant graviton will appear as a stable spherical D2-brane
in String Theory No.~I with a nonzero light cone momentum $p_- =
N/R$. Using
\ben
T_2 = \frac{1}{4\pi^2 \ell_P^3} = \frac{1}{4\pi^2 g_s \ell_s^3},
\een
one sees that the radius of the D2-brane is
\ben
r_0 = \frac{\mu g_s \ell_s^3}{R} \frac{\pi N}{3} = \ell_s \frac{Mg_s}{3} \pi
N.
\label{eq:beleven2}
\een
As explained above, although there is a nonzero $p_-$, the brane is
actually static. 

In String Theory No.~II the giant graviton also appears as 
a static D2-brane, but with a D0-brane 
charge $N$ and no other momentum. This D2-brane has a compact
transverse direction. T-dualizing along the compact direction yields
a D3-brane in String Theory No.~III 
having a shape $S^2 \times S^1$. This is the continuum
analog of the ``fuzzy cylinder'' configuration of matrix string
theory discussed in this paper.

\subsection{Fluctuations} \label{sec:giantfluc}

The fluctuation spectrum around the giant graviton may be obtained by
an analysis similar to~\cite{djm}. For our purposes, however, it is 
sufficient to examine some general features. The effective action 
for some mode of 
transverse fluctuation $\Phi$ of the giant graviton would be given, in the
linearized approximation, by
\ben
S = \int d^3\xi {\sqrt{-g}} g^{ab}\partial_a \Phi \partial_b \Phi,
\label{eq:cone}
\een
where $g_{ab}$ denotes the induced metric on the brane worldvolume,
whose components are given by
\ben
g_{ab}  = G_{\mu\nu}(X)\partial_a X^\mu \partial_b X^\nu,
\een
The target space metric $G$ has to be evaluated {\em on the classical
solution\/}. In the light cone gauge used above this solution is
described by
\begin{equation} \label{eq:ctwo}
\begin{aligned}
\theta &= \sigma, \qquad \phi = \rho, \\
X^+ & = \tau, \\
r & = r_0 = \frac{\mu}{12\pi T_2} P_-, \\
X^{a'} & =  0, \\
X^- & =  \text{constant},
\end{aligned}
\end{equation}
where $\sigma$ and $\rho$ denote the two angular spatial coordinates on
the worldvolume.
Thus on the giant graviton solution one has
\begin{equation} \label{eq:cthree}
\begin{aligned}
g_{\tau\tau} & = -(\frac{\mu}{3})^2 r_0^2, \\
g_{\sigma\sigma} & = r_0^2, \\
g_{\rho\rho} & = r_0^2 \sin^2 \theta,
\end{aligned}
\end{equation}
so that the action for fluctuations (\ref{eq:cone}) becomes
\ben
S \sim \frac{\mu}{3}r_0 \int dt d\theta d\phi \sin \theta 
\left[-(\frac{3}{\mu})^2 (\partial_\tau \Phi)^2 + (\partial_\theta \Phi)^2
+ \frac{1}{\sin^2\theta} (\partial_\phi \Phi)^2\right].
\label{eq:cfour}
\een
It is now clear that the frequencies of oscillation $\omega$ 
are independent of
$r_0$ and the scale is set entirely by $\mu$. In fact from
eq.~\eqref{eq:cfour}
one would get
\ben
\omega = (\frac{\mu}{3})\ell (\ell+1),
\label{eq:cfive}
\een
where $\ell$ is the angular momentum quantum number. 
It would be interesting to verify the above argument by a detailed
calculation of the fluctuation spectrum.
It is
clear that while the details of the spectrum would not agree precisely with 
(\ref{eq:cfive}), the fact that it is independent of $r_0$ and hence
$p_-$ would persist.

In the following sections we set up the matrix string theory for this
problem and study its vacua and fluctuations. 

\section{The Matrix String Action} \label{sec:action}

We start with the matrix theory action of BMN~\cite{bmn},
modulo some conventions; see also~\cite{dsv1,kp}.
Specifically, we normalize the worldline coordinate
$\tau$ as in BMN and our $R$ differs from their $R_{11}$ by a factor of
2.%
\footnote{Also, we use the convention that complex conjugation
interchanges the order of Grassmann variables.
The gauge theory conventions are \hbox{$D_\mu = \p_\mu + i
\com{A_\mu}{\cdot}$}; \hbox{$F_{\mu\nu}=2\p_{[\mu} A_{\nu]} + i
\com{A_\mu}{A_\nu}$}.  Under gauge transformations, \hbox{$A_\mu\rightarrow
-i U \p_\mu U^{-1} + U A_\mu U^{-1}$}; \hbox{$X^{i} \rightarrow U X^i
 U^{-1}$}; \hbox{$\Psi \rightarrow U \Psi U^{-1}$}.
Unless stated otherwise, $\Psi^\transpose$ denotes the transpose of
the spinor $\Psi$, without affecting the U$(N)$ matrices.  So under
 gauge transformations, \hbox{$\Psi^\transpose\rightarrow U
 \Psi^\transpose U^{-1}$}.
\label{ft:isinaction}}
We also follow the usual convention of setting the 11-dimensional
Planck length, $\ell_P=1$, (for
now).
\iftoomuchdetail
\begin{detail}%
We have stolen some factors of $i$ in the fermionic
terms from~\cite{dsv1}; see footnote~\ref{ft:isinaction}.
\end{detail}%
\fi
The conventions for indices are $i=1\cdots9$, $a=1,2,3$, $a'=4\cdots9$
and (later)
$a''=4\cdots7$.
\begin{multline} \label{MQM}
S = R \int d\tau \Tr \left\{
\frac{1}{2 R^2} (D_\tau X^i)^2
+ \frac{i}{R} \Psi^\transpose D_{\tau} \Psi
+ \Psi^\transpose \Gamma^i \com{X^i}{\Psi}
+ \frac{1}{4} \com{X^i}{X^j}^2
\right.\\* \left.
- \frac{1}{2} \bigl(\frac{\mu}{3 R}\bigr)^2 (X^a)^2
- \frac{1}{2} \bigl(\frac{\mu}{6 R}\bigr)^2 (X^{a'})^2
- i \frac{\mu}{4 R} \Psi^\transpose \Gamma^{123} \Psi
- i \frac{\mu}{3 R} \epsilon_{abc} X^a X^b X^c
\right\}.
\end{multline}
\iftoomuchdetail
\begin{detail}%
The action~\eqref{MQM} is invariant under the 16 linearly realized
supersymmetries generated
by~\cite{bmn,dsv1,dsv2}
\begin{subequations} \label{QMSUSY}
\begin{gather} \label{QMSUSYb}
\begin{align}
\delta X^i &= i \Psi^\transpose \Gamma^i \epsilon(\tau), &
\delta A_\tau &= i R \Psi^\transpose \epsilon(\tau), 
\end{align} \\
\label{QMSUSYf}
\delta \Psi = \frac{1}{2 R} D_\tau X^i \Gamma^i \epsilon(\tau)
   + \frac{\mu}{6 R} X^a \Gamma^a \Gamma^{123} \epsilon(\tau)
   - \frac{\mu}{12 R} X^{a'} \Gamma^{a'} \Gamma^{123} \epsilon(\tau)
   + \frac{i}{4} \com{X^i}{X^j} \Gamma^{ij} \epsilon(\tau),
\end{gather}
\end{subequations}%
where, as usual, $i=1\cdots9, a=1\cdots3$ and $a'=4\cdots9$.  Also,
\begin{equation}
\epsilon(\tau) = e^{-\frac{\mu}{12}\Gamma^{123} \tau} \epsilon_0,
\end{equation}
where $\epsilon_0$ is a constant, Majorana spinor.
\end{detail}%
\fi
The first part of this section will resemble ref.~\cite{sy} (see
also~\cite{hs}) and the latter is similar to~\cite{dvv}.

The first step toward making this a matrix string action is to make
the {\em field redefinition\/} ({\em cf.\/}~eq.~\eqref{eq:atwo})
\begin{subequations}
\begin{gather} \label{frX}
\begin{align} 
X^8 &= \hat{X}^8 \cos \frac{\mu}{6} \tau + \hat{X}^9 \sin \frac{\mu}{6} \tau,
&
X^9 &= -\hat{X}^8 \sin \frac{\mu}{6} \tau + \hat{X}^9 \cos \frac{\mu}{6} \tau
,
\end{align} \\ 
\Psi = e^{\frac{\mu}{12} \Gamma^{89} \tau} \Hat{\Psi}.\label{frP}
\end{gather}
\end{subequations}%
The field redefinition~\eqref{frX} constitutes part of the coordinate
transformation (the full coordinate transformation involves $X^-$,
which has already been eliminated via use of the infinite momentum
frame) to make $\frac{\p}{\p\Hat{X}^9}$ manifestly
Killing~\cite{jm1,sy}.
This, of
course, is just a $\tau$-dependent rotation of the $(X^8, X^9)$ plane,
and so motivates the additional redefinition~\eqref{frP}.  Then the
interaction term $\Psi^\transpose \Gamma^i \com{X^i}{\Psi}$ is invariant.
Upon dropping the hats, the action is
\begin{multline}
S = R \int d\tau \Tr \left\{
\frac{1}{2 R^2} (D_\tau X^i)^2
+ \frac{i}{R} \Psi^\transpose D_{\tau} \Psi
+ \Psi^\transpose \Gamma^i \com{X^i}{\Psi}
+ \frac{1}{4} \com{X^i}{X^j}^2
\right.\\* \left.
- \frac{1}{2} \bigl(\frac{\mu}{3 R}\bigr)^2 (X^a)^2
- \frac{1}{2} \bigl(\frac{\mu}{6 R}\bigr)^2 (X^{a''})^2
- \frac{\mu}{3 R^2} X^8 D_{\tau} X^9
- i \frac{\mu}{3 R} \epsilon_{abc} X^a X^b X^c
\right.\\* \left.
- i \frac{\mu}{4 R} \Psi^\transpose (\Gamma^{123} 
        -\frac{1}{3} \Gamma^{89}) \Psi
\right\},
\end{multline}
where we have dropped a total derivative term $\frac{\mu R}{3} \int d\tau
\Tr\left[D_{\tau} (X^8 X^9)\right]$.
We see that $X^9$ only appears differentiated in the action, and so we
can consider compactifying the $X^9$ direction to a circle of radius
$\hat{R}$.%
\footnote{The very observant reader will note that choosing to
subtract instead of add the total derivative would lead to $\hat{X}^8$
rather than $\hat{X}^9$ as the direction of the circle.  Equivalently, we
could have chosen the rotation in the field redefinition to
be in the opposite direction, and still have obtained an isometry
along $\hat{X}^9$.  At first sight, this is discomforting.
The latter statement is, in fact, true, though in the full
coordinate transformation it would require $X^- = \hat{X}^- +
\frac{\mu}{6} \hat{X}^8 \hat{X}^9$ rather than the $X^- = \hat{X}^- -
\frac{\mu}{6} \hat{X}^8 \hat{X}^9$ in
eq.~\eqref{eq:atwo}.  In other words, M-theory is parity invariant
($\hat{X}^9\rightarrow -\hat{X}^9$).}

We compactify and T-dualize simultaneously, in the usual way~\cite{wtl}.
This
means that we insert an $\int d\sigma$, with \hbox{$0\leq\sigma<2\pi$}~%
\iftoomuchdetail
\begin{detail}%
\footnote{DVV\cite{dvv} include a factor of $\frac{1}{\hat{R}}$ with the
integral;
however, Taylor\cite{wtl} only includes this factor
because his $\sigma$ has dimensions of length.  We are following
Taylor, but using DVV's $\sigma$ and therefore the factor cancels out
of the redefinition of $\sigma$.\label{ft:Wati}}
\end{detail}%
\fi
and replace
\begin{equation}
X^9 = -i \hat{R} D_\sigma,
\end{equation}
where $D_\sigma$ is the gauge covariant derivative.  This
implies that
\begin{equation}
D_\tau X^9 = \hat{R} F_{\tau \sigma}.
\end{equation}
\iftoomuchdetail
\begin{detail}%
The reason for the factor of $i \hat{R}$ in these expressions is slightly
involved.  With Taylor's conventions (footnote~\ref{ft:Wati}), it is
(fairly) clear that $X^9_{\text{W.T.}} = i \p_{\sigma_{\text{W.T.}}} + A$;
this is
essentially just T-duality.  However, if $\sigma$ has dimensions of
length, then this equation only makes sense if $X^9$ has dimensions of
mass; this is the familiar rescaling of $X^i$ by $\alpha'$
to convert mass dimension into length dimension.  Moreover, as
mentioned in the footnote,
Taylor's $\sigma$ runs over the physical circle, not the rescaled
circle.  Since we've T-dualized, this means that
$0\leq \sigma_{\text{W.T.}}< \apr/R_{9}$.  That is,
$\sigma_{\text{W.T.}} = -\frac{\apr}{\hat{R}} \sigma$, and so
\begin{equation}
X^9 = \apr X^9_{\text{W.T.}} = i \apr D_{\sigma_{\text{W.T.}}}
 = -i \apr \frac{\hat{R}}{\apr} D_{\sigma}.
\end{equation}
\end{detail}%
\fi

So now the action reads%
\iftoomuchdetail
\begin{detail}%
\footnote{The absence of a minus sign in the
  $F^2$ term of the action, is just the fact that we have explicitly
  written $F_{\tau\sigma}^2$ instead of
  $-F_{\alpha\beta}F^{\alpha\beta}$.  This also explains the numerical factor.}
\end{detail}%
\fi
\begin{multline}
S = R \int d\tau d\sigma \Tr \left\{
\frac{\hat{R}^2}{2 R^2} F_{\tau \sigma}^2
+ \frac{1}{2 R^2} (D_\tau X^i)^2 
- \frac{\hat{R}^2}{2} (D_\sigma X^i)^2
+ \frac{i}{R} \Psi^\transpose D_{\tau} \Psi
- i \hat{R} \Psi^\transpose \Gamma^9 D_\sigma \Psi
\right. \\* \left.
+ \Psi^\transpose \Gamma^i \com{X^i}{\Psi}
+ \frac{1}{4} \com{X^i}{X^j}^2
-\frac{1}{2} \bigl(\frac{\mu}{3 R}\bigr)^2 (X^a)^2
-\frac{1}{2} \bigl(\frac{\mu}{6 R}\bigr)^2 (X^{a''})^2
\right. \\* \left.
- \frac{\mu \hat{R}}{3 R^2} X^8 F_{\tau\sigma}
-i \frac{\mu}{4 R} \Psi^\transpose (\Gamma^{123}
        - \frac{1}{3} \Gamma^{89}) \Psi
- i \frac{\mu}{3 R} \epsilon_{abc} X^a X^b X^c
\right\}.
\end{multline}
The plan is now to put this in a canonical form.

To do this, rescale%
\iftoomuchdetail
\begin{detail}%
\ ({\em cf.\/}~DVV, but note that they do not
rescale $\tau$; Kim and Plefka~\cite{kp} do a different rescaling of
$\tau$ but
are only interested in eliminating $R$ from the M-theory action)
\end{detail}%
\fi
\begin{equation} \label{rescale}
\tau \rightarrow \frac{1}{\hat{R} R} \tau
\quad
X^i \rightarrow \frac{1}{\sqrt{\hat{R}}} X^i.
\end{equation}
After taking the transformation properties of $D_\tau$ and
$F_{\tau\sigma}$ into account and reinserting 
appropriate powers of $\ell_P$
(so as to ensure that the new $X^i$ is dimensionless, as are $\tau,\sigma$), 
we obtain
\begin{multline}
S = \int d\tau d\sigma \Tr \left\{
\frac{1}{2} \bigl(\frac{\hat{R}}{\ell_P}\bigr)^3 F_{\tau \sigma}^2
+ \frac{1}{2} (D_\tau X^i)^2 
- \frac{1}{2} (D_\sigma X^i)^2
+ i \Psi^\transpose D_{\tau} \Psi
- i \Psi^\transpose \Gamma^9 D_\sigma \Psi
\right. \\* \left.
+ \bigl(\frac{\ell_P}{\hat{R}}\bigr)^{3/2} 
            \Psi^\transpose \Gamma^i \com{X^i}{\Psi}
+ \frac{1}{4} \bigl(\frac{\ell_P}{\hat{R}}\bigr)^3 \com{X^i}{X^j}^2
-\frac{1}{2} \bigl(\frac{\mu \ell_P^2}{3 R}\bigr)^2
          \bigl(\frac{\ell_P}{\hat{R}}\bigr)^2 (X^a)^2
-\frac{1}{2} \bigl(\frac{\mu \ell_P^2}{6 R}\bigr)^2
          \bigl(\frac{\ell_P}{\hat{R}}\bigr)^2 (X^{a''})^2
\right. \\* \left.
- \frac{\mu \ell_P^2}{3 R} \sqrt{\frac{\hat{R}}{\ell_P}} X^8 F_{\tau\sigma}
-i\frac{\mu \ell_P^2}{4 R}\frac{\ell_P}{\hat{R}} \Psi^\transpose (\Gamma^{123}
        - \frac{1}{3} \Gamma^{89}) \Psi
- i \frac{\mu \ell_P^2}{3 R} \bigl(\frac{\ell_P}{\hat{R}}\bigr)^{5/2}
            \epsilon_{abc} X^a X^b X^c
\right\}.
\end{multline}

Observe that, with the 9/11 flip, all coefficients may be expressed 
in terms of the quantities
\begin{equation} \label{defM}
\frac{\mu \ell_P^2}{R} = \frac{\mu \apr}{R} g_s^{2/3} \equiv
M g_s^{2/3}, \quad \text{and} \quad
\frac{\hat{R}}{\ell_P} = g_s^{2/3} .
\end{equation}
(Note also that, as defined, $M$ is {\em dimensionless\/}.)
Our final expression is then
\begin{multline} \label{useS}
S = \int d\tau d\sigma \Tr \left\{
\frac{1}{2} g_s^2 F_{\tau \sigma}^2
+ \frac{1}{2} (D_\tau X^i)^2 
- \frac{1}{2} (D_\sigma X^i)^2
+ i\Psi^\transpose D_{\tau} \Psi
- i\Psi^\transpose \Gamma^9 D_\sigma \Psi
\right. \\* \left.
+ \frac{1}{g_s}
            \Psi^\transpose \Gamma^i \com{X^i}{\Psi}
+ \frac{1}{4 g_s^2} \com{X^i}{X^j}^2
-\frac{1}{2} \bigl(\frac{M}{3}\bigr)^2 (X^a)^2
-\frac{1}{2} \bigl(\frac{M}{6}\bigr)^2 (X^{a''})^2
\right. \\* \left.
- \frac{M}{3} g_s X^8 F_{\tau\sigma}
-i\frac{M}{4} \Psi^\transpose (\Gamma^{123} - \frac{1}{3} \Gamma^{89}) \Psi
- i \frac{M}{3 g_s} \epsilon_{abc} X^a X^b X^c
\right\}.
\end{multline}

Note that all powers of $g_s$ are integer, and that the mass terms
have no powers of $g_s$.  Finally, each power of $F_{\tau\sigma}$
carries a power of $g_s$.

\section{Properties of the Matrix String Action} \label{sec:prop}

\subsection{Supersymmetry and Vacua} \label{sec:soln} \label{sec:susy}

The Matrix quantum mechanics~\eqref{MQM} admits the
classical solutions%
\footnote{These are related to solutions which appeared
in~\cite{bmn}.  While they seemed to only consider either fuzzy
spheres ($X^8_0=0$) or rotating solutions ($J^a=0$), the BPS
conditions allow them to be
superposed under certain conditions (namely, $\com{X^8}{J^a}=0$).
Bak~\cite{db}, considered rotating fuzzy
spheres, but in that work some of the supersymmetries 
were broken by adding time
dependence to, and deforming, the $\su(2)$ generators, so as to
obtain rotating ellipsoids.  There are much simpler states with the
same supersymmetry as Bak's,
for which one adds an extra term to $X^a$ that commutes with the
$\su(2)$ generators, and which adds a rotation in a plane.  
These are the oscillations that we believe were misidentified
in~\cite{dsv1} as being associated with Bak's states.  The additional
rotations---that we just described as being in a plane that intersects
the fuzzy spheres---can also be done in other directions, such as the $X^8,
X^9$-directions that we use here.
\label{ft:baketc}}
\begin{subequations}
\begin{gather} \label{bmnrotfuzzy}
\begin{align}
X^a &= \frac{\mu}{3 R} J^a, &
X^8 &= X^8_0 \cos \frac{\mu}{6} \tau, &
X^9 &= -X^8_0 \sin \frac{\mu}{6} \tau,
\end{align} \\
\begin{align}
\com{J^a}{J^b} &= i \epsilon_{abc} J^c, &
\com{X^8_0}{J^a} &= 0,
\end{align}
\end{gather}
\end{subequations}
with all other fields vanishing.
Here $J^a$ is an $N$-dimensional representation of the $\su(2)$
algebra, and $X^8_0$ is a constant matrix.  When $J^a$ is nontrivial,
these describe a fuzzy sphere,
rotating in a circle in the $X^8$-$X^9$ plane.  When $X^8_0=0$ these
are fully supersymmetric vacua with vanishing vacuum energy.  For
$X^8_0 \neq 0$, these are half-supersymmetric, saturating a BPS
bound which relates the energy to the angular momentum in the
$X^8$-$X^9$ plane.

Under the reduction to the matrix string~\eqref{useS}, the
solution~\eqref{bmnrotfuzzy} becomes
\begin{align} \label{MSvacua}
X^a &= \frac{M g_s}{3} J^a, &
X^8 &= X^8_0, &
\com{J^a}{X^8_0} &=0,
\end{align}
with all other fields (such as $A_\sigma$) vanishing.
The ``rotation'' that is included in the
field-redefinition~\eqref{frX} modifies the Hamiltonian by an angular
momentum term~\cite{jm1}.  Also, the supersymmetries broken by the
angular momentum are precisely those that are broken by the
compactification.  Thus, in the matrix string theory,
all fuzzy spheres are fully supersymmetric and have vanishing
energy for all values of $X^8_0$.  The fact that $X^8_0$ must commute
with the $\su(2)$ generators implies that the fuzzy spheres have
definite position in the $8$-direction.  We will eventually see, and
explain, that despite this translation invariance, there is no
associated massless mode in the matrix string spectrum.

Explicitly,
it can be checked that the action~\eqref{useS} is invariant under the (4,4)
supersymmetries, ({\em cf.\/}~\cite{bmn,dsv1,hs} and {\em
e.g.\/}~\cite{bbn1} for related \ppwave\ and matrix string, supersymmetry
transformations)
\begin{subequations} \label{SUSY}
\begin{gather} \label{SUSYb}
\begin{align}
\delta X^i &= i\Psi^\transpose \Gamma^i \epsilon, &
\delta A_\tau &= \frac{i}{g_s} \Psi^\transpose \epsilon, &
\delta A_\sigma &= \frac{i}{g_s} \Psi^\transpose \Gamma^9 \epsilon,
\end{align} \\ \label{SUSYf}
\begin{split}
\delta \Psi &= \frac{1}{2} g_s F_{\tau\sigma} \Gamma^9 \epsilon
   + \frac{1}{2} D_\tau X^i \Gamma^i \epsilon
   - \frac{1}{2} D_\sigma X^i \Gamma^i \Gamma^9 \epsilon
   + \frac{M}{6} X^a \Gamma^a \Gamma^{123} \epsilon
   + \frac{i}{4 g_s} \com{X^i}{X^j} \Gamma^{ij} \epsilon
\\ & \qquad
   - \frac{M}{12} X^{a''} \Gamma^{a''123} \epsilon,
\end{split}
\end{gather}
\end{subequations}%
where $\epsilon$ obeys the constraint
\begin{equation} \label{8susy}
\epsilon = \Gamma^{12389} \epsilon  \Leftrightarrow 
\epsilon = \Gamma^{4567} \epsilon.
\end{equation}
The action is also invariant under the 16 nonlinearly realized
supersymmetries
\begin{equation} \label{nlSUSY}
\delta \Psi = e^{\frac{M}{4} (\Gamma^{123} - \frac{1}{3} \Gamma^{89}) \tau}
   \chi,
\end{equation}
where $\chi$ is an unconstrained Majorana spinor.  We will not discuss
the nonlinearly
realized supersymmetries further.

It is straightforward to check that the vacua~\eqref{MSvacua}
preserve all eight (linearly realized) supersymmetries.
Conversely, by linear independence of antisymmetrized products of
$\Gamma$-matrices (though in principle
it should have been necessary to take~\eqref{8susy} into account)
a fully supersymmetric background requires
\begin{equation}
\begin{gathered}
F_{\tau\sigma} = D_\tau X^i = D_\sigma X^i = X^{a''} =
\bigl[{X^a},{X^{a''}}\bigr] = \bigl[X^a,X^8\bigr] = \bigl[X^{a''},X^8\bigr] 
= \bigl[{X^{a''}},{X^{b''}}\bigr] = 0, \\
\com{X^a}{X^b} = i \frac{M g_s}{3} \epsilon_{abc} X^c.
\end{gathered}
\end{equation}
Note that, this does not imply $X^8=0$, but allows for a constant
value of $X^8$ that commutes with $J^a$.  This is precisely
eq.~\eqref{MSvacua}.
Moreover the $X^a$ generically
higgses the U($N$) gauge group
to U$(1)^r$, where $r$ is the number of irreducible representations
in $J^a$.  Then $X^8$ has $r$ components that can be turned on, and
this (roughly) induces $r$ ``$\theta$-angles'' for the corresponding U$(1)$s.

After a tedious calculation, one finds the on-shell algebra,
\begin{subequations} \label{s1s2}
\begin{align} \label{s1s2Xi}
\begin{split}
\com{\delta_1}{\delta_2} X^i &= 
i (\epsilon_1^\transpose \epsilon_2) D_\tau X^i
- i (\epsilon_1^\transpose \Gamma^9 \epsilon_2) D_\sigma X^i
+ i (\epsilon_1^\transpose \Gamma^a \epsilon_2) 
     \left\{-\frac{M}{3} \delta_b^i \epsilon_{abc} X^c 
         - \frac{i}{g_s} \com{X^a}{X^i} \right\}
\\ & \qquad
- i \frac{i}{g_s} (\epsilon_1^\transpose \Gamma^8 \epsilon_2)
     \com{X^8}{X^i}
+ i \frac{M}{6} (\epsilon_1^\transpose \Gamma^{a''b''89} \epsilon_2)
     \delta_{a''}^i X^{b''},
\end{split} \\ \label{s1s2At}
\com{\delta_1}{\delta_2} A_\tau &=
i (\epsilon_1^\transpose \Gamma^9 \epsilon_2) F_{\tau\sigma}
+ \frac{i}{g_s} (\epsilon_1^\transpose \Gamma^a \epsilon_2) 
     D_\tau X^a
+ \frac{i}{g_s} (\epsilon_1^\transpose \Gamma^8 \epsilon_2)
     D_\tau X^8, \\ \label{s1s2As}
\com{\delta_1}{\delta_2} A_\sigma &=
i (\epsilon_1^\transpose \epsilon_2) F_{\tau\sigma}
+ \frac{i}{g_s} (\epsilon_1^\transpose \Gamma^a \epsilon_2) 
     D_\sigma X^a
+ \frac{i}{g_s} (\epsilon_1^\transpose \Gamma^8 \epsilon_2)
     D_\sigma X^8, \\
\begin{split} \label{s1s2Psi}
\com{\delta_1}{\delta_2} \Psi &=
i (\epsilon_1^\transpose \epsilon_2) D_\tau \Psi
- i (\epsilon_1^\transpose \Gamma^9 \epsilon_2) D_\sigma \Psi
+ i (\epsilon_1^\transpose \Gamma^a \epsilon_2) 
     \left\{-\frac{M}{12} \epsilon_{abc} \Gamma^{bc} \Psi 
         - \frac{i}{g_s} \com{X^a}{\Psi} \right\}
\\ & \qquad
+ \frac{1}{g_s} (\epsilon_1^\transpose \Gamma^8 \epsilon_2)
     \com{X^8}{\Psi}
+ i \frac{M}{24} (\epsilon_1^\transpose \Gamma^{a''b''89} \epsilon_2)
     \Gamma^{a''b''} \Psi.
\end{split}
\end{align}
\end{subequations}%
We can summarize these equations as%
\begin{multline}
\label{QQ}
\anti{Q_\alpha}{Q_\beta} = 
\proj_{\alpha\beta} \bigl[H - G(A_\tau)\bigr]
+ (\Gamma^9\proj)_{\alpha\beta} \bigl[P + G(A_\sigma)\bigr]
- (\Gamma^a\proj)_{\alpha\beta} \bigl[\frac{M}{3} \ang^a 
       - \frac{1}{g_s} G(X^a)\bigr]
\\*
+ (\Gamma^8\proj)_{\alpha\beta} \bigl[-\frac{1}{g_s} G(X^8)\bigr]
+ \frac{M}{12} (\Gamma^{a''b''89}\proj)_{\alpha\beta} \ang_-^{a''b''};
\qquad
\proj = \frac{1}{2}\left(\one+\Gamma^{12389}\right),
\end{multline}
where $G(\Lambda)$ is the Hermitian generator of infinitesimal
gauge transformations with parameter $\Lambda$,%
\footnote{Explicitly,
$\com{G(\Lambda)}{A_\mu} = i \p_\mu \Lambda + \com{\Lambda}{A_\mu}$,
 and {\em e.g.\/}~$\com{G(\Lambda)}{X^i} = \com{\Lambda}{X^i}$.}
$H$ is the Hamiltonian, $P$ generates translations in $\sigma$, $\ang^a$
is the SU(2) rotational generator in the 123-directions and
$\ang_-^{a''b''}$ is the anti-selfdual SU(2) rotational generator in the
4567-directions.
The projection operator, $\proj$, enforces the constraint~\eqref{8susy}.
This explains why only $\ang_-^{a''b''}$ appears, and not the full SO(4), as
\begin{equation} \label{j-}
\Gamma^{a''b''89} \proj
= \frac{1}{2} \Gamma^{c''d''89}
  \Bigl(\delta_{a''c''}\delta_{b''d''}
                -\frac{1}{2} \epsilon_{a''b''c''d''}\Bigr),
\end{equation}
which projects out the self-dual SU(2).%
\footnote{Incidentally, eq.~\eqref{j-} is crucial for eliminating
extraneous terms that would otherwise have appeared in~\eqref{s1s2Psi}.}
Thus we see that---as for the 11-dimensional Matrix
Theory~\cite{dsv1}---the supersymmetry algebra contains angular
momentum.

Of course, it is well known (see {\em e.g.\/}~\cite{jl}) that the
supersymmetry algebra cannot contain angular momentum.  This follows
by first considering the $(P,P,Q)$ Jacobi identity, which (in flat space)
implies that $Q$ commutes with all momentum generators.  Then
the $(Q,Q,P)$ Jacobi identity implies that the supersymmetry
algebra cannot contain angular momentum.  However, in the
\ppwave\ background, the spacetime momentum
and boost/rotational generators are
replaced by Heisenberg generators which do not commute.  In
particular, the Hamiltonian does not commute with ``momentum''.  So, there
is no contradiction; indeed the anti-de Sitter superalgebra is a
well-known example in which the ``theorem'' is ``violated.''  Other
explicit examples of \ppwave\ supersymmetry algebras which contain angular
momentum in the anticommutation relations can be found in~\cite{jm2}.

Although so far we have isolated the gauge transformations in the
supersymmetry algebra from the usual generators, in the fuzzy sphere
background, the gauge transformations are crucial to rotational
invariance.  Let us define the operator
\begin{equation}
\hat{\ang}^a \equiv \ang^a - \frac{3}{M g_s} G(X_0^a),
\end{equation}
where $X_0^a$ is the background value of $X^a$, eq.~\eqref{MSvacua}.
It is clear that, for the fuzzy sphere background~\eqref{MSvacua}, $\hat{\ang}$
also obeys the \su(2) algebra,
\begin{equation}
\com{\hat{\ang}^a}{\hat{\ang}^b} = i \epsilon_{abc} \hat{\ang}^c.
\end{equation}
Moreover, it is a symmetry of the background:
\begin{equation}
\com{\hat{\ang}^a}{X_0^b} = \com{\ang^a}{X_0^b} - \com{J^a}{X_0^b}
= i\epsilon_{abc} X_0^c - i\epsilon_{abc} X_0^c = 0.
\end{equation}
Thus, we rewrite eq.~\eqref{QQ} as
\begin{multline} \label{useQQ}
\anti{Q_\alpha}{Q_\beta} = 
\proj_{\alpha\beta} \bigl[H - G(A_\tau)\bigr]
+ (\Gamma^9\proj)_{\alpha\beta} \bigl[P + G(A_\sigma)\bigr]
- \frac{M}{3} (\Gamma^a\proj)_{\alpha\beta} 
      \bigl[\hat{\ang}^a - \frac{3}{M g_s} G(X^a-X^a_0)\bigr]
\\*
+ (\Gamma^8\proj)_{\alpha\beta} \bigl[-\frac{1}{g_s} G(X^8)\bigr]
+ \frac{M}{12} (\Gamma^{a''b''89}\proj)_{\alpha\beta} \ang_-^{a''b''}.
\end{multline}
From the algebra~\eqref{useQQ}, we see that there are short BPS
multiplets of mass $m=\frac{M}{3} n$, with $n$ an integer.
More precisely, multiplets with quantum numbers $(\ell_1,\ell_2)$ under
$(\hat{\ang}^a,\ang^{a''b''}_-)$ can be short if the mass obeys
one of
\begin{equation} \label{ultraBPS}
\text{mass}^2 = \begin{cases} \bigl(\frac{M}{3}\bigr)^2 (\ell_1+\ell_2)^2, \\
   \bigl(\frac{M}{3}\bigr)^2 (\ell_1-\ell_2+1)^2, \\
   \bigl(\frac{M}{3}\bigr)^2 (\ell_1-\ell_2-1)^2, \\
   \bigl(\frac{M}{3}\bigr)^2 (\ell_1+\ell_2+2)^2. \end{cases}
\end{equation}
This follows from eq.~\eqref{JonSjlm}%
\iftoomuchdetail;\else.\fi%
\footnote{For the anti-selfdual rotations, this follows by decomposing
the four-dimensional spinors into two-dimensional Weyl spinors.  Then
the two-dimensional spinors can be decomposed in terms of spinor
spherical harmonics, and eq.~\eqref{JonSjlm} is easily applied.}
\iftoomuchdetail
\begin{detail}%
the removal of a factor of 2
for $\ell_2$ follows from exchanging $a''\leftrightarrow b''$ or
replacing the pair by
$(c'',d'')$, with $\epsilon_{a''b''c''d''}=1$.%
\end{detail}%
\fi\ 
The result~\eqref{ultraBPS}
agrees with the spectrum we give in Sec.~\ref{sec:spectrum}.
Namely, we have four short
multiplets: two with
\hbox{$\ell_1 = \ell$}, \hbox{$\ell_2=0$}, and two with \hbox{$\ell_1=\ell$},
\hbox{$\ell_2=\frac{1}{2}$}.
It is interesting---though required by counting of supermultiplet
components---that the \hbox{$X^{a''}\in (1/2,1/2)$~of~SO(4)} form two
degenerate multiplets, and not a longer multiplet.

It is tempting to speculate that the other gauge transformations in
the action are related to other BPS conditions.  The $G(A_\mu)$ terms are
simply the gauge completions of the Hamiltonian
($i\frac{d}{d\tau}$) and the momentum generator ($P=-i
\frac{d}{d\sigma}$), but can play an important role if there are
Wilson lines.  The presence of
$G(X^8)$, and the fact that there are vacua with
non-trivial values of $X^8$, is
particularly intriguing.

\subsection{Perturbation Theory and Fuzzy Spheres} \label{sec:gM}

In eq.~\eqref{useS}, we wrote down an action in which the fields
were scaled in such a way that the matrix string action had mass
parameters where expected, and coupling constant dependence that is
familiar from the flat-space matrix string.
There is another useful rescaling of the fields and coordinates, which
isolates the combination $M g_s$.  Specifically, let
\begin{align}
\tau &= \frac{\hat{\tau}}{M}, &
\sigma &= \frac{\hat{\sigma}}{M}, &
A_\mu &= M \hat{A}_\mu, &
X^i &= M g_s \hat{X}^i, &
\Psi &= M^{\frac{3}{2}} g_s \hat{\Psi},
\end{align}
Then,
\begin{multline} \label{MgS}
S = (M g_s)^2 \int d\hat{\tau} d\hat{\sigma} \Tr \left\{
\frac{1}{2} \hat{F}_{\hat{\tau} \hat{\sigma}}^2
+ \frac{1}{2} (\hat{D}_{\hat{\tau}} \hat{X}^i)^2 
- \frac{1}{2} (\hat{D}_{\hat{\sigma}} \hat{X}^i)^2
+ i\hat{\Psi}^\transpose \hat{D}_{\hat{\tau}} \hat{\Psi}
- i\hat{\Psi}^\transpose \Gamma^9 \hat{D}_{\hat{\sigma}} \hat{\Psi}
\right. \\* \left.
+ \Hat{\Psi}^\transpose \Gamma^i \com{\hat{X}^i}{\hat{\Psi}}
+ \frac{1}{4} \com{\Hat{X}^i}{\Hat{X}^j}^2
-\frac{1}{18} (\Hat{X}^a)^2
-\frac{1}{72} (\Hat{X}^{a''})^2
- \frac{1}{3} \Hat{X}^8 \Hat{F}_{\hat{\tau}\hat{\sigma}}
\right. \\* \left.
-i\frac{1}{4} \Hat{\Psi}^\transpose (\Gamma^{123} - \frac{1}{3}
    \Gamma^{89}) \Hat{\Psi}
- i \frac{1}{3} \epsilon_{abc} \Hat{X}^a \Hat{X}^b \Hat{X}^c
\right\},
\end{multline}
where
$\Hat{D}_\mu = \hat{\p}_\mu + i [{\hat{A}_{\mu}},{\cdot}]$ and
$\hat{F} = d\hat{A} + i \hat{A}\wedge \hat{A}$.
In this expression $M g_s$ has been completely factored out of the
action, which is otherwise independent of both $M$ and $g_s$.
Thus, the dimensionless $M g_s$ acts as an inverse coupling constant,
and perturbation theory in $(M g_s)^{-1}$ is valid.

Since, for the validity of the matrix string theory, we should
take $g_s$ small, we
take $M \gg g_s^{-1}$ for the perturbative realm.  Alternatively, the regime
of $M g_s \ll 1$ requires a strong coupling expansion.

\subsubsection{Strong Coupling and Perturbative IIA Strings} \label{sec:pert}
For $M g_s \ll 1$, the Yang-Mills theory is at strong coupling, and we
expect to be able to compare to type IIA perturbation theory.  In
fact, type IIA perturbation theory is an expansion around $g_s=0$, and
is therefore strictly at $M g_s=0$.  For this value, there are no
fuzzy spheres; the fuzzy sphere radii are strictly zero and the
background values are $X^a = \frac{M g_s}{3} J^a = 0$.

Moreover, in the limit $M g_s \to 0$, finiteness of the energy
(derived from~\eqref{useS}) sets the commutator terms in the
action to zero~\cite{hms,dvv}.  Thus one is left with a quadratic (but
massive) action for the elements of the Cartan subalgebra---that is,
$N$ copies of the IIA string on this background.

There are actually some subtleties with this discussion.
Let us consider
the action for a single element of the Cartan subalgebra.  It is
\begin{multline} \label{cms}
S = (M g_s)^2 \int d\hat{\tau} d\hat{\sigma} \Tr \left\{
\frac{1}{2} \left(
   \hat{F}_{\hat{\tau} \hat{\sigma}} - \frac{1}{3} \Hat{X}^8
   \right)^2
+ \frac{1}{2} (\Dot{\Hat{X}}^i)^2 
- \frac{1}{2} ({\Hat{X}}{'}{^i})^2
+ i\Hat{\Psi}^\transpose \Dot{\Hat{\Psi}}
- i\Hat{\Psi}^\transpose \Gamma^9 \Hat{\Psi}'
\right. \\* \left.
-\frac{1}{18} (\Hat{X}^a)^2
-\frac{1}{18} (\Hat{X}^8)^2
-\frac{1}{72} (\Hat{X}^{a''})^2
-i\frac{1}{4} \Hat{\Psi}^\transpose (\Gamma^{123} - \frac{1}{3}
    \Gamma^{89}) \Hat{\Psi}
\right\}.
\end{multline}
If we ignore the first term---this is justified in flat space for standard
$D=2$ gauge-theoretic reasons~\cite{bs,dvv}---then we have
written the
action for the IIA perturbative string, albeit with nonstandard
normalizations.
Equivalently, we could treat
$\hat{F}_{\hat{\tau}\hat{\sigma}}$ as an auxiliary field and integrate
it out. However, neither procedure can be justified.  

Instead, we should
integrate out the gauge potential.  
In order to do this, we introduce an auxiliary scalar field $\phi$, 
and replace the term $\frac{1}{2}\hat{F}^2_{\hat{\tau}\hat{\sigma}}$ 
inside the brackets in~\eqref{cms} by 
\begin{equation} \label{replacephi} 
-\frac{1}{2} \phi^2 + \phi \hat{F}_{\hat{\tau}\hat{\sigma}}.
\end{equation}
The resulting action is equivalent to~\eqref{cms}, 
but now the gauge potential appears only linearly.  
Varying $\hat{A}_{\hat{\mu}}$ gives the constraint
\begin{equation} \label{constraintA}
\partial_{\hat{\mu}} (\phi - \frac{1}{3}\Hat{X}^8)=0,
\end{equation}
which together with the subsequent equation of motion for $\phi$ 
tells us that $\phi=\frac{1}{3}(\Hat{X}^8 -  \Hat{X}^8_0)$. 
Substituting this for $\phi$ in~\eqref{replacephi} gives 
the equivalent action 
\begin{multline} \label{cmsphi}
S = (M g_s)^2 \int d\hat{\tau} d\hat{\sigma} \Tr \left\{
\frac{1}{2} (\Dot{\Hat{X}}^i)^2 
- \frac{1}{2} ({\Hat{X}}{'}{^i})^2
+ i\Hat{\Psi}^\transpose \Dot{\Hat{\Psi}}
- i\Hat{\Psi}^\transpose \Gamma^9 \Hat{\Psi}'
\right. \\* \left.
-\frac{1}{18} (\Hat{X}^a)^2
-\frac{1}{18} ( {\Hat X}^8 - \Hat{X}^8_0 )^2
-\frac{1}{72} (\Hat{X}^{a''})^2
-i\frac{1}{4} \Hat{\Psi}^\transpose (\Gamma^{123} - \frac{1}{3}
    \Gamma^{89}) \Hat{\Psi}
\right\}.
\end{multline}

The result---consistent with eq.~\eqref{MSvacua}---is
that $\Hat{X}^8$ can take any constant value $\Hat{X}^8_0$.  
Clearly, the spectrum is independent of
$\Hat{X}^8_0$.  
Thus, we do obtain the IIA string;
however, there are an infinite number of choices, with identical
physics, parametrized in the matrix string theory by
$\Hat{X}^8_0$, although this value is invisible to the
perturbative string theory.
Furthermore, fluctuations of $\Hat{X}^8$
about this value are massive; the spectrum contains no Nambu-Goldstone
boson.

This perhaps surprising result is reminiscent of the bosonized massive
Schwinger model~\cite{ks}.  More generally, recall that a
massless scalar field $\varphi$ in $D$ dimensions has an infinite number of
vacua parametrized by the vacuum expectation value of $\varphi$.  This
is just spontaneous symmetry breaking, where $\varphi$ is its own
Nambu-Goldstone boson, and the vacuum expectation value characterizes a
superselection sector (for $D>2$).  Of course, for $D=2$, the typical
story is that fluctuations of the Nambu-Goldstone boson permeate space
and destroy the spontaneous symmetry breaking.

However, spontaneous symmetry breaking does occur in the
two-dimensional
bosonized massive Schwinger model~\cite{ks}.  For identical reasons,
spontaneous symmetry
breaking via the background value of $\hat{X}^8$ also occurs in
the action~\eqref{cms}.  The
symmetry which is
broken by the vacuum expectation value for $\Hat{X}^8$ is (classically)
anomalous.  More precisely, the conserved current is
\begin{equation}
J_\mu = \p_\mu \Hat{X}^8 + \frac{1}{6} \epsilon_\mu{^\nu} A_{\nu},
\end{equation}
which is gauge-variant.  So, there is no (physical) Nambu-Goldstone
boson in the spectrum to mix up the vacua, and the vacuum
expectation value of
$\Hat{X}^8$ characterizes a superselection sector of otherwise
identical vacua.  The eventual Type IIA Lagrangian is independent of
$X^8_0$, although the IIA string obtained is only one element of an
infinite family.

Nevertheless,
since $\Hat{X}^8$ has a target space interpretation, we should expect
that, in a partition function, we should sum over all vacua, which
means integrating over the superselection sectors, thereby obtaining
a factor of the coordinate length of $\Hat{X}^8$.  This does not
appear to be visible in perturbation theory.

We expect that an argument parallel to those in~\cite{gs} should yield
a thermodynamic behaviour identical to that for the perturbative IIA
string on this background, and we further expect that thermodynamics to
be qualitatively identical to that already worked out~\cite{gspp} for the IIB
string on the maximally supersymmetric \ppwave.  Indeed the recent
work~\cite{hpy} explicitly gave the expected thermodynamics of the IIA
string.  For earlier work on
\ppwave\ thermodynamics, see~\cite{gss,pv,s1,s2}.

\subsubsection{Weak Coupling and Fuzzy Spheres} \label{sec:strong}

If $M g_s$ is large, then no degrees of
freedom are suppressed.  Instead, [in the original
variables~\eqref{useS}] the fuzzy sphere vacua
\begin{equation} \label{firstfuzzysphere}
\com{X^a}{X^b} = i \frac{M g_s}{3} \epsilon_{abc} X^c,
\end{equation}
solve the equations of motion.
This is solved by
\begin{equation} \label{firstJ}
X^a = \frac{M g_s}{3} J^a,
\end{equation}
where $J^a$ is an $N$-dimensional representation of \su(2).  
One might expect that since no degrees of freedom are suppressed, 
the weakly-coupled matrix model could display
a drastically different thermodynamics compared to $M g_s \ll 1$.
In fact, this appears not to be the case,
as is further discussed in Sec.~\ref{sec:thermo}.

In Sec.~\ref{sec:spectrum}, we study the
fluctuations around these vacua.

\section{The Matrix String Spectrum} \label{sec:spectrum}

In this section we compute the perturbative spectrum of the
matrix string theory.  In Sec.~\ref{sec:bosspec}, we give the
bosonic spectrum about an irreducible vacuum; the corresponding
fermionic spectrum is degenerate, as shown in
Sec.~\ref{sec:fermspec}.  General vacua are discussed in
Sec.~\ref{sec:redfluc}.  Nontrivial boundary conditions are applied
and discussed in Sec.~\ref{sec:long}, but to facilitate the
discussion, we start in Sec.~\ref{sec:bdy} with a discussion of
boundary conditions in a general gauge theory.

\subsection{Boundary Conditions in Gauge Theory} \label{sec:bdy}

The most general boundary condition one can imagine writing for a
U($N$) gauge theory on a circle%
\footnote{This is easily extended to a general topology in an
  arbitrary dimension
  following~\cite{hos}.  However, though ref.~\cite{hos} claims that
  homogeneity of the space restricts the boundary conditions to be
  independent of position, we do not assume that here.  
  In any case, we will shortly show that our boundary conditions 
  can be taken to be constant.  Ref.~\cite{hos}
  also allows for the matter fields to change by a constant phase after going
  around a nontrivial cycle; however, except for a sign this is
  incompatible with hermiticity of an adjoint field, and the sign is
  generically not compatible with the symmetries of the action.
  Of course, we could, more generally, consider {\em
  e.g.\/}~\hbox{$X^{a''}(\sigma+2\pi) = {\mathcal O}^{a''}{_{b''}} U
  X^{b''}(\sigma) U^{-1}$} with ${\mathcal O}$ a constant SO(4)
  matrix.  Such boundary conditions appear in twisted sectors of \ppwave\ 
  {\em orbifolds\/}.}
only requires the fields to return to themselves up to a gauge
transformation $U(\tau,\sigma)$,
\begin{equation} \label{genbdy}
\begin{aligned}
A_\mu(\tau,\sigma+2\pi) &= U(\tau,\sigma) A_\mu(\tau,\sigma) U(\tau,\sigma)^{-1}
  - i U(\tau,\sigma) \p_\mu U(\tau,\sigma)^{-1}, \\
X^i(\tau,\sigma+2\pi) &= U(\tau,\sigma) X^i(\tau,\sigma)
U(\tau,\sigma)^{-1}.
\end{aligned}
\end{equation}
Of course, a (not necessarily periodic) gauge transformation
$\Omega$ need not preserve these boundary
conditions.  The boundary conditions of the fields
\begin{equation} \label{gengt}
\begin{aligned}
A'_\mu &= \Omega A_\mu \Omega^{-1} - i \Omega \p_\mu \Omega^{-1}, \\
{X'}{^i} &= \Omega X^i \Omega^{-1},
\end{aligned}
\end{equation}
are of the form~\eqref{genbdy}, but with $U$ replaced by $U'$ where
\begin{equation} \label{gengtbdy}
U'(\tau,\sigma) = \Omega(\tau,\sigma+2\pi) U(\tau,\sigma)
   \Omega(\tau,\sigma)^{-1}.
\end{equation}
\iftoomuchdetail
\begin{detail}%
Explicitly,
\begin{equation}
\begin{split}
A'_\mu(\tau,\sigma+2\pi) 
&= \Omega(\tau,\sigma+2\pi) A_\mu(\tau,\sigma+2\pi)
      \Omega^{-1}(\tau,\sigma+2\pi) 
   - i \Omega(\tau,\sigma+2\pi) \p_\mu \Omega^{-1}(\tau,\sigma+2\pi) \\
&= \Omega(\tau,\sigma+2\pi) U(\tau,\sigma) A_\mu(\tau,\sigma)
      U(\tau,\sigma)^{-1} \Omega(\tau,\sigma+2\pi)^{-1}
\\ & \qquad
   -i \Omega(\tau,\sigma+2\pi) U(\tau,\sigma) \p_\mu U(\tau,\sigma)^{-1}
      \Omega(\tau,\sigma+2\pi)^{-1}
\\ & \qquad
   - i \Omega(\tau,\sigma+2\pi) \p_\mu \Omega^{-1}(\tau,\sigma+2\pi) \\
&= \Omega(\tau,\sigma+2\pi) U(\tau,\sigma) \Omega(\tau,\sigma)^{-1}
    A'_\mu(\tau,\sigma) \Omega(\tau,\sigma) U(\tau,\sigma)^{-1} 
   \Omega(\tau,\sigma+2\pi)^{-1} 
\\ & \qquad
   + i \Omega(\tau,\sigma+2\pi) U(\tau,\sigma) \p_\mu \Omega(\tau,\sigma)^{-1}
       \Omega(\tau,\sigma) U(\tau,\sigma)^{-1}  \Omega(\tau,\sigma+2\pi)^{-1}
\\ & \qquad
   -i \Omega(\tau,\sigma+2\pi) U(\tau,\sigma) \p_\mu \left[U(\tau,\sigma)^{-1}
      \Omega(\tau,\sigma+2\pi)^{-1}\right] \\
&= \left[\Omega(\tau,\sigma+2\pi) U(\tau,\sigma)
         \Omega(\tau,\sigma)^{-1}\right] A'_\mu(\tau,\sigma)
   \left[\Omega(\tau,\sigma+2\pi) U(\tau,\sigma) 
         \Omega(\tau,\sigma)^{-1}\right]^{-1}
\\ & \qquad
  -i \left[\Omega(\tau,\sigma+2\pi) U(\tau,\sigma) 
           \Omega(\tau,\sigma)^{-1}\right] \p_\mu
     \left[\Omega(\tau,\sigma+2\pi) U(\tau,\sigma) 
           \Omega(\tau,\sigma)^{-1}\right]^{-1}.
\end{split}
\end{equation}
\end{detail}%
\fi
Thus, the {\em field redefinition\/}~\eqref{gengt} preserves the
action, and changes the boundary conditions according to
eq.~\eqref{gengtbdy}.

Also, eq.~\eqref{gengtbdy} tells us that the unbroken gauge
{\em symmetries\/} ({\em i.e.,\/}  gauge transformations which are periodic), 
for a given boundary condition $U(\tau,\sigma)$,
consist of those $\Omega(\tau,\sigma)$ which commute with $U$.  This
follows since a gauge symmetry is necessarily single-valued.

Now, following~\cite{hos}, suppose the vacuum is such that
$F_{\mu\nu}=0$.  Then,
there is a unitary matrix $V(\tau,\sigma)$ (not necessarily periodic) such that
\begin{equation} \label{pg}
\vev{A_\mu} = -i V^{-1} \p_\mu V.
\end{equation}
Here, $\vev{A_\mu}$ simply means the background, or vacuum, value for $A_\mu$.
For a gauge field of the form~\eqref{pg}, the boundary
condition~\eqref{genbdy} is
equivalent to
\iftoomuchdetail
\begin{detail}%
\begin{equation}
-i V(\tau,\sigma+2\pi)^{-1} \p_\mu V(\tau,\sigma+2\pi)
 = -i U(\tau,\sigma) V(\tau,\sigma)^{-1} \p_\mu V(\tau,\sigma)
      U(\tau,\sigma)^{-1}
   - i U(\tau,\sigma) \p_\mu U(\tau,\sigma)^{-1},
\end{equation}
which we can rewrite as ({\em cf.\/}~\cite[eq.~(3.2)]{hos})
\begin{equation} \label{flatAbdy}
V(\tau,\sigma+2\pi)^{-1} \p_\mu V(\tau,\sigma+2\pi) U(\tau,\sigma) 
 = U(\tau,\sigma) V(\tau,\sigma)^{-1} \p_\mu V(\tau,\sigma)
   - \p_\mu U(\tau,\sigma),
\end{equation}
or better yet,
\begin{equation}
V(\tau,\sigma) U(\tau,\sigma)^{-1} V(\tau,\sigma+2\pi)^{-1}
\p_\mu \left[V(\tau,\sigma+2\pi) U(\tau,\sigma) V(\tau,\sigma)^{-1} \right]
= 0,
\end{equation}
which implies,
\end{detail}%
\fi
\begin{equation} \label{constup}
\p_\mu \left[V(\tau,\sigma+2\pi) U(\tau,\sigma) V(\tau,\sigma)^{-1}\right]
= 0.
\end{equation}
Under the field redefinition~\eqref{gengt} with~$\Omega=V$,
\begin{equation} \label{gotusebdy}
\begin{aligned}
\vev{A'_\mu} &= 0, \\
U' &= V(\tau,\sigma+2\pi) U(\tau,\sigma) V(\tau,\sigma)^{-1}.
\end{aligned}
\end{equation}
Observe that, because of eq.~\eqref{constup}, $U'$ is {\em
constant\/}.  
Moreover, both the vanishing gauge field and the
constant boundary conditions are clearly preserved by constant
(or global) ``gauge'' transformations.

To summarize, it follows that, without loss of generality, one can take
$\vev{A_\mu}=0$ and boundary conditions
\begin{equation} \label{usebdy}
\begin{aligned}
A_\mu(\tau,\sigma+2\pi) &= U A_\mu(\tau,\sigma) U^{-1}, &
X^i(\tau,\sigma+2\pi) &= U X^i(\tau,\sigma) U^{-1},
\end{aligned}
\end{equation}
where $U$ is a {\em constant\/} matrix in the gauge group.  Moreover,
for any constant $\Omega$ in the gauge group, $U$ and $\Omega U
\Omega^{-1}$ are equivalent.   Thus, the set of boundary conditions
consist of conjugacy classes of the gauge group.  For U$(N)$, this is
the maximal torus $U(1)^N/S_N$, where $S_N$ is the group of
permutations on the $N$ eigenvalues of the unitary matrix.

Thus, without loss of generality, we can take $\vev{A_\mu}=0$ and
boundary conditions
\begin{equation} \label{Uexpl}
U = \begin{pmatrix} e^{i \theta_1} \\ & e^{i \theta_2} \\ & & \ddots
    \end{pmatrix}
\end{equation}
where the $\theta_i$'s are constant and ordered.  Note that this
explicit form fixes the gauge.

Conversely, suppose we have $\vev{A_\mu}=0$ and a boundary condition
given by a constant $U$, possibly---though this is not necessary for
the following result---of the form~\eqref{Uexpl}.  Under the
field-redefinition~\eqref{gengt}, with unitary $\Omega =
e^{\frac{\sigma}{2\pi}\ln U^{-1}}$, the theory is equivalently
described by
\begin{equation} \label{wilson}
\begin{aligned}
\vev{A'_\sigma} &= -\frac{i}{2\pi} \ln U^{-1}, \\
U' &= \one.
\end{aligned}
\end{equation}
Therefore, a nontrivial boundary condition but vanishing gauge field
is equivalent to a
nontrivial (constant) Wilson line along the circle, and periodic
boundary conditions.  In particular, this shows that, in the path
integral, fixing the
boundary conditions and summing over
Wilson lines is equivalent to fixing $\vev{A_\mu}=0$ and summing over
boundary conditions.%
\footnote{This also explains the empirical observation of~\cite{gs}
that (flat space) IIA thermodynamics is reproduced by matrix string
theory under the prescription that boundary conditions are summed with
unit weight.}
This point has been previously emphasized by~\cite{ah,tw1,tw2}.

For example, consider strongly coupled $d{=}2,\ {\mathcal N}{=}(8,8)$ super
Yang-Mills
theory---that is, flat-space matrix string theory.  
We start with the fact that for the
strongly-coupled two-dimensional super Yang-Mills, the gauge field
kinetic term is irrelevant~\cite{bjsv} and, at each point, the matter fields
live in a Cartan subalgebra, $\com{X^i}{X^j}=0$~\cite{hms,bjsv}. 
Because the gauge field
kinetic term is
an irrelevant operator, the gauge field equation of motion---or equivalently,
the constraint equation for having gone to $\vev{A_\mu}=0$ gauge---is
\begin{equation} \label{cur}
\com{X^i}{D_\mu X^i} = 0.
\end{equation}
This is just the vanishing of the charge current~\cite{hms}. 

Since this equation involves a sum over $i$,  we cannot immediately
conclude that the Cartan subalgebra does not rotate.  That is, just
because the $X^i$ commute at every point does not (immediately) imply
that $X^i\,$'s at different points commute.
However,
eq.~\eqref{cur} implies that
\begin{equation}
\begin{split}
0 &= \Tr \com{X^i}{D_\mu X^i} \com{X^j}{D_\mu X^j}, \\
&= \Tr \com{D_\mu X^j}{X^i}\com{D_\mu X^i}{X^j}
     + \Tr \com{X^i}{X^j}\com{D_\mu X^i}{D_\mu X^j}, \\
&= \Tr \com{X^i}{D_\mu X^j}^2.
\end{split}
\end{equation}
There need not be a sum over $\mu$ here.  In the first step the Jacobi
identity with cyclicity of the trace was used; since, at
each $(\tau,\sigma)$, the $X^i$ live in a Cartan subalgebra, the
second term vanishes. This also implies that \hbox{$\com{D_\mu X^i}{X^j} =
D_\mu\com{X^i}{X^j} - \com{X^i}{D_\mu X^j} = -\com{X^i}{D_\mu X^j}$},
which was used in the second step.  Since the result is (minus) a sum
of squares, each term in the sum must separately vanish, and since
U$(N)$ is compact we can
conclude that
\begin{equation} \label{cartanconst}
\com{X^i}{D_\mu X^j} = 0.
\end{equation}

Let us now take, without loss of generality as shown in the discussion
which led to eq.~\eqref{gotusebdy}, $\vev{A_\mu}=0$ and a
constant boundary condition $U$ of the form~\eqref{Uexpl}.
Although the $X^i$ can be diagonalized, it is not
generically true that the boundary condition $U$ and the fields $X^i$
can be simultaneously diagonalized.  What we do know is that there is
a gauge in which $\vev{A_\mu} = 0$, and for which the boundary
conditions are constant.  We also know that the degrees of freedom in
the $X^i$ are constrained
to a Cartan subalgebra~\cite{hms,bjsv}, and by
eq.~\eqref{cartanconst}, this Cartan subalgebra does not
rotate.  If this Cartan subalgebra can be
simultaneously diagonalized with the boundary condition $U$, then this
is the sector of short strings.  In particular, the $U(1)^N$ commutes
with the Cartan subalgebra, and so the maximal torus of boundary
conditions is invisible.  Otherwise, consistency demands that
the matrix $U$ permutes the elements of the Cartan subalgebra; this
leads to the sectors of long strings.  So, for flat-space matrix
string theory, we reproduce the result that
there are $N!$ (the number of elements of $S_N$)
sectors of boundary conditions, each of which contains, in principle,
a maximal torus of boundary conditions, but this $U(1)^N$ acts
trivially on the restricted fields, and so can be neglected.

It is useful to see all this explicitly for the case $N=2$. The usual
description is to take the fields $X^i$ to be diagonal and a vanishing
Wilson line. Let us denote any one of these fields by $\Phi$. Then
\begin{equation}\label{utwoone}
\Phi = \begin{pmatrix}\phi_1 \\ & \phi_2
\end{pmatrix}
\end{equation}
For the trivial boundary condition, the fields $\phi_1$ and $\phi_2$
are periodic, while for the twisted boundary condition one has
\begin{equation}\label{utwotwo}
\phi_1(\tau,\sigma + 2\pi) = \phi_2 (\tau,\sigma) \qquad
\phi_2(\tau,\sigma + 2\pi) = \phi_1 (\tau,\sigma)
\end{equation}
This means
\begin{equation}\label{utwothree}
\Phi (\tau,\sigma + 2\pi) = U \Phi (\tau,\sigma) U^\dagger
\end{equation}
where
\begin{equation}\label{utwofour}
U = \begin{pmatrix} 0  & 1 \\ 1 & 0 \end{pmatrix}
\end{equation}
By a global gauge rotation one may bring the field $\Phi$ to the form
\begin{equation}\label{utwofive}
\Phi = \frac{1}{2}\begin{pmatrix} \phi_1 + \phi_2 & \phi_2 - \phi_1 \\
\phi_2 - \phi_1 & \phi_1 + \phi_2 \end{pmatrix}
\end{equation}
As argued above, the general form of the matrix which implements
boundary conditions is now given by
\begin{equation}\label{utwosix}
U' = \begin{pmatrix} e^{i\theta_1}  \\ & e^{i\theta_2} \end{pmatrix}
\end{equation}
which means that $\phi_1{+} \phi_2$ is periodic while $\phi_2{-}\,\phi_1$
acquires a phase $e^{i(\theta_1 -\, \theta_2)}$. However since both the
fields $\phi_1, \phi_2$ are real the only nontrivial value of
$\theta_1 {-}\, \theta_2$ which is allowed is $\pi$. This gives the
twisted boundary condition above.

It is clear, however, that acting on a general $2 \times 2$ hermitian
matrix the boundary condition $U'$ in (\ref{utwosix}) puts in
arbitrary phases in the off-diagonal elements.

\subsection{The Bosonic Spectrum in the Irreducible Vacuum} \label{sec:bosspec}

For now we will ignore the fermions.
Starting with the matrix string action~\eqref{useS},
and completing a square gives ($a=1,2,3$, $a''=4,5,6,7$)
\begin{multline} \label{bms}
S = \int d\tau d\sigma \Tr \left\{
\frac{1}{2} \biggl(g_s F_{\tau \sigma} - \frac{\mass}{3} X^8\biggr)^2
+ \frac{1}{2} (D_\tau X^i)^2 
- \frac{1}{2} (D_\sigma X^i)^2
-\frac{1}{2} \bigl(\frac{\mass}{6}\bigr)^2 (X^{a''})^2
\right. \\* \left.
- \frac{1}{2} \bigl(\frac{\mass}{3}\bigr)^2 (X^8)^2
- \frac{1}{2 g_s^2} \left(\frac{\mass g_s}{3} X^a 
     + \frac{i}{2} \epsilon_{abc} \com{X^b}{X^c}\right)^2
+ \frac{1}{2 g_s^2} \bigl[{X^a},{X^{a''}}\bigr]^2
+ \frac{1}{2 g_s^2} \bigl[{X^a},{X^8}\bigr]^2
\right. \\* \left.
+ \frac{1}{2 g_s^2} \bigl[{X^{a''}},{X^8}\bigr]^2
+ \frac{1}{4 g_s^2} \bigl[{X^{a''}},{X^{b''}}\bigr]^2
\right\}.
\end{multline}
One might expect to be able to set the first term to zero, by treating
$F_{\tau\sigma}$ as an auxiliary field.\cite{sy}  However,
as in Sec.~\ref{sec:pert}, we will be more
careful.  In particular, we need to take account of the gauge field in
the covariant derivatives.  This would not be an issue (or at least
not a serious issue) were $M=0$, as then the equation of motion would
(essentially) set $A_\mu=0$.  However, here as the na\"{\i}ve equation
of motion sets $F_{\tau\sigma} = \frac{\mass}{3} X^8$, and therefore turns
on the gauge field, we will immediately have an interaction between
the gauge field and $X^i$ in the $X^i$ kinetic term.

It is convenient to decompose the fields into a sum over a basis
of U$(N)$ matrices.  For now, we will only treat the irreducible
vacuum---that is, the vacuum~\eqref{firstJ} for which $J^a$ generate
an irreducible representation of \su(2).
So, let us write
\begin{subequations} \label{expXA}
\begin{gather}
\begin{align} 
X^{a''}(\tau,\sigma) &= \sum_{\ell=0}^{N-1} \sum_{m=-\ell}^\ell
      X^{a''}_{\ell,m}(\tau,\sigma) Y_{\ell m}, &
X^{8}(\tau,\sigma) &= \sum_{\ell=0}^{N-1} \sum_{m=-\ell}^\ell
      X^{8}_{\ell,m}(\tau,\sigma) Y_{\ell m}, 
\\
A_\mu(\tau,\sigma) &= \sum_{\ell,m} A_{\mu \ell,m}(\tau,\sigma) Y_{\ell,m},&
X^a &= \frac{M g_s}{3} J^a + 
    \sum_{\ell=0}^{N-1} \sum_{j=\abs{\ell-1}}^{\ell+1}
    \sum_{m=-\ell}^\ell X_{j\ell m} Y^a_{j \ell m},
\end{align} \\
\begin{gathered}
      X^{a''}_{\ell,-m} = (-1)^m X^{a''*}_{\ell,m}, \quad
            X^{8}_{\ell,-m} = (-1)^m X^{8*}_{\ell,m},
      \quad A_{\ell,-m} = (-1)^m A_{\mu \ell m}^*, \\
      \quad X_{j,\ell, -m} = (-1)^{m+(j-\ell)} X_{j \ell m}^*.
\end{gathered}
\end{gather}
\end{subequations}%
Here we have used the scalar and vector spherical
harmonics; see Appendix~\ref{app:ylm}.
Plugging eq.~\eqref{expXA} into eq.~\eqref{bms} gives, to
quadratic order,
\begin{multline} \label{bmsylm}
S = \int d^2\sigma
\frac{N}{2} \sum_{\ell,m} \left\{
\abs{ g_s \dot{A}_{\sigma\ell m} - g_s A'_{\tau\ell m} 
   - \frac{M}{3} X^8_{\ell m} }^2
+ \abs{ \dot{X}_{\ell \ell m} 
     - i \frac{M g_s}{3} \sqrt{\ell(\ell+1)} A_{\tau \ell m} }^2
\right. \\* \left.
+ \abs{ \dot{X}_{\ell-1,\ell m} }^2
+ \abs{ \dot{X}_{\ell+1,\ell m} }^2
+ \abs{ \dot{X}^{a''}_{\ell m} }^2
+ \abs{ \dot{X}^8_{\ell m} }^2
- \abs{ X'_{\ell \ell m} 
     - i \frac{M g_s}{3} \sqrt{\ell(\ell+1)} A_{\sigma \ell m} }^2
\right. \\* \left.
- \abs{ X'_{\ell-1,\ell m} }^2
- \abs{ X'_{\ell+1,\ell m} }^2
- \abs{ {X'}^{a''}_{\ell m} }^2
- \abs{ {X'}^8_{\ell m} }^2
- \bigl( \frac{M}{3} \bigr)^2 (\ell+1)^2 \abs{X_{\ell+1,\ell, m}}^2
\right. \\* \left.
- \bigl( \frac{M}{3} \bigr)^2 \ell^2 \abs{X_{\ell-1,\ell, m}}^2
- \bigl( \frac{M}{3} \bigr)^2 (\ell^2+\ell+\frac{1}{4})
      \abs{X^{a''}_{\ell m}}^2
- \bigl( \frac{M}{3} \bigr)^2 (\ell^2+\ell+1) \abs{X^8_{\ell m}}^2
\right\},
\end{multline}
where the overdot (prime) denotes a $\tau$ ($\sigma$) derivative.

Clearly, the fields $X_{\ell-1,\ell,m}$, $X^{a''}_{\ell m}$ and
$X_{\ell+1,\ell,m}$ have respective masses $\frac{M}{3} \ell$, $\frac{M}{3}
(\ell+\frac{1}{2})$, and $\frac{M}{3} (\ell+1)$.  To diagonalize the
action for the remaining fields $X^8_{\ell m}, X_{\ell \ell m}$ and
$A_{\mu \ell m}$, the first step is to introduce auxiliary (scalar) fields
$\phi_{\ell m}=(-1)^m \phi^*_{\ell m}$, as was done in Sec.~\ref{sec:pert}
\begin{multline} \label{Swithphi}
S = \frac{N}{2} \sum_{\ell m} \int d^2 \sigma \left\{
\abs{ g_s \dot{A}_{\sigma\ell m} - g_s A'_{\tau\ell m} 
   - \frac{M}{3} X^8_{\ell m} }^2
+ \abs{ \dot{X}_{\ell \ell m} 
     - i \frac{M g_s}{3} \sqrt{\ell(\ell+1)} A_{\tau \ell m} }^2
\right. \\* \left.
- \abs{ X'_{\ell \ell m} 
     - i \frac{M g_s}{3} \sqrt{\ell(\ell+1)} A_{\sigma \ell m} }^2
- \abs{ w_{\ell} \phi_{\ell m} 
   + g_s \dot{A}_{\sigma\ell m} - g_s A'_{\tau\ell m} 
   - \frac{M}{3} X^8_{\ell m} }^2
\right. \\* \left.
+ \abs{ \dot{X}^8_{\ell m} }^2
- \abs{ {X'}^8_{\ell m} }^2
- \bigl( \frac{M}{3} \bigr)^2 (\ell^2+\ell+1) \abs{X^8_{\ell m}}^2
+ {\mathcal L}(X^{a''},X_{\ell\pm1,\ell,m})
\right\},
\end{multline}
where ${\mathcal L}(X^{a''},X_{\ell\pm1,\ell,m})$ is the
Lagrangian density for
those massive fields---we will drop this in the following---and
$w_{\ell}$ is an arbitrary normalization.
Obviously, integrating $\phi_{\ell m}$ out of the
action~\eqref{Swithphi} gives back the action~\eqref{bmsylm}.
However, after integrating by parts, the action~\eqref{Swithphi} is
independent of any derivatives of the gauge field.

In fact, $A_{\mu 0 0}$ only appears linearly in the action; thus
integrating out $A_{\mu 0 0}$ produces the constraints $\p_\mu
\phi_{00} = 0$. The normalizations $w_{\ell}$ are arbitrary. However
it is convenient to choose $w_{0} = \frac{M}{3}$. 
Then the action for the $\ell=0$ fields becomes
\begin{equation}
S_{\ell = 0} = \frac{N}{2} \int d^2 \sigma \left\{
\left(\dot{X}^8_{00}\right)^2
- \left({X'}^8_{00}\right)^2
- \bigl( \frac{M}{3} \bigr)^2 \left(X^8_{00}-\phi_{00}\right)^2
\right\}.
\end{equation}
Since $\phi_{00}$ is an arbitrary constant, this shows that, at
$\ell=0$, there is a
single degree of freedom with mass $\frac{M}{3}$.  Nevertheless, as
was also seen in Sec.~\ref{sec:prop}, the 
constant mode of $X^8_{00}$ is arbitrary.
Note that the counting of degrees of freedom at $\ell=0$ is correct;
there is no $X_{\ell\ell m}$ for $\ell=0$, so the only fields
considered here are the gauge field, which contributes no degrees of
freedom in two dimensions, and $X^8_{00}$ which has one degree of
freedom. Also, the specific choice of $w_{0}$ is not essential.
For some arbitrary $w_0$ the dynamical combination which appears
in the action is $(w_0 \phi_{00} - \frac{M}{3} X^8_{00})$ and the
spectrum is of course unchanged.

For $\ell \neq 0$, we can still integrate out the gauge field.  Now
the convenient choice is \hbox{$w_{\ell} = \frac{M}{3} \sqrt{\ell(\ell+1)}$}.
After an integration by parts, the resulting action is
\begin{multline} \label{Slnot0}
S_{\ell \neq 0} = \frac{N}{2} \int d^2 \sigma \left\{
   \abs{\dot{X}^8_{\ell m}}^2 - \abs{{X'}^8_{\ell m}}^2
 + \abs{\dot{\phi}_{\ell m}}^2 - \abs{\phi'_{\ell m}}^2
\right. \\* \left.
 - \bigl(\frac{M}{3}\bigr)^2 (\ell^2+\ell+1) \abs{X^8_{\ell m}}^2
 - \bigl(\frac{M}{3}\bigr)^2 \ell(\ell+1) \abs{\phi_{\ell m}}^2
 + 2 \bigl(\frac{M}{3}\bigr)^2 \sqrt{\ell(\ell+1)} 
     \real X^{8*}_{\ell m} \phi_{\ell m}
\right\}.
\end{multline}
Diagonalizing the mass matrix in this action~\eqref{Slnot0} gives
masses $\frac{M}{3} \ell$ and $\frac{M}{3} (\ell+1)$,
which is precisely the spectrum of $X_{\ell\pm 1,\ell, m}$! Once
again one could have chosen any other $w_{\ell}$.

Counting, for each $\ell$ we have $4(\ell+1)$ bosons of mass-squared
$\bigl(\frac{M}{3}\bigr)^2 (\ell+1)^2$; $4 \ell$ bosons of mass-squared
$\bigl(\frac{M}{3}\bigr)^2 \ell^2$;
and (from $X^{a''}$) $4(2\ell+1)$ bosons of mass-squared
$\bigl(\frac{M}{3}\bigr)^2 (\ell+\frac{1}{2})^2$.  This is summarized
in table~\ref{tab:bosirred}.

This result for the spectrum is not surprising, as we expected $X^8$
and $X^a$ to be in the same supermultiplet, and to therefore have
related masses.  What is more surprising is that there is,
nevertheless a splitting within this supermultiplet.
Since the linearly realized (4,4)
worldsheet supersymmetries commute with the Hamiltonian, one might
have expected the
spectrum to be degenerate.  
In Sec.~\ref{sec:susy}, we examined the superalgebra and observed
that the angular momentum generators appear in the square of the
supercharge in such a way as to account for this splitting.  This is
similar to the M-theory BPS analysis of~\cite{dsv1,dsv2}.
(In the case of the
M-theory Matrix quantum mechanics, $X_{\ell\pm1,\ell m}$ had an
identical split spectrum, but the supercharges did not commute with
the Hamiltonian in precisely a way that produced the splitting.)

Observe that the field $X_{\ell \ell m}$ is not dynamical and can be
removed by a gauge transformation.  That is, to linearized order,
$X_{\ell \ell m}$ does not appear in the action~\eqref{Slnot0}
because, to this order, $X_{\ell \ell m}$ is a pure gauge mode.
A gauge transformation is generated by the scalar
\hbox{$\Lambda = \sum_{\ell m}
\Lambda_{\ell m} Y_{\ell m}$}.  Under a gauge transformation, $X^i
\rightarrow X^i + i \com{\Lambda}{X^i}$.
Because, to zeroth order, $X^a = \frac{M g_s}{3} J^a$,
\begin{equation} \label{gt}
X_{\ell \ell m} \rightarrow X_{\ell \ell m} - i
   \frac{M g_s}{3} \sqrt{\ell (\ell+1)} \Lambda_{\ell m}
   + \order{(M g_s)^0},
\end{equation}
So, to leading order in perturbation theory%
\footnote{The subleading term is bilinear in $\Lambda$ and $X_{jlm}$,
with coefficients extracted from eq.~\eqref{comylmyjlm}.}
in $\frac{1}{M g_s}$,
we see that by setting \hbox{$\Lambda_{\ell m} =
\frac{i}{\bigl(\frac{M g_s}{3}\bigr) \sqrt{\ell(\ell+1)}}
\left(X^{\text{(desired)}}_{\ell \ell m}
 - X^{\text{(old)}}_{\ell \ell m}\right)$}, 
we can set $X_{\ell \ell m}$ to any convenient value.
Recall that $X_{\ell\ell m}$ exist
only for $\ell\geq 1$; thus this gauge choice does not fix the U(1)
degree of freedom.  Otherwise, this fixes $\Lambda$ and so there are no
residual gauge transformations other than the U(1).
The discussion above would have been simplified very
slightly by taking $X_{\ell \ell m}=0$, but the more general
presentation is instructive.

\TABLE[t]{\begin{tabular}{||c||c|c|c||}
\hline \hline 
origin & degeneracy & mass$^2$ & range \\
\hline 
$X^a, X^8, A_\mu$ & $\begin{array}{c} 4 (\ell+1) \\ 4 \ell \end{array}$ &
 $\begin{array}{c} \bigl(\frac{M}{3}\bigr)^2(\ell+1)^2 \\
                \bigl(\frac{M}{3}\bigr)^2 \ell^2 \end{array} $ &
   $\begin{array}{c} 0 \leq \ell \leq N-1 \\ 1 \leq \ell \leq N-1
    \end{array}$ \\
\hline
$X^{a''}$ & $4 (2\ell+1)$ & $\bigl(\frac{M}{3}\bigr)^2(\ell+\frac{1}{2})^2$ &
   $0 \leq \ell \leq N-1$ \\
\hline \hline
\end{tabular}
\caption{The bosonic spectrum for the irreducible vacuum.\label{tab:bosirred}}
}

For example, we could alternatively choose a gauge such that
\hbox{$X^3 = J^3 + (\text{fluctuations})$} is
diagonal.  This is achieved by first choosing to use
standard $\su(2)$ representations, so that $J^3$ is diagonal, and then
only allowing the diagonal elements of $X^3$ to fluctuate.  In other
words, we demand that
\begin{equation}
\sum_{j,\ell,m} X_{j\ell m} Y^3_{j \ell m} = \text{diagonal}.
\end{equation}
Using the explicit form for the vector spherical harmonics in terms of
scalar spherical harmonics,
eq.~\eqref{myjlm}, this is
\begin{equation}
\sum_{\ell,m} \left[ \sqrt{\frac{(\ell+1)^2-m^2}{(\ell+1) (2 \ell+1)}}
   X_{\ell+1,\ell,m}
  + \frac{m}{\sqrt{\ell(\ell+1)}} X_{\ell \ell m}
  - \sqrt{\frac{\ell^2-m^2}{\ell (2 \ell+1)}} X_{\ell-1,\ell,m}
  \right] Y_{\ell m} = \text{diagonal}.
\end{equation}
Since the $Y_{\ell m}$'s are linearly independent, and diagonal
precisely for $m=0$, this is
solved by setting
\begin{equation} \label{x3diag}
X_{\ell \ell m} = \frac{1}{m} \sqrt{\frac{(\ell+1)(\ell^2-m^2)}{2 \ell+1}}
    X_{\ell-1,\ell,m}
  - \frac{1}{m}\sqrt{\frac{\ell [(\ell+1)^2-m^2]}{2\ell+1}} X_{\ell+1,\ell,m},
\quad m \neq 0.
\end{equation}
The gauge freedom~\eqref{gt} allows this, and further allows
us to set the otherwise unconstrained
\begin{equation} \label{x3stilldiag}
X_{\ell \ell 0} = 0.
\end{equation}
The overall U$(1)$ is fixed via $A_{\tau 00} = 0$.
\iftoomuchdetail
\begin{detail}%
This does not affect the derivation above as $\dot{\phi}_{00}=0$ would
still be imposed as a gauge-fixing constraint equation.
\end{detail}%
\fi

\subsection{The Fermionic Spectrum in the Irreducible Vacuum}
\label{sec:fermspec}

Now let us look at the fermions.  Heretofore, we have implicitly
worked in a Majorana basis so that $\Psi = \Psi^*$ and the 
$\Gamma$-matrices were real and symmetric.  It is now convenient to abandon
this basis in favour of one in which the $3+4+1=\{a,a'',8\}$ splitting of the
$\Gamma$-matrices is more transparent.  In terms of the Pauli matrices
$\sigma^a$ and the 4-dimensional $\gamma$-matrices, we write
\begin{align}
\Gamma^a &= \sigma^a \otimes \one \otimes \sigma^3, &
\Gamma^{a''} &= \one \otimes \gamma^{(a''-3)} \otimes \sigma^1, &
\Gamma^8 &= \one \otimes \gamma^5 \otimes \sigma^1, &
\Gamma^9 &= -\one \otimes \one \otimes \sigma^2.
\end{align}
Here $\gamma^5=\gamma^{1234}$ and so indeed $\Gamma^9=\Gamma^{12345678}$.
In this basis, the charge conjugation matrix $C$ is
\begin{equation}
C = \sigma^2 \otimes C \otimes \sigma^1
\end{equation}
where we have also called the 4-dimensional charge conjugation matrix
$C$.
As
\begin{align}
\Gamma^{123} &= i \one \otimes \one \otimes \sigma^3, &
\Gamma^{89} &= -i \one \otimes \gamma^5 \otimes \sigma^3,
\end{align}
the Weyl representation for the 4-dimensional $\gamma$-matrices is
convenient.  In particular, we will use a representation in which
$\gamma^5=\diag(-1,1,-1,1)$.

We now expand $\Psi$ using the spinor spherical harmonics (see
Appendix~\ref{sec:ylm}) as
\begin{subequations}%
\begin{gather}
\Psi = \sum_{\ell, m} \sum_{I=1}^8 \sum_{\Lambda=1}^2 
 \left[ \psi^{I\Lambda}_{\ell+\frac{1}{2},\ell,m}
   \boldsymbol{S}_{\ell+\frac{1}{2},\ell,m} \otimes \boldsymbol{b}_I 
   \otimes \boldsymbol{b}_\Lambda
+ \psi^{I\Lambda}_{\ell-\frac{1}{2},\ell,m}
   \boldsymbol{S}_{\ell-\frac{1}{2},\ell,m} \otimes \boldsymbol{b}_I 
   \otimes \boldsymbol{b}_\Lambda
\right], \\
(-1)^m C_{IJ} \sigma^1_{\Lambda\Sigma} 
\psi^{J\Sigma*}_{\ell+\frac{1}{2},\ell,-m} = 
\psi^{I\Lambda}_{\ell+\frac{1}{2},\ell,m}, \qquad
(-1)^{m+1} C_{IJ} \sigma^1_{\Lambda\Sigma} 
\psi^{J\Sigma*}_{\ell-\frac{1}{2},\ell,-m} = 
\psi^{I\Lambda}_{\ell-\frac{1}{2},\ell,m}.
\end{gather}
\end{subequations}%
Here $\boldsymbol{b}_I$ and $\boldsymbol{b}_\Lambda$ are 
(commutative) orthonormal
basis spinors in their respective subspaces, namely $b_I^J = \delta_I^J$ and
$b_\Lambda^\Sigma=\delta_\Lambda^\Sigma$.
The reality condition on the (Grassmann) coefficients is the Majorana condition
$\Psi = C \Psi^*$, where the complex conjugation includes Hermitian
conjugation of the U$(N)$ matrix.

The fermionic quadratic action is now reasonably simple,
\begin{multline} \label{faction}
S = N \sum_{\ell, m, I, \Lambda} \left[
  i \psi^{I\Lambda*}_{\ell+\frac{1}{2}, \ell, m} 
  \dot{\psi}^{I\Lambda}_{\ell+\frac{1}{2}, \ell, m}
+ i \psi^{I\Lambda*}_{\ell-\frac{1}{2}, \ell, m} 
  \dot{\psi}^{I\Lambda}_{\ell-\frac{1}{2}, \ell, m}
+ i \psi^{I\Lambda*}_{\ell+\frac{1}{2}, \ell, m} \sigma^2_{\Lambda\Sigma}
  \psi^{'I\Sigma}_{\ell+\frac{1}{2}, \ell, m}
\right. \\* \left.
+ i \psi^{I\Lambda*}_{\ell-\frac{1}{2}, \ell, m} \sigma^2_{\Lambda\Sigma}
  \psi^{'I\Sigma}_{\ell-\frac{1}{2}, \ell, m}
- \frac{M}{4} (-1)^\Lambda [1+\frac{1}{3}(-1)^I]
  \psi^{I\Lambda*}_{\ell+\frac{1}{2}, \ell, m} 
  \psi^{I\Lambda}_{\ell+\frac{1}{2}, \ell, m}
\right. \\* \left.
- \frac{M}{4} (-1)^\Lambda [1+\frac{1}{3}(-1)^I]
  \psi^{I\Lambda*}_{\ell-\frac{1}{2}, \ell, m} 
  \psi^{I\Lambda}_{\ell-\frac{1}{2}, \ell, m}
- \frac{M}{3} (-1)^\Lambda \ell \psi^{I\Lambda*}_{\ell+\frac{1}{2}, \ell, m} 
   \psi^{I\Lambda}_{\ell+\frac{1}{2}, \ell, m}
\right. \\* \left.
+ \frac{M}{3} (-1)^\Lambda (\ell+1) 
   \psi^{I\Lambda*}_{\ell-\frac{1}{2}, \ell, m} 
   \psi^{I\Lambda}_{\ell-\frac{1}{2}, \ell, m}
\right].
\end{multline}
It is easy to read off the mass-squareds of the fermions, namely
\begin{equation} \label{fmass}
\begin{aligned}
j=\ell+\frac{1}{2}, I \text{ even}:& \bigl(\frac{M}{3}\bigr)^2 (\ell+1)^2, \\
j=\ell\pm\frac{1}{2}, I \text{ odd}:& 
   \bigl(\frac{M}{3}\bigr)^2 (\ell+\frac{1}{2})^2, \\ 
j=\ell-\frac{1}{2}, I \text{ even}:& \bigl(\frac{M}{3}\bigr)^2 \ell^2.
\end{aligned}
\end{equation}

\TABLE[t]{\begin{tabular}{||c||c|c|c||}
\hline \hline 
origin & degeneracy & mass$^2$ & range \\
\hline 
$\frac{\one+\Gamma^{12389}}{2} \Psi$ & 
 $\begin{array}{c} 8 (\ell+1) \\ 8 \ell \end{array}$ &
 $\begin{array}{c} \bigl(\frac{M}{3}\bigr)^2(\ell+1)^2 \\
                \bigl(\frac{M}{3}\bigr)^2 \ell^2 \end{array} $ &
   $\begin{array}{c} 0 \leq \ell \leq N-1 \\ 1 \leq \ell \leq N-1
    \end{array}$ \\
\hline
$\frac{\one-\Gamma^{12389}}{2} \Psi$ & 
$8 (2\ell+1)$ & $\bigl(\frac{M}{3}\bigr)^2(\ell+\frac{1}{2})^2$ &
   $0 \leq \ell \leq N-1$ \\
\hline \hline
\end{tabular}
\caption{The fermionic spectrum for the irreducible
vacuum.  The degeneracy of the fermions is twice that of the bosons,
because we include the number of components of the
fermions.\label{tab:fermirred}}
}

This is precisely the spectrum of the bosons.  In particular, for each
$\ell$, there are $8(2 \ell+1)$ fermions of mass-squared
$\bigl(\frac{M}{3}\bigr)^2 (\ell+\frac{1}{2})^2$; $8 \ell$ fermions of
mass-squared $\bigl(\frac{M}{3}\bigr)^2 \ell^2$; and $8 (\ell+1)$
fermions of mass-squared $\bigl(\frac{M}{3}\bigr)^2 (\ell+1)^2$.  
This is precisely (double) the spectrum of the bosons!

Also, recall that whether $I$ is even or odd is correlated with the
$\gamma^5$-chirality.  Furthermore,
\begin{equation}
\Gamma^{12389} = \one\otimes \gamma^5 \otimes \one.
\end{equation}
Therefore, we see that all fermions of negative $\Gamma^{12389}$
chirality have the same mass as $X^{a''}$, and that all fermions of
positive $\Gamma^{12389}$ have a mass splitting, but with masses
identical to those of the remaining bosons.  This has been tabulated
in table~\ref{tab:fermirred}.

\subsection{Reducible Vacua} \label{sec:redfluc}

For the simplest reducible vacua,
\begin{equation}
X^a = X^a_0 = \begin{pmatrix} J^a_{N_1} & & 0 \\
                      & J^a_{N_2} &  \\
                      0 & & \ddots \end{pmatrix},
\end{equation}
with all other fields vanishing,
the fluctuations are again written in terms of spherical harmonics.
Explicitly, for example,
\begin{align} \label{preXaredfluc}
X^a &= X^a_0 + \begin{pmatrix} 
\sum X^{(1,1)}_{j\ell m} Y^{(1,1)a}_{j\ell m} &
\sum X^{(1,2)}_{j\ell m} Y^{(1,2)a}_{j\ell m} &
\dots \\
\sum X^{(1,2)*}_{j\ell m} Y^{(1,2)a\dagger}_{j\ell m} &
\sum X^{(2,2)}_{j\ell m} Y^{(2,2)a}_{j\ell m} &
\dots \\
\vdots &\ddots \end{pmatrix}.
\
\end{align}
For notational purposes, we write
\begin{equation}
\begin{aligned}
X^i &= \begin{pmatrix} X^{(1,1)i} & X^{(1,2)i} & \dots \\
                       X^{(2,1)i} & X^{(2,2)i} & \dots \\ 
                       \vdots & \ddots &
       \end{pmatrix}, &
X^{(x,y)i\dagger} &= X^{(y,x)i}, \\
A_\mu &= \begin{pmatrix} A_\mu^{(1,1)} & A_\mu^{(1,2)} & \dots \\
                       A_\mu^{(2,1)} & A_\mu^{(2,2)} & \dots \\ 
                       \vdots & \ddots &
         \end{pmatrix}, &
A_\mu^{(x,y)\dagger} &= A_\mu^{(y,x)},
\end{aligned}
\end{equation}
where $x,y$ run over the blocks which have dimension $N_x,N_y, \sum_x
N_x = N$.  Then,
\begin{subequations} \label{redfluc}
\begin{align} \label{Xa''redfluc}
X^{(x,y)a''} &= 
\sum_{j=\frac{\abs{N_x-N_y}}{2}}^{\frac{N_x+N_y}{2}-1} 
\sum_{m=-j}^j
X^{(x,y)a''}_{j m} Y^{(N_x,N_y)}_{j m}, &
X^{(x,y)a''*}_{j m} &= (-1)^{m-\frac{N_x-N_y}{2}} 
   X^{(y,x)a''}_{j,-m}, \\
\label{X8redfluc}
X^{(x,y)8} &= 
\sum_{j=\frac{\abs{N_x-N_y}}{2}}^{\frac{N_x+N_y}{2}-1} 
\sum_{m=-j}^j
X^{(x,y)8}_{j m} Y^{(N_x,N_y)}_{j m}, &
X^{(x,y)8*}_{j m} &= (-1)^{m-\frac{N_x-N_y}{2}} 
   X^{(y,x)8}_{j,-m}, \\
A^{(x,y)}_\mu &= 
\sum_{j=\frac{\abs{N_x-N_y}}{2}}^{\frac{N_x+N_y}{2}-1} 
\sum_{m=-j}^j
A^{(x,y)}_{\mu j m} Y^{(N_x,N_y)}_{j m}, &
A^{(x,y)*}_{\mu j m} &= (-1)^{m-\frac{N_x-N_y}{2}} 
   A^{(y,x)}_{\mu j,-m},
\end{align}
\begin{multline} \label{Xaredfluc}
X^{(x,y)a} = \frac{M g_s}{3} \delta_{xy} J^a_{N_x} + 
\sum_{\ell=\frac{\abs{N_x-N_y}}{2}}^{\frac{N_x+N_y}{2}-1} 
\sum_{j=\ell-1}^{\ell+1} \sum_{m=-\ell}^\ell
X^{(x,y)}_{j\ell m} Y^{(N_x,N_y)a}_{j\ell m}, \\*
X^{(x,y)*}_{j\ell m} = (-1)^{\ell+m+1-j-\frac{N_x-N_y}{2}} 
   X^{(y,x)}_{j,\ell,-m}.
\end{multline}
\end{subequations}%

Because the properties [eqs.~\startylmprops--\enylmprops,
\startyvecprops--\enyvecprops\ and~\startSjlmprops--\enSjlmprops]
of the spherical harmonics are 
essentially independent of
the dimension---or even squareness---of the blocks, we can
immediately copy the results of the spectrum of the irreducible
vacuum, and apply it to the reducible vacua, albeit for a different
range of $\ell$s.  That is, the spectrum is:
\begingroup \small
\begin{equation}
\begin{alignedat}{4}
X^a, X^8:\ && \text{mass}^2&= \bigl(\frac{M}{3}\bigr)^2 j^2,\ &
   \text{degeneracy }& 4j,\ & 
   \text{with }&\frac{\abs{N_x-N_y}}{2} \leq j \leq \frac{N_x+N_y}{2}-1. \\
X^a, X^8:\ && \text{mass}^2&= \bigl(\frac{M}{3}\bigr)^2 (j+1)^2,\ &
   \text{degeneracy }&4(j+1),\ & 
   \text{with }&\frac{\abs{N_x-N_y}}{2} \leq j \leq \frac{N_x+N_y}{2}-1. \\
X^{a''}:\ && \text{mass}^2&= \bigl(\frac{M}{3}\bigr)^2 (j+\frac{1}{2})^2,\ &
   \text{degeneracy }&4(2j+1),\ & 
   \text{with }&\frac{\abs{N_x-N_y}}{2} \leq j \leq
   \frac{N_x+N_y}{2}-1.
\end{alignedat}
\end{equation}
\endgroup
with an identical spectrum for the fermions.

However, an analysis in Sec.~\ref{sec:susy} showed that
more general supersymmetric vacua exist with constant $X^8$ subject to
the condition
\begin{equation} \label{x8diag}
\com{X^a}{X^8}=0.
\end{equation}
For the irreducible vacuum, this requires that
$X^8$ be proportional to the identity; the constant is then
effectively just a $\theta$-angle for the U$(1)$ part of
the gauge group.
However, for a reducible vacuum consisting of $r>1$ irreducible
representations, the story is more complicated.  By
eq.~\eqref{x8diag}, $X^8$ and $X^3$ are simultaneously
diagonalizable; thus, by Schur's lemma,
$X^8$ consists (up to a similarity transformation)
of $r$ blocks, each proportional
to the identity matrix.
Thus, eq.~\eqref{X8redfluc} is
generalized to%
\iftoomuchdetail
\begin{detail}%
\footnote{One might want a more general expression when two or more
representations are identical; however, we have already explained that
a diagonal expression is sufficiently general.}
\end{detail}%
\fi
\begin{equation}
X^{(x,y)8} = \wastheta{(x)} \delta_{xy} \one_{N_x}
  + \sum_{j,m} X^{(x,y)8}_{jm} Y^{(N_x,N_y)}_{jm},
\end{equation}
where $\wastheta{(x)}$ is real.

In Sec.~\ref{sec:soln} we explained that these solutions arise from
time-dependent, 1/2 supersymmetric solutions of the 11-dimensional
matrix quantum mechanics~\eqref{MQM}.  There are similarly other
solutions that correspond to rotating fuzzy spheres, and are 1/2
supersymmetric in both the matrix quantum mechanics and the matrix
string theory.  (For example, in matrix string theory,
a rotation in the $X^6$-$X^7$ plane
preserves the 4 supersymmetries preserved by $\Gamma^{12389}
\epsilon = \epsilon = \Gamma^{12367} \epsilon$.  This lifts to a
Matrix Theory solution that preserves the 8 supersymmetries preserved
by $\Gamma^{12367} \epsilon=\epsilon$.)  However, we do not discuss
these further.

Rather than work with the action, let us work directly with the
equations of motion.%
\footnote{It is possible to obtain the spectrum using the same
auxiliary fields as in Sec.~\ref{sec:bosspec}, but the algebra is
more complicated.}
The exact equations of motion which follow from
the action~\eqref{bms} are
\begin{subequations}%
\begin{align}
\label{Aeom}
g_s D^\nu &\left(g_s F_{\mu\nu} - \frac{M}{3} \epsilon_{\mu\nu} X^8 \right)
+ i \com{X^i}{D_\mu X^i} = 0, \\
\label{Xaeom}
\begin{split}
-D^2 X^a 
&+ \frac{M}{3} \left( \frac{M}{3} X^a 
     + \frac{i}{2 g_s} \epsilon_{abc} \com{X^b}{X^c} \right)
+ \frac{i}{g_s} \epsilon_{abc} \com{X^b}{\left( \frac{M}{3} X^c 
     + \frac{i}{2 g_s} \epsilon_{cde} \com{X^d}{X^e} \right)}
\\ &
+ \frac{1}{g_s^2} \com{X^{a''}}{\com{X^{a''}}{X^{a}}}
+ \frac{1}{g_s^2} \com{X^8}{\com{X^8}{X^a}} = 0, 
\end{split} \\
\label{Xa''eom}
\begin{split}
-D^2 X^{a''} &+ \bigl(\frac{M}{6}\bigr)^2 X^{a''}
+ \frac{1}{g_s^2} \com{X^a}{\com{X^a}{X^{a''}}}
+ \frac{1}{g_s^2} \com{X^8}{\com{X^8}{X^{a''}}} \\ &
+ \frac{1}{g_s^2} \com{X^{b''}}{\com{X^{b''}}{X^{a''}}} = 0,
\end{split} \\
\label{X8eom}
-D^2 X^8 &+ \frac{M}{3} g_s F_{\tau \sigma}
+ \frac{1}{g_s^2} \com{X^a}{\com{X^a}{X^8}}
+ \frac{1}{g_s^2} \com{X^{a''}}{\com{X^{a''}}{X^8}} = 0.
\end{align}
\end{subequations}%
We note that the mass term for $X^8$ was replaced by $F_{\tau\sigma}$.
Moreover, the equation of motion for the gauge field,~\eqref{Aeom},
only requires (assuming the current vanishes) \hbox{$g_s
F_{\tau\sigma} - \frac{M}{3} X^8 = \text{constant}$}; this is why we can set
$X^8$ to a constant matrix and $F_{\tau\sigma}=0$ while still obeying the
equations of motion, though taking the interaction terms into account
shows that $X^8$ should commute with $X^a$, in agreement with the
supersymmetry conditions.

Plugging the expansion~\eqref{redfluc} into the equations of motion,
and linearizing, gives
\begin{subequations} \label{lineom}
\begin{align} \label{linAeom}
\begin{split}
0 &= g_s \p_\mu \epsilon^{\nu\rho} \p_\nu A^{(x,y)}_{\rho j m}
-\left[\bigl(\frac{M}{3}\bigr)^2 g_s j(j+1) 
    + \frac{1}{g_s}(\wastheta{(x)}-\wastheta{(y)})^2 \right]
   \epsilon_\mu{^\nu} A^{(x,y)}_{\nu j m}
+ \frac{M}{3} \p_\mu X^{(x,y)8}_{jm}
\\ &
-i \frac{M}{3} (\wastheta{(x)}-\wastheta{(y)}) A^{(x,y)}_{\mu j m}
- \frac{i}{g_s} (\wastheta{(x)}-\wastheta{(y)}) 
     \epsilon_\mu{^\nu} \p_\nu X^{(x,y)8}_{jm}
- i \frac{M}{3} \sqrt{j(j+1)} \epsilon_\mu{^\nu} \p_\nu X^{(x,y)}_{jjm},
\end{split} \\
\label{linXjjeom}
\begin{split}
0 &= \left[-\p^2 + \frac{1}{g_s^2} (\wastheta{(x)}-\wastheta{(y)})^2 \right]
   X^{(x,y)}_{jjm}
+ i \frac{M g_s}{3} \sqrt{j(j+1)} \p^\mu A^{(x,y)}_{\mu j m}
\\ & \qquad
- \frac{M}{3} \frac{1}{g_s} (\wastheta{(x)}-\wastheta{(y)})
    \sqrt{j(j+1)} X^{(x,y)8}_{jm},
\end{split} \\
\label{linX8eom}
\begin{split}
0 &= \left[-\p^2 + \bigl(\frac{M}{3}\bigr)^2 j(j+1) \right]
   X^{(x,y)8}_{jm}
+ i (\wastheta{(x)}-\wastheta{(y)}) \p^\mu A^{(x,y)}_{\mu j m}
- \frac{M g_s}{3} \epsilon^{\mu\nu} \p_\mu A^{(x,y)}_{\nu j m} \\ & \qquad
- \frac{M}{3}\frac{1}{g_s} \sqrt{j(j+1)} (\wastheta{(x)}-\wastheta{(y)})
   X^{(x,y)}_{jjm},
\end{split} \\
\label{linXjpeom}
0 &= \left[-\p^2 + \bigl(\frac{M}{3}\bigr)^2 (j+1)^2
   + \frac{1}{g_s^2} (\wastheta{(x)}-\wastheta{(y)})^2 \right]
   X^{(x,y)}_{j+1,j,m}, \\
\label{linXjmeom}
0 &= \left[-\p^2 + \bigl(\frac{M}{3}\bigr)^2 j^2
   + \frac{1}{g_s^2} (\wastheta{(x)}-\wastheta{(y)})^2 \right]
   X^{(x,y)}_{j-1,j,m}, \\
\label{linXa''eom}
0 &= \left[-\p^2 + \bigl(\frac{M}{3}\bigr)^2 (j+\frac{1}{2})^2
   + \frac{1}{g_s^2} (\wastheta{(x)}-\wastheta{(y)})^2 \right]
   X^{(x,y)a''}_{jm}.
\end{align}
\end{subequations}%
The last three equations show that the effect of adding a constant to
$X^8$ is to shift the spectrum by \hbox{$\frac{1}{g_s^2}
(\wastheta{(x)}-\wastheta{(y)})^2$}.  In particular, since the shift is by
a difference of $\wastheta{}$\,'s, the spectrum for the
irreducible vacuum---and also the fluctuations in blocks on the
diagonal---is unaffected.

Let us confirm this result by examining the remainder of the spectrum
from the gauge field and $X^8$.
For now let us choose the
gauge $X^{(x,y)}_{j j m}=0$;
as in Sec.~\ref{sec:bosspec}, the precise value of
$X^{(x,y)}_{jjm}$ eventually drops out anyway.
Then eq.~\eqref{linXjjeom}
becomes the constraint equation
\begin{equation} \label{cswth}
\p^\mu A^{(x,y)}_{\mu j m}
= -\frac{i}{g_s^2} (\wastheta{(x)}-\wastheta{(y)}) X^{(x,y)8}_{jm}, \quad
j\neq 0,
\end{equation}
which generalizes the Lorentz gauge condition.%
\footnote{Thus, fixing a gauge fixes a gauge!}
Next take the divergence of the gauge field equation of
motion~\eqref{linAeom}.  Using the constraint~\eqref{cswth} we find
\begin{equation} \label{gotF}
g_s \epsilon^{\mu\nu} \p_\mu A^{(x,y)}_{\nu j m}
= -\frac{M}{3} \frac{-\p^2+\frac{1}{g_s^2}(\wastheta{(x)}-\wastheta{(y)})^2}{%
      -\p^2 + \bigl(\frac{M}{3}\bigr)^2 j(j+1) 
      + \frac{1}{g_s^2} (\wastheta{(x)}-\wastheta{(y)})^2}
  X^{(x,y)8}_{jm}.
\end{equation}
Eqs.~\eqref{cswth} and~\eqref{gotF} completely determine the
gauge field in terms of $X^8$.

Inserting eqs.~\eqref{cswth} and~\eqref{gotF} into the $X^8$
equation of motion~\eqref{linX8eom} gives
\begin{multline} \label{X8wth}
\frac{\left[-\p^2 + \bigl(\frac{M}{3}\bigr)^2 j^2 
      + \frac{1}{g_s^2} (\wastheta{(x)}-\wastheta{(y)})^2\right]
      \left[-\p^2 + \bigl(\frac{M}{3}\bigr)^2 (j+1)^2 
      + \frac{1}{g_s^2} (\wastheta{(x)}-\wastheta{(y)})^2\right]}{%
\left[-\p^2 + \bigl(\frac{M}{3}\bigr)^2 j(j+1)^2 
   + \frac{1}{g_s^2} (\wastheta{(x)}-\wastheta{(y)})^2\right]} X^{(x,y)8}_{jm}
\\* = 0,
\end{multline}
from which%
\iftoomuchdetail
\begin{detail}%
, setting the numerator to zero,
\end{detail}%
\else
\ 
\fi
we read off the same mass
spectrum as
$X^{(x,y)}_{j\pm1,j,m}$.
In particular,
the two higher derivatives
in the numerator of eq.~\eqref{X8wth} imply that there is one extra
degree of freedom in $X^8$.  As we started with a two-dimensional
gauge field, $X^8$ and one-third of $X^a$ (the last is $X_{\ell \ell m}$)
for two bosonic degrees of freedom, this is precisely correct.

This derivation is, strictly, only true for $j\neq 0$; for
$j=0$, $Y^a_{000}=0$ and so its coefficient in
eq.~\eqref{Xaeom}---namely eq.~\eqref{linXjjeom}---need not
vanish.  This is not an issue unless $N_x=N_y$.  Moreover, in that
case we can still fix eq.~\eqref{cswth} as
our gauge choice; then the first factor of eq.~\eqref{X8wth}
cancels the denominator and we find a single massive mode.
Alternatively, we can choose the gauge $A^{(x,y)}_{\tau 00}=0$; an
analysis of the equations of motion again yields a single massive
mode, with mass-squared \hbox{$\bigl(\frac{M}{3}\bigr)^2 + \frac{1}{g_s^2}
(\wastheta{(x)}-\wastheta{(y)})^2$}, as given by eq.~\eqref{X8wth},
although in this case, it is convenient to assign this mass to
$A^{(x,y)}_{\sigma 00}$.
It is somewhat surprising that, although $\wastheta{(x)}$ can have
arbitrary values, there are no massless modes associated with them; instead
the would-be massless mode has been eaten by the now-massive gauge
field.
In fact, as we discussed in Sec.~\ref{sec:pert}, the massless modes
required by Goldstone's theorem are
gauge-variant and so do not appear in the physical spectrum~\cite{ks}.  

We should also note that the shift in the spectrum
implies that $\wastheta{(x)}$ are not periodic.  Indeed, although we
have occasionally referred to them as $\theta$-angles,
$\wastheta{(x)}$ are not periodic because there are no
fundamental charges in the theory~\cite{sc}.  In fact, we have already
commented that $\wastheta{(x)}$ is simply the location of the
$x^{\text{th}}$ fuzzy sphere, in the $X^8$-direction, which
is noncompact.

Incidentally, there is a sense in which
$X^8$ can be compactified, with arbitrary radius.
In the full M-theory solution~\eqref{eq:athree}, this corresponds to
an additional compactification along the Killing vector
\begin{equation} \label{X8kill}
-\frac{\mu}{3} \hx^8 \cos \frac{\mu}{3} \hx^+ \frac{\p}{\p \hx^-}
   + \sin \frac{\mu}{3} \hx^+ \frac{\p}{\p \hx^8}
   - \cos \frac{\mu}{3} \hx^+ \frac{\p}{\p \hx^9}.
\end{equation}
In the matrix string theory, this corresponds to allowing for a {\em
time-dependent\/} identification of $X^8$, \hbox{$X^8 \sim X^8 + 2 \pi R_8 \sin
\frac{M}{3} \tau$}.  For example, formally, the configuration
\begin{equation}
\begin{aligned}
g_s A_\sigma &= -w R_8\, \sigma \cos \frac{M}{3} \tau, \\
X^8 &= w R_8\, \sigma \sin \frac{M}{3} \tau,
\end{aligned}
\qquad w \in \ZZ,
\end{equation}
is a solution to the matrix string theory equations of motion.
However, an analysis following~\cite{jm1} shows that such a
compactification of $X^8$ breaks all the linearly realized
supersymmetries of the matrix string theory.

The fermions are decomposed in blocks of spinor spherical harmonics.
The analysis is then virtually identical to both that for the irreducible
vacuum and that for the reducible bosons; the additional contribution
to the spectrum from having non-zero $\wastheta{(x)}$ comes from the term
\begin{equation} \label{newpsifromtheta}
\frac{1}{g_s} \Psi^\transpose \Gamma^8 \com{X^8}{\Psi}
= \sum_{x,y} \sqrt{N_x N_y} \sum_{j\ell m} \sum_{I,\Lambda,\Sigma} (-1)^I
\frac{1}{g_s} (\wastheta{(x)}-\wastheta{(y)})
    \psi^{(x,y)I\Lambda*}_{j\ell m}  (\sigma^1)_{\Lambda\Sigma}
    \psi^{(x,y)I\Sigma}_{j\ell m}.
\end{equation}
The $\sigma^1_{\Lambda \Sigma}$ in eq.~\eqref{newpsifromtheta}
anticommutes with
the $\sigma^3_{\Lambda \Sigma} =(-1)^{\Lambda+1} \delta_{\Lambda \Sigma}$
in eq.~\eqref{faction} and so we see
that, compared to eq.~\eqref{fmass}, the fermion spectrum is
also shifted by $\frac{1}{g_s^2} (\wastheta{(x)}-\wastheta{(y)})^2$.
The spectrum has been summarized in Table~\ref{tab:redspectrum}.

\TABLE[t]{
\begin{tabular}{||c||c|c|c||}
\hline \hline 
origin & degeneracy & mass$^2$ & range \\
\hline 
$X^a, X^8, A_\mu$ & $\begin{array}{c} 4 (j+1) \\ 4 j \end{array}$ &
 $\begin{array}{c} \bigl(\frac{M}{3}\bigr)^2(j+1)^2 
       + \frac{1}{g_s^2}\bigl(\wastheta{(x)}-\wastheta{(y)}\bigr)^2 \\
                \bigl(\frac{M}{3}\bigr)^2 j^2
       + \frac{1}{g_s^2}\bigl(\wastheta{(x)}-\wastheta{(y)}\bigr)^2
   \end{array} $ &
   $\begin{array}{c} \frac{\abs{N_x-N_y}}{2} \leq j \leq \frac{N_x+N_y}{2}-1 \\
    \frac{\abs{N_x-N_y}}{2}\skipthis{+\delta_{N_x,N_y}}
             \leq j \leq \frac{N_x+N_y}{2}-1
    \end{array}$ \\
\hline
$X^{a''}$ & $4 (2j+1)$ & $\bigl(\frac{M}{3}\bigr)^2(j+\frac{1}{2})^2
   + \frac{1}{g_s^2}\bigl(\wastheta{(x)}-\wastheta{(y)}\bigr)^2$ &
   $\frac{\abs{N_x-N_y}}{2} \leq j \leq \frac{N_x+N_y}{2}-1$ \\
\hline
$\frac{\one+\Gamma^{12389}}{2} \Psi$ & 
 $\begin{array}{c} 8 (j+1) \\ 8 j \end{array}$ &
 $\begin{array}{c} \bigl(\frac{M}{3}\bigr)^2(j+1)^2
       + \frac{1}{g_s^2}\bigl(\wastheta{(x)}-\wastheta{(y)}\bigr)^2 \\
                \bigl(\frac{M}{3}\bigr)^2 j^2
       + \frac{1}{g_s^2}\bigl(\wastheta{(x)}-\wastheta{(y)}\bigr)^2
  \end{array} $ &
   $\begin{array}{c} \frac{\abs{N_x-N_y}}{2} \leq j 
                         \leq \frac{N_x+N_y}{2}-1 \\
                     \frac{\abs{N_x-N_y}}{2}\skipthis{+\delta_{N_x,N_y}}\leq j 
                         \leq \frac{N_x+N_y}{2}-1
    \end{array}$ \\
\hline
$\frac{\one-\Gamma^{12389}}{2} \Psi$ & 
$8 (2j+1)$ & $\bigl(\frac{M}{3}\bigr)^2(j+\frac{1}{2})^2
   + \frac{1}{g_s^2}\bigl(\wastheta{(x)}-\wastheta{(y)}\bigr)^2$ &
   $\frac{\abs{N_x-N_y}}{2} \leq j \leq \frac{N_x+N_y}{2}-1$ \\
\hline \hline
\end{tabular}
\caption{The spectrum for the general
vacuum.  The indices $x,y$ run over all irreducible representations in
the vacuum.  The degeneracy of the fermions is twice that for the
bosons, because we count the components of the fermions.  The number
of degrees of freedom is the same between bosons and
fermions.\label{tab:redspectrum}}
}

\subsection{Twisted Strings} \label{sec:long}

We have now discussed the nontrivial (reducible) vacua.  
As we reviewed in Sec.~\ref{sec:bdy},
in flat
space, the only relevant degrees
of freedom for the matrix string, in the $g_s\rightarrow 0$ limit,
are those that live in the Cartan subalgebra~\cite{lm,bs,dvv}.  By a
gauge choice
we can take the degrees of freedom to be the diagonal elements of the
matrix; then, the residual gauge
transformations are the elements of the Weyl group $S_N$.  Or
equivalently, as presented in Sec.~\ref{sec:bdy}, the boundary
condition can be diagonalized, by a constant gauge transformation, but
then the Cartan subalgebra consists of off-diagonal matrices that are
permuted by the diagonal boundary conditions.  Thus the
boundary conditions are%
\footnote{Similar expressions hold for the gauge fields and fermions;
  we use ``$X$'' to denote a generic field here.}
\begin{equation} \label{fs:P}
X(\tau,\sigma+2\pi) = U_P X(\tau,\sigma) U_P^{-1}, \quad M g_s \ll 1
\end{equation}
where $P$ is the element of the Weyl group, and $U_P$ is the matrix in
the regular representation.  That is, the fields are periodic up to a
residual gauge transformation.

However, for the \ppwave,
we have seen that all matrix elements of
the fields are important, and there is no restriction to the Cartan
subalgebra.  
An expansion around the vacuum is an
expansion about $D_\mu X^i=0$ and $F_{\tau\sigma}=0$.
The vacua are representations of SU(2), and we choose the gauge
in which the SU(2) representations are manifestly reduced, decomposed
into the standard irreducible representations such that shortest
representations are top-leftmost.
Within a given SU(2) representation, one can further order the
representations with
respect to the value of $\wastheta{(x)}$.
Since the representations are constant matrices, $D_\mu X^i=0$ implies
that $\com{A_\mu}{X^a}=0$.  So, $A_\mu$ is
pure gauge and an element of $\oplus_n {\mathfrak u}(N_n)$.  So, without
loss of generality, we can take $\vev{A_\mu}=0$ and the boundary
condition  is
\begin{equation} \label{pp:P}
\begin{aligned}
A_\mu(\tau,\sigma+2\pi) &= U_P A_\mu(\tau,\sigma) U_P^{-1}, \\
X(\tau,\sigma+2\pi) &= U_P X(\tau,\sigma) U_P^{-1},
\end{aligned}
\qquad M g_s \gg 1,
\end{equation}
but now, to be consistent with the form of $X^a$, $P$ is an element of
the group $\prod_{n,x} \text{U}(N_{n,x})$,
where $N_{n,x}$ is the number of $n$-dimensional SU(2) representations in
the vacuum, for which $\wastheta{}=x$.  Clearly, 
$\sum_{n,x} n N_{n,x}=N$.  We therefore write
\begin{equation} \label{pp:defP}
P \equiv \underset{n=1}{\overset{N}{\otimes}} \underset{x}{\otimes} P_{n,x} \in 
    \prod_{n=1}^N \prod_x \text{U}(N_{n,x}).
\end{equation}%

Turned around, generic boundary
conditions permit only the trivial vacua; for example, the irreducible
representation exists only in the sector for which the boundary
conditions are trivial ($U$ a matrix proportional to the identity).
Moreover, $U$ can be diagonalized without destroying the gauge, and so we
only need to consider $U$ diagonal, proportional to the identity in
blocks, and constant.  
Thus, in terms of the blocks, the boundary conditions are
\begin{equation} \label{blockbdy}
\begin{aligned}
X^{(x,y)i}(\tau,\sigma+2\pi) &= 
   e^{i (\phi^{(x)}-\phi^{(y)})} X^{(x,y)i}(\tau,\sigma), \\
A_\mu^{(x,y)}(\tau,\sigma+2\pi) &= 
   e^{i (\phi^{(x)}-\phi^{(y)})} A_\mu^{(x,y)}(\tau,\sigma), \\
\Psi^{(x,y)}(\tau,\sigma+2\pi) &=
   e^{i (\phi^{(x)}-\phi^{(y)})} \Psi^{(x,y)i}(\tau,\sigma).
\end{aligned}
\end{equation}

There is still further gauge
freedom---namely, the unbroken
$\prod_{n,x} \text{U}(N_{n,x})$ which allows us to,
for example, either set $X^{(x,y)}_{j j m}$
to a convenient value or keep $X^3$ diagonal.  Setting
$X^{(x,y)}_{jjm}$ to a convenient value requires an infinitesimal
gauge transformation of a generic fluctuation, but rotating $X^3$ to
make it
diagonal generically requires a finite gauge transformation.
Explicitly,
achieving the gauge for which $X^3$
is diagonal
requires setting \hbox{$X^{(x,y)3} = (\text{diagonal}) \delta_{x,y}$}.
It is clear that the analog of eq.~\eqref{x3diag} is
\begin{equation} \label{genx3diag}
X^{(x,y)}_{j j m} = \frac{1}{m} \sqrt{\frac{(j+1)(j^2-m^2)}{2 j+1}}
    X^{(x,y)}_{j-1,j,m}
  - \frac{1}{m}\sqrt{\frac{j [(j+1)^2-m^2]}{2j+1}} X^{(x,y)}_{j+1,j,m},
\quad m \neq 0,
\end{equation}
and---when $j$ takes integer values---we can still further declare
\begin{equation} \label{genx3stilldiag}
X^{(x,y)}_{jj0} = 0.
\end{equation}
This uses up all but some U(1) degrees of gauge freedom,
but results only in
\begin{equation}
X^{(x,y)3} = \begin{cases}
\sum_j \left[ \sqrt{\frac{j+1}{2j+1}} X^{(x,y)}_{j+1,j,0}
  - \sqrt{\frac{j}{2j+1}} X^{(x,y)}_{j-1,j,0} \right] Y^{(N_x,N_y)}_{j 0},
& N_x = N_y \bmod 2, \\
0, & \text{otherwise}.
\end{cases}
\end{equation}
Making $X^3$ purely diagonal would require a relation between the
fields $X^{(x,y)}_{j\pm1,j,0}$, which requires the higher order terms
in the gauge transformation.
Thus the $X^3$-diagonal
gauge is incompatible with $\vev{A_\mu}=0$.  So although the
gauge in which $X^3$ is diagonal appears to reduce the allowed
boundary conditions to $\prod_{n,x} S_{N_{n,x}}$, in fact, the full $\prod_{n,x}
\text{U}(N_{n,x})$ has simply been moved into Wilson lines.

Moreover, the
$X^3$-diagonal gauge is calculationally inconvenient.
To achieve it required a relation between
$X^{(x,y)}_{j\pm1,j,0}$,
and these have different masses.
Thus, although $X^3$ is Hermitian, and so, of course, can be chosen to
be diagonal, such a choice does not appear to be
compatible with diagonalization of
the action.  At best we can only make each $N_x \times N_y$ block
matrix in $X^3$ ``diagonal''.  Of course, this is just a manifestation
that in this gauge the gauge field is generically nontrivial, and so
the linearization of the gauge field was unjustified.  In other words,
finding the spectrum in this gauge is hard.

Finally, one might object, with regards to our argument that we only
have to consider
$\prod_{n,x} \text{U}(N_{n,x})$ instead of U$(N)$ or $S_N$, that the
ordering of vacua
was also possible in flat space; on
ordering the eigenvalues from smallest to largest, the residual gauge
transformations appear to be much smaller than $S_N$ (in fact, appear
to be
nontrivial only on a set of vacua of measure zero!).  However, this
neglects the fact that the eigenvalues can vary continuously with
$\sigma$.  Therefore, although one might order the eigenvalues at
$\sigma=0$, by $\sigma=2\pi$ they need not be ordered; the
transformation $U_P\in S_N$ is then necessary to reorder them, and
conversely, for
any element of $U_P \in S_N$, one can always arrange for $U_P$ to be
the necessary transformation.  For
the \ppwave, however, the vacua are discrete and cannot vary
continuously with $\sigma$.
This is true even for values of $\wastheta{(x)}$, which, as we
discussed in sections~\ref{sec:pert} and~\ref{sec:redfluc}, define
superselection sectors.
Thus, our discussion is sensible.

We conclude that the boundary conditions are given by eq.~\eqref{blockbdy}.
Thus the spectrum of the diagonal blocks
strings is unaffected, and the off-diagonal blocks are
``twisted''.  That is, the moding of
the $\sigma$-momentum is fractional
($n+\frac{\phi^{(x)}-\phi^{(y)}}{2\pi}$, where $n$ is integer), and
generically the fields have no periodicity.

\section{Thermodynamics} \label{sec:thermo}

In this section we make some qualitative speculations about the
thermodynamics of the {\em pp\/} matrix string theory.  More rigorous
statements, along with calculational details, are deferred
to~\cite{dms2}.

We have seen that for $M g_s \gg 1$, all of the matrix
elements are important, and not just the diagonal ones.  Thus, there
are $N^2$ degrees of freedom, even for the trivial vacuum.  
One might think that this renders the partition function for the canonical
ensemble divergent, for any temperature.
Specifically, the extra factor of $N$ degrees of freedom makes
the density of states appear to have, roughly, $e^{N}$ behaviour
instead of the $e^{\sqrt{N}}$ behaviour of Cardy's formula.  If true,
the temperature is always above the Hagedorn temperature,
and the canonical ensemble
is ill-defined for $M g_s \gg 1$.

However, the boundary condition~\eqref{blockbdy} modifies
this.  Including the thermal circle in the $\tau$-direction, the
boundary conditions are
\begin{equation} \label{blockTbdy}
\begin{aligned}
X^{(x,y)i}(\tau,\sigma+2\pi) &=
   e^{i (\phi^{(x)}-\phi^{(y)})} X^{(x,y)i}(\tau,\sigma), &
X^{(x,y)i}(\tau,\sigma+2\pi) &=
   e^{i (\theta^{(x)}-\theta^{(y)})} X^{(x,y)i}(\tau,\sigma),\\
A_\mu^{(x,y)}(\tau,\sigma+2\pi) &=
   e^{i (\phi^{(x)}-\phi^{(y)})} A_\mu^{(x,y)}(\tau,\sigma), &
A_\mu^{(x,y)}(\tau,\sigma+2\pi) &=
   e^{i (\theta^{(x)}-\theta^{(y)})} A_\mu^{(x,y)}(\tau,\sigma),\\
\Psi^{(x,y)}(\tau,\sigma+2\pi) &= 
   e^{i (\phi^{(x)}-\phi^{(y)})} \Psi^{(x,y)i}(\tau,\sigma), &
\Psi^{(x,y)}(\tau,\sigma+2\pi) &= 
   -e^{i (\theta^{(x)}-\theta^{(y)})} \Psi^{(x,y)i}(\tau,\sigma).
\end{aligned}
\end{equation}
In principle the ``$U_Q$'' for the thermal circle could be an
arbitrary U($N$) matrix; however, consistency at
$(\sigma+2\pi,\tau+2\pi)$ requires $U_P$ and $U_Q$ to commute and
therefore be simultaneously diagonalizable.

The partition function includes an integral over the boundary
conditions $\phi^{(x)}, \theta^{(x)}$; as explained in
Sec.~\ref{sec:bdy}, this is equivalent to summing over Wilson
lines.  It turns out that this integral over phases for the
off-diagonal matrix elements
suppresses their contribution to the partition function.  So, in fact,
the Hagedorn behaviour appears to persist, even in the presence of fuzzy
spheres.

\section{Conclusions} \label{sec:conc}

In this paper we have used a matrix theory formulation to study
aspects of the nonperturbative behaviour of strings in a \ppwave\ background
with a compact lightlike direction. The matrix string theory may be
characterized by a dimensionless parameter $Mg_s$. For small $Mg_s$
one has a strongly coupled two-dimensional Yang-Mills theory whose IR
limit consists of $N$ degrees of freedom and 
reproduces the standard perturbative string in the appropriate
\ppwave\ background, in a way entirely analogous to flat space.

However when $Mg_s\gg 1$ the Yang-Mills theory is essentially
perturbative.  The underlying string theory is at {\em finite} but
small string coupling. Nevertheless, this regime probes essentially
nonperturbative properties. In particular there are degenerate vacua
of the light cone Hamiltonian corresponding to BPS states representing
(multiple) fuzzy spheres of various sizes.

Perhaps the most significant result of our analysis is that the
physical fluctuations around the fuzzy sphere vacua consist of $N^2$
degrees of freedom rather than $N$ degrees of freedom which appear at
small $Mg_s$ limit. The mass spectra of these fluctuations have 
spacings which are independent of the sizes of the fuzzy spheres.
Furthermore, in a gauge where all the fields are periodic along the
string, the Wilson line degrees of freedom are now characterized by
continuous rather than discrete parameters.  Equivalently if one
transforms to a description where the Wilson lines are trivial, the
``matter'' degrees of freedom generically do not return to their
original values as one goes around the string finite number of
times. Thus as we change the parameters of the theory, strings cease
to be stringlike.

We expect these results to have non-trivial consequences for
thermodynamics.  In particular the Hagedorn behaviour should be
quantitatively, though perhaps not qualitatively,
modified. We hope to report on these aspects soon.

It would be interesting to construct and study matrix string theories
in other types of \ppwave\ backgrounds with different amounts of
supersymmetry. For example, matrix string theory for 
\ppwave{s} in IIB theory 
should allow us to study the nonperturbative behaviour of large
angular momentum states of string theory in the underlying $AdS$
background.

\acknowledgments
We thank many people for very inspiring conversations.
They include A.~Adams, P.~Argyres, V.~Balasubramanian, 
R.~Gopakumar, J.~Harvey, M.~Headrick, F.~Larsen, H.~Liu, 
G.~Mandal, S.~Mathur, H.~Morales,
P.~Mukhopadhayay, M.~Peskin and S.~Trivedi. 
S.R.D. would like to thank Tata Institute of Fundamental
Research, Mumbai 
and Indian Association for the Cultivation of Science, Kolkata,
for hospitality during the preparation of this manuscript.
J.M. is very grateful to M.~Spradlin for preserving some old
notes that otherwise would have been lost in a hard drive crash.
This work was supported in part by Department of Energy contract
\#DE-FG01-00ER45832 and NSF grant \#PHY-0071312. 

\appendix
\renewcommand\theequation{\thesection.\arabic{equation}}

\section{(Fuzzy) Spherical and Vector Spherical Harmonics} \label{sec:ylm}

\subsection{Ordinary Spherical Harmonics}

We start by reviewing some facts about ordinary spherical harmonics on
$S^2$.  

\subsubsection{Scalar Spherical Harmonics}

The scalar spherical harmonics, $Y_{\ell m}$, are
well-understood.
The property we wish to emphasize, is the one-to-one correspondence between
scalar spherical harmonics and homogeneous, harmonic polynomials on
$\ZR^3$.  Specifically,
\begin{equation}
H_{\ell m}(\boldsymbol{X}) = r^\ell Y_{\ell m}(\theta,\phi)
\end{equation}
are homogeneous, harmonic polynomials of degree $\ell$:
\begin{align}
\boldsymbol{X} \cdot \boldsymbol{\p} H_{\ell m}(\boldsymbol{X}) 
   &= \ell H_{\ell m}(\boldsymbol{X}), &
\boldsymbol{\p}^2 H_{\ell m}(\boldsymbol{X}) = 0.
\end{align}
It is straightforward to see that this is equivalent to the well-known
property
\hbox{$-\boldsymbol{\nabla}^2 Y_{\ell m}(\theta,\phi) 
= \ell(\ell+1) Y_{\ell m}(\theta,\phi)$}.  Furthermore, one can show
({\em e.g.\/}~\cite{ed}) that the
number of linearly independent harmonic polynomials that are
homogeneous of degree $\ell$ is precisely \hbox{$2 \ell+1$}.

\subsubsection{Vector Spherical Harmonics}

Vector spherical harmonics can be defined by~\cite{mathews}
\begin{equation} \label{vecyjlm}
\boldsymbol{Y}_{j\ell m}(y) = \sum_{m'=-1}^1 
  \cg{\ell}{m-m'}{1}{m'}{j}{m}
Y_{\ell,m-m'} \boldsymbol{\hat{e}}_{m'},
\end{equation}
where $\cg{j_1}{m_1}{j_2}{m_2}{j_3}{m_3}$
are Clebsch-Gordan coefficients%
\footnote{We use a notation in which Clebsch-Gordan coefficients could be
easily confused with the more symmetric 3-$j$ symbols
\hbox{$\thj{j_1}{m_1}{j_2}{m_2}{j_3}{m_3} \equiv 
\frac{(-1)^{j_1-j_2-m_3}}{\sqrt{2j_3+1}}\cg{j_1}{m_1}{j_2}{m_2}{j_3}{-m_3}$},
but the presence or absence of the vertical bar should alleviate the
confusion.\label{ft:3j}}
and
\begin{align} \label{unitvectors}
\boldsymbol{\hat{e}}_{-1} & = \frac{\boldsymbol{\hat{X}}-i
    \boldsymbol{\hat{Y}}}{\sqrt{2}} &
\boldsymbol{\hat{e}}_{0} & = \boldsymbol{\hat{Z}} &
\boldsymbol{\hat{e}}_{1} & = -\frac{\boldsymbol{\hat{X}}+i
    \boldsymbol{\hat{Y}}}{\sqrt{2}};
\end{align}
compare eq.~(\ref{unitvectors}) with expressions for
$r Y_{1\beta}(\boldsymbol{X})$.
As the Clebsch-Gordan coefficients vanish unless $j=\ell$ or
$j=\ell\pm 1$, there are three families of vector spherical harmonics.
These are often written~\cite{hill}
\begin{align}
\boldsymbol{V}_{\ell m} &= -\boldsymbol{Y}_{\ell,\ell+1,m}, &
\boldsymbol{X}_{\ell m} &= \boldsymbol{Y}_{\ell,\ell,m}, &
\boldsymbol{W}_{\ell m} &= \boldsymbol{Y}_{\ell,\ell-1,m}.
\end{align}
We will use both notations here, but in the main text we use
$\boldsymbol{Y}_{j \ell m}$ to avoid confusing the matrix string field
with a vector spherical harmonic.
They obey the reality property
\begin{equation}
\boldsymbol{Y}_{j \ell m}^* = (-1)^{m+\ell+1-j} \boldsymbol{Y}_{j,\ell, -m}
\end{equation}
and are orthonormal
\begin{equation}
\int d\Omega_2 \boldsymbol{Y}_{j \ell m} \cdot \boldsymbol{Y}_{j' \ell' m'}^*
 = \delta_{j j'} \delta_{\ell \ell'} \delta_{m m'}.
\end{equation}

Alternatively, there are essentially three ``natural'' ways of
generating vectors out of the scalar spherical harmonics, namely
\begin{equation}
\boldsymbol{J} H_{\ell m},
\boldsymbol{\p} H_{\ell m},
\boldsymbol{\hat{r}} H_{\ell m}.
\end{equation}
These are related to the vector spherical harmonics via~\cite{hill}%
\footnote{We have included a radial factor for later convenience.%
  \iftoomuchdetail
  \begin{detail} 
  This requires the additional relation
  \hbox{$\boldsymbol{\hat{r}} \ell r^{\ell-1} Y_{\ell m} 
  = \frac{\ell}{\sqrt{2\ell+1}} r^{\ell-1} [\sqrt{\ell} \boldsymbol{W}_{\ell m}
       - \sqrt{\ell+1} \boldsymbol{V}_{\ell m}]$}.%
  \end{detail}%
  \fi
}
\begin{subequations} \label{natYjlm}
\begin{align}
\boldsymbol{J} H_{\ell m} &= \sqrt{\ell(\ell+1)}
    r^\ell \boldsymbol{X}_{\ell m}, \\
\boldsymbol{\p} H_{\ell m} &= \sqrt{\ell(2\ell+1)} r^{\ell-1}
   \boldsymbol{W}_{\ell m}, \\
\boldsymbol{X} H_{\ell m} &= -\sqrt{\frac{\ell+1}{2\ell+1}} 
r^{\ell+1} \boldsymbol{V}_{\ell m} 
+ \sqrt{\frac{\ell}{2\ell+1}} r^{\ell+1} \boldsymbol{W}_{\ell m}
 = -\sqrt{\frac{\ell+1}{2\ell+1}} 
r^{\ell+1} \boldsymbol{V}_{\ell m} + \frac{r^2}{2\ell+1}
   \boldsymbol{\p} H_{\ell m}.
\end{align}
\end{subequations}
From either~\eqref{natYjlm} or~\eqref{vecyjlm}, we can compute
$\boldsymbol{V}_{\ell m}$ and
$\boldsymbol{W}_{\ell m}$ explicitly.  We find,
\begin{align}
\begin{split}
\boldsymbol{W}_{\ell m} &= \frac{1}{\sqrt{\ell(2\ell-1)}} \left\{
  -\boldsymbol{e}_{-1} \sqrt{\frac{(\ell-m)(\ell-m-1)}{2}} Y_{\ell-1,m+1}
\right. \\ & \qquad \qquad \left.
  -\boldsymbol{e}_{1}
       \sqrt{\frac{(\ell+m)(\ell+m-1)}{2}} Y_{\ell-1,m-1}
  +\boldsymbol{e}_0 \sqrt{\ell^2-m^2} Y_{\ell-1,m} \right\},
\end{split} \\
\begin{split}
\boldsymbol{V}_{\ell m} &= \frac{1}{\sqrt{(\ell+1)(2\ell+3)}} \left\{
  \boldsymbol{e}_{-1} \sqrt{\frac{(\ell+m+1)(\ell+m+2)}{2}} 
         Y_{\ell+1,m+1}
\right. \\ & \qquad \qquad \left.
  +\boldsymbol{e}_{1} \sqrt{\frac{(\ell+m)(\ell+m-1)}{2}}
         Y_{\ell+1,m-1}
  +\boldsymbol{e}_0 \sqrt{(\ell+1)^2-m^2} Y_{\ell+1,m} \right\}.
\end{split}
\end{align}

\subsubsection{Spinor Spherical Harmonics}

By analogy with eq.~\eqref{vecyjlm}, we can define spinor
spherical harmonics in terms of the scalar spherical harmonics via
\begin{equation} \label{defsylm}
S^\alpha_{j \ell m} = \cg{\ell}{m-\alpha}{\frac{1}{2}}{\alpha}{j}{m}
   Y_{\ell,m-\alpha}, -j \leq m \leq j.
\end{equation}
Here $\alpha=\pm \frac{1}{2}$ is a spinor index, and clearly
$j=\ell\pm \frac{1}{2}$ and $m$ is a half-integer.
Explicitly,~\cite{merz}
\begin{align} \label{sylm}
\boldsymbol{S}_{\ell+\frac{1}{2},\ell,m} &=
   \begin{pmatrix} \sqrt{\frac{\ell+m+\frac{1}{2}}{2\ell+1}}
         Y_{\ell,m-\frac{1}{2}} \\
      \sqrt{\frac{\ell-m+\frac{1}{2}}{2\ell+1}}
         Y_{\ell,m+\frac{1}{2}} \end{pmatrix}, &
\boldsymbol{S}_{\ell-\frac{1}{2},\ell,m} &=
   \begin{pmatrix} -\sqrt{\frac{\ell-m+\frac{1}{2}}{2\ell+1}}
         Y_{\ell,m-\frac{1}{2}} \\
      \sqrt{\frac{\ell+m+\frac{1}{2}}{2\ell+1}}
         Y_{\ell,m+\frac{1}{2}} \end{pmatrix}.
\end{align}
These are clearly orthonormal
\begin{equation}
\int d\Omega_2 \boldsymbol{S}_{j \ell m}^\dagger 
  \boldsymbol{S}_{j'\ell' m'} = \delta_{jj'} \delta_{\ell \ell'}
  \delta_{m m'}.
\end{equation}
Also, they obey the reality condition
\begin{equation}
\sigma^2 \boldsymbol{S}_{j,\ell,m}^* 
= -i (-1)^{j+\ell+m} \boldsymbol{S}_{j,\ell,-m};
\end{equation}
$\sigma^2$ appears as it is the charge conjugation matrix.

\subsection{Fuzzy Spherical Harmonics} \label{app:ylm}

\subsubsection{Scalar Spherical Harmonics}

The coordinates of the fuzzy sphere are $\boldsymbol{J}$.  So, while
it is difficult to generalize the spherical harmonics themselves, it
is trivial to generalize $H_{\ell m}(\boldsymbol{X})$.  The only
subtlety is ordering, but since the commutator of two $J$'s gives a
$J$, it is clear that one should use the symmetric ordering.  Up to
normalization, this defines the matrix scalar spherical harmonics,
which we also denote $Y_{\ell m}$, hopefully without confusion.

Equivalently, in an $N$-dimensional irreducible representation of SU(2),
we define
$Y_{\ell,m}$ recursively as
\begin{equation} \label{defYlm}
Y_{\ell m} 
= \frac{1}{\sqrt{(\ell+m+1)(\ell-m)}} \com{J^-}{Y_{\ell,m+1}},
\quad
Y_{\ell\ell} 
= (-1)^\ell \frac{\sqrt{N}}{\ell!}
\sqrt{\frac{(2\ell+1)!(N-\ell-1)!}{(N+\ell)!}} (J^+)^\ell.
\end{equation}
It is sometimes convenient---especially for multiple commutators---to
use the Lie derivative to denote a
commutator
;
we also introduce the normalization
constant
\begin{equation}
N_\ell = (-1)^\ell \frac{\sqrt{N}}{\ell!}
\sqrt{\frac{(2\ell+1)!(N-\ell-1)!}{(N+\ell)!}},
\end{equation}
so that
\begin{equation}
Y_{\ell m} = \frac{1}{\sqrt{(\ell+m+1)(\ell-m)}} \lie{J^-}{Y_{\ell,m+1}},
\qquad
Y_{\ell\ell} = N_\ell (J^+)^\ell = (-1)^\ell \abs{N_\ell} (J^+)^\ell.
\end{equation}
The sign ensures that, written as polynomials in $\boldsymbol{X}$,
$Y_{\ell m}^{\text{fuzzy}} ``{=}\text{''} H_{\ell m}^{\text{classical}}$.

It is straightforward to check that the definition~\eqref{defYlm}
ensures the usual properties,
\begin{gather} 
\begin{align*}
\com{J^\pm}{Y_{\ell m}} &= 
   \sqrt{(\ell\mp m)(\ell \pm m + 1)} Y_{\ell m\pm1}, &
\com{J^3}{Y_{\ell m}} &= m Y_{\ell m},
\end{align*} \\ \label{jony}
\com{J^a}{\com{J^a}{Y_{\ell m}}} = \ell(\ell+1) Y_{\ell m}, \\
Y^\dagger_{\ell m} = (-1)^m Y_{\ell, -m}.
\end{gather}
\iftoomuchdetail
\begin{detail}%
(The relation $Y_{\ell, -m} = (-1)^m Y_{\ell m}^\dagger$, follows from
eqs.~\eqref{jony}, and the fact that normalization must imply
that $Y_{\ell,-\ell} = \pm \abs{N_\ell} (J^-)^\ell$.  The sign follows
from
$\com{J^-}{\com{J^-}{J^+}} = -2 J^-$.)
\end{detail}%
\fi
$N_\ell$ has been chosen for the normalization
\begin{equation}
\Tr\biggl[Y_{\ell m}^\dagger Y_{\ell' m'}\biggr] 
    = N \delta_{\ell\ell'}\delta_{m m'}.
\end{equation}
The consistency of this normalization as $m$ varies is
easy to check, and the orthogonality follows from the commutation
relations with $J^3$ (for $m$) and $(\lie{J^a}{})^2$ (for $\ell$).  In
particular, this implies that the $Y_{\ell m}$'s are linearly independent.
However, it is worth mentioning that the normalization, $N_\ell$,
depends on $N$ as well as on $\ell$ in a nontrivial way.

\iftoomuchdetail
\begin{detail}%
It is straightforward to compute $N_\ell$.  Specifically,
\begin{equation}
\begin{split}
\Tr (J^-)^\ell (J^+)^\ell 
&= \sum_{m=-{\frac{N-1}{2}}}^{\frac{N-1}{2}}
\frac{(\frac{N-1}{2}+\ell+m)! (\frac{N-1}{2}-m)!}{%
   (\frac{N-1}{2}+m)!(\frac{N-1}{2}-m-\ell)!}
\\ 
&= \sum_{m=0}^{N-1-\ell} \frac{(\ell+m)!(N-1-m)!}{m! (N-1-\ell-m)!} \\
&= \frac{\ell! (N-1)!}{(N-\ell-1)!}
  \sum_{m=0}^{N-1-\ell} \frac{1}{m!} \frac{(\ell+1)_m (\ell-N+1)_m}{(1-N)_n}
\\
&= \frac{\ell! (N-1)!}{(N-\ell-1)!} {_2F_1}(\ell+1,\ell+1-N;1-N;1) \\
&= \frac{\ell!^2 (N+\ell)!}{(2\ell+1)!(N-\ell-1)!}.
\end{split}
\end{equation}
In the last step we have analytically continued in the arguments of
the hypergeometric function to apply a standard formula outside its
realm of validity.  (This is justified by the fact that the final
expression is well-defined.) Then we see that
\begin{equation}
\abs{N_\ell} = \frac{\sqrt{N}}{\ell!}
\sqrt{\frac{(2\ell+1)!(N-\ell-1)!}{(N+\ell)!}}.
\end{equation}
\end{detail}%
\fi

More generally, since the spherical harmonics are acted on by
angular momentum generators, it is natural to consider the $N_1 \times
N_2$ matrix
whose matrix elements are
\begin{equation} \label{Ylm12}
[Y^{(N_1,N_2)}_{j m}]_{M_1 M_2} 
= (N_1 N_2)^{1/4} (-1)^{\frac{N_2-1}{2}-M_2}
\cg{\frac{N_1-1}{2}}{M_1}{\frac{N_2-1}{2}}{-M_2}{j}{m},
\qquad \begin{matrix}
\frac{\abs{N_1-N_2}}{2} \leq j \leq \frac{N_1+N_2}{2}+1, \\
-\frac{N_1-1}{2} \leq M_1 \leq \frac{N_1-1}{2}, \\
-\frac{N_2-1}{2} \leq M_2 \leq \frac{N_2-1}{2}.
\end{matrix}
\end{equation}
Using the standard representation of the angular momentum generators
$J^a$, and the properties of the Clebsch-Gordan coefficients,~\cite{merz}
\begin{gather}
\cg{j_1}{m_1}{j_2}{m_2}{j_3}{m_3}
= \cg{j_2}{-m_2}{j_1}{-m_1}{j_3}{-m_3}, \\
\begin{split} \label{recurcg}
\sqrt{(j_1\mp m_1)(j_1 \pm m_1+1)} &\cg{j_1}{m_1\pm 1}{j_2}{m_2}{j_3}{m_3}
+ \sqrt{(j_2\mp m_2)(j_2 \pm m_2+1)} \cg{j_1}{m_1}{j_2}{m_2 \pm 1}{j_3}{m_3}
\\ &= \sqrt{(j_3 \pm m_3)(j_3 \mp m_3+1)} 
   \cg{j_1}{m_1}{j_2}{m_2}{j_3}{m_3 \mp 1},
\end{split} \\
\sum_{m_1,m_2} \cg{j_1}{m_1}{j_2}{m_2}{j_3}{m_3}
  \cg{j_1}{m_1}{j_2}{m_2}{j'_3}{m'_3} = \delta_{j_3,j'_3} \delta_{m_3,m'_3},
\end{gather}
we easily deduce that (here, we loosely write
$\com{J^a}{\cdot}$ for the difference between
left multiplication 
and right multiplication 
with representations of the appropriate dimension)
\begin{gather} \label{startylmprops}
\begin{align}
\com{J^\pm}{Y^{(N_1,N_2)}_{j m}} &= 
   \sqrt{(j\mp m)(j \pm m + 1)} Y^{(N_1,N_2)}_{j m\pm1}, &
\com{J^3}{Y^{(N_1,N_2)}_{j m}} &= m Y^{(N_1,N_2)}_{j m},
\end{align} \\ \label{jonalty}
\com{J^a}{\com{J^a}{Y^{(N_1,N_2)}_{j m}}} = j(j+1) Y^{(N_1,N_2)}_{j m}, \\
Y^{(N_1,N_2)\dagger}_{j m} 
= (-1)^{m-\frac{N_1-N_2}{2}} Y^{(N_2,N_1)}_{j, -m}, \\
\Tr Y^{(N_1,N_2)\dagger}_{j m} Y^{(N_1,N_2)}_{j' m'}
= \sqrt{N_1 N_2} \delta_{j j'} \delta_{m m'}.
\label{endylmprops}
\end{gather}
In particular, we see that%
\footnote{The equality follows from linear independence and
  normalization, but only up to a phase.  We can check the phase by
  examining the sign of the non-zero matrix element
  \hbox{$[Y_{\ell \ell}]_{\frac{N-1}{2},\frac{N-1}{2}-\ell} = (-1)^\ell
  \abs{N_\ell} \sqrt{\frac{(N-1)! \ell!}{(N-1-\ell)!}} \stackrel{?}{=}
  \sqrt{N} (-1)^\ell \cg{\frac{N-1}{2}}{\frac{N_1}{2}}{
     \frac{N-1}{2}}{\ell-\frac{N-1}{2}}{\ell}{\ell}.$}
  By definition~\cite{merz}, this Clebsch-Gordan coefficient is real
  and positive, and so we confirm that the phase is correct.}
\begin{equation} \label{altYlm}
[Y_{\ell m}]_{MM'} = [Y^{(N,N)}_{\ell m}]_{MM'}
= \sqrt{N} (-1)^{\frac{N-1}{2}-M'}
      \cg{\frac{N-1}{2}}{M}{\frac{N-1}{2}}{-M'}{\ell}{m}.
\end{equation}
Whether to use eq.~\eqref{defYlm} or eq.~\eqref{altYlm} as
the definition of the fuzzy spherical harmonics is according to
taste; however, we have seen that the definition~\eqref{altYlm} immediately
generalizes to non-square matrices~\eqref{Ylm12}.  These are used for
nontrivial, non-irreducible representations~\cite{dsv1}.

We can also observe that $J^a$ commutes with $\sum_{m=-\ell}^\ell
Y_{\ell m} Y_{\ell m}^\dagger$. By Schur's lemma, this implies that
\begin{equation}
\sum_{m=-j}^j
Y^{(N_1,N_2)}_{j m} Y^{(N_1,N_2)\dagger}_{j m}
   = (2j+1) \sqrt{\frac{N_2}{N_1}} \one,
\end{equation}
where the proportionality constant follows upon taking the trace and
using the aforementioned normalization.%
\footnote{This is reminiscent of the classical ``sum rule''
$\sum_{m=-\ell}^\ell \abs{Y_{\ell m}(\theta, \phi)}^2 = \frac{2 \ell+1}{4\pi}$.}

Note an important difference between the classical and fuzzy spherical
harmonics.  Because $(J^+)^N=0$%
\iftoomuchdetail
\begin{detail}%
\ (labeling the columns and rows of the $N\times N$ matrices by the
value of $m$ for this $N$-dimensional representation of SU(2),
$(J^+)^N$ would take the $m=-\frac{N-1}{2}$ state to
\hbox{$m=\frac{N+1}{2}>j=\frac{N-1}{2}$})%
\end{detail}
\else
,
\fi
there are only a finite number of fuzzy spherical harmonics, namely
$\ell\leq N-1$, for a total of $N^2$.  Thus, the fuzzy spherical
harmonics form an orthogonal basis for the $N\times N$ matrices.
[Alternatively, and more generally, the Clebsch-Gordan coefficients
  restrict $j$ as
  quoted in eq.~\eqref{altYlm}; thus the number of
  $Y^{(N_1,N_2)}_{jm}$ is $N_1 N_2$, and they form an orthogonal
  basis for the $N_1 \times N_2$ matrices.]
Note that $Y_{00} = \one$.

For computations involving interactions and higher order terms in the
gauge transformations, one
needs a formula for the products of spherical harmonics.  Since the
spherical harmonics form a basis, a product of spherical harmonics is a
sum of spherical harmonics.  Specifically, we have found
\begin{multline} \label{prody}
Y_{j_1 m_1}^{(N_1,N_2)} Y_{j_2 m_2}^{(N_2,N_3)}
= \sqrt{N_2} \sqrt{(2j_1+1)(2j_2+1)}
  (-1)^{2j_1-\frac{N_1-N_3}{2}+N_2-1}
\\* \times
  \sum_{j_3} (-1)^{j_3} \cg{j_1}{m_1}{j_2}{m_2}{j_3}{m_1+m_2}
  \sixj{j_1}{j_2}{j_3}{\frac{N_3-1}{2}}{\frac{N_1-1}{2}}{\frac{N_2-1}{2}}
  Y_{j_3, m_1+m_2}^{(N_1,N_3)},
\end{multline}
where $\sixj{j_1}{j_2}{j_3}{j_4}{j_5}{j_6}$ is the 6-$j$~symbol;
see {\em e.g.\/}~\cite{ed}.
The Clebsch-Gordan coefficient in
eq.~\eqref{prody} is easy to understand since the coefficients on
the right-hand side are given by $(N_1 N_3)^{-1/2} \Tr Y_{j_1 m_1}
Y_{j_2 m_2} Y^\dagger_{j_3 m_3}$.  Inserting $\lie{J^\pm}{}$ into the
trace, and using the fact that the trace of a commutator is zero,
results in recursion relations for these coefficients which are
identical to those [eq.~\eqref{recurcg}]
for the Clebsch-Gordan coefficients.  Thus, the
$m_1, m_2$-dependence of the right-hand side is determined, up to a
$j_1,j_2,j_3$-dependent normalization.  The normalization is most
easily found by comparing the explicit trace written
using the expression~\eqref{altYlm} for the spherical harmonics, with
{\em e.g.\/}~eq.~(6.2.8) of~\cite{ed} which expresses a sum of three
Clebsch-Gordan coefficients in terms of the 6-j symbol.  This method,
of course, reproduces
the Clebsch-Gordan coefficient.

\subsubsection{Vector Spherical Harmonics}

We can now define the vector spherical harmonics in parallel to the
ordinary sphere.  We set, with $-j \leq m \leq j$,
\begin{subequations} \label{myjlm}
\begin{align}
\boldsymbol{Y}^{(N_1,N_2)}_{j j m} &= 
\boldsymbol{X}^{(N_1,N_2)}_{j m} = \frac{1}{\sqrt{j(j+1)}}
   \com{\boldsymbol{J}}{Y^{(N_1,N_2)}_{j m}}, 
   \tfrac{\abs{N_1-N_2}}{2} \leq j \leq \tfrac{N_1+N_2}{2}-1; j \neq 0, \\
\begin{split}
\boldsymbol{Y}^{(N_1,N_2)}_{j, j-1, m} &= 
\boldsymbol{W}^{(N_1,N_2)}_{j m} = \frac{1}{\sqrt{j(2j-1)}} \left\{
   \boldsymbol{e}_{-1} \sqrt{\frac{(j-m)(j-m-1)}{2}} 
   Y^{(N_1,N_2)}_{j-1,m+1}
\right. \\ &  \left.
  +\boldsymbol{e}_{1}
       \sqrt{\frac{(j+m)(j+m-1)}{2}} Y^{(N_1,N_2)}_{j-1,m-1}
  +\boldsymbol{e}_0 \sqrt{j^2-m^2} Y^{(N_1,N_2)}_{j-1,m} \right\},
\\ & \qquad
   \tfrac{\abs{N_1-N_2}}{2}+1\leq j \leq \tfrac{N_1+N_2}{2},
\end{split} \\
\begin{split}
\boldsymbol{Y}^{(N_1,N_2)}_{j, j+1, m} &= 
-\boldsymbol{V}^{(N_1,N_2)}_{j m} = \frac{1}{\sqrt{(j+1)(2j+3)}}
 \left\{
  \boldsymbol{e}_{-1} \sqrt{\frac{(j+m+1)(j+m+2)}{2}} 
         Y^{(N_1,N_2)}_{j+1,m+1}
\right. \\ &  \left. \negthickspace\negthickspace 
  +\boldsymbol{e}_{1} \sqrt{\frac{(j-m+1)(j-m+2)}{2}}
         Y^{(N_1,N_2)}_{j+1,m-1}
  -\boldsymbol{e}_0 \sqrt{(j+1)^2-m^2} Y^{(N_1,N_2)}_{j+1,m} \right\},
\\ & \qquad
   \tfrac{\abs{N_1-N_2}}{2}-1 \leq j \leq \tfrac{N_1+N_2}{2}-2; j\geq 0.
 \raisetag{3\baselineskip}
\end{split}
\end{align}
\end{subequations}
The restrictions on $j$ follow from the range of $j$ in 
$Y^{(N_1,N_2)}_{j m}$, plus the fact that because $Y_{00}=\one$, it necessarily
commutes with $\boldsymbol{J}$.  There are therefore 
\hbox{$[N_1 N_2-\delta_{N_1,N_2}]$}
$\boldsymbol{X}_{j m}$'s, \hbox{$[\min(N_1,N_2)(\max(N_1,N_2)+2)]$}
$\boldsymbol{W}_{j m}$'s and
\hbox{$[\min(N_1,N_2)(\max(N_1,N_2)-2)+\delta_{N_1,N_2}]$}
$\boldsymbol{V}_{j m}$'s, for a grand total of $3 N_1 N_2$, the same as
the number of linearly independent triplets of $N_1\times N_2$
matrices.%
\iftoomuchdetail
\begin{detail}%
\footnote{For the counting of $\boldsymbol{V}_{jm}$s, note that
for $\abs{N_1-N_2}=\frac{1}{2}$, the illegal $j=-\frac{1}{2}$ are
$2j+1=0$ in number.}
\end{detail}%
\fi

Some useful identities are
(we drop the dimension-specifying superscripts where they are trivial)
\begin{gather} \label{startyvecprops} 
\boldsymbol{Y}_{j \ell m}^{(N_1,N_2)\dagger}
= (-1)^{j-\ell+m+1-\frac{N_1-N_2}{2}} \boldsymbol{Y}_{j,\ell,-m}^{(N_2,N_1)},
\\
\begin{gathered}
\begin{aligned}
\epsilon_{abc}\com{J^b}{Y^c_{j j m}} &= i
   Y^a_{j j m}, &
&&
\epsilon_{abc}\com{J^b}{Y^c_{j-1, j m}} &= i (j+1)
   Y^a_{j-1,j, m},
\end{aligned} \\
\epsilon_{abc}\com{J^b}{Y^c_{j+1,j, m}} = -i j
   Y^a_{j+1,j, m},
\end{gathered} \\
\begin{align}
\com{J^a}{X^a_{j m}} &= \sqrt{j(j+1)} Y_{j m}, &
\com{J^a}{W^a_{j m}} &= 0, &
\com{J^a}{V^a_{j m}} &= 0,
\end{align} \\
\com{J^a}{\com{J^a}{\boldsymbol{Y}_{j \ell m}}} 
   = \ell(\ell+1) \boldsymbol{Y}_{j \ell m}, \\
\label{vnorm}
\Tr \boldsymbol{Y}^{(N_1,N_2)}_{j \ell m} \cdot 
    \boldsymbol{Y}_{j' \ell' m'}^{(N_1,N_2)\dagger}
= \sqrt{N_1 N_2} \delta_{j j'} \delta_{\ell \ell'} \delta_{m m'}.
\end{gather}
In particular, the inner product~\eqref{vnorm} shows that these vector
spherical harmonics are linearly independent, and therefore form a
complete, orthogonal basis for vectors of $N\times N$ matrices.
In the main text we exclusively use the
$\boldsymbol{Y}_{j \ell m}$ notation, to avoid confusing the matrix
string field with a vector spherical harmonic.

It is useful to be able to write products in terms of the basis.
Using eq.~\eqref{prody} and the definition~\eqref{vecyjlm}, one
finds that
\begin{multline} \label{ylmyjlm}
Y^{(N_1,N_2)}_{j_1 m_1} Y^{(N_2,N_3)a}_{j_2 \ell_2 m_2}
= \sqrt{N_2} \sqrt{(2 j_1+1)(2 \ell_2+1)(2 j_2+1)}
  (-1)^{j_1-\ell_2-\frac{N_1-N_3}{2}+N_2}
\\* \times
  \sum_{j_3,\ell_3} (-1)^{j_3-\ell_3} \sqrt{2 \ell_3+1}
  \cg{j_1}{m_1}{j_2}{m_2}{j_3}{m_1+m_2}
  \sixj{j_1}{j_3}{j_2}{1}{\ell_2}{\ell_3}
  \sixj{j_1}{\ell_2}{\ell_3}{\frac{N_3-1}{2}}{\frac{N_1-1}{2}}{\frac{N_2-1}{2}}
  Y^{(N_1,N_3)a}_{j_3,\ell_3,m_1+m_2},
\end{multline}
and
\begin{multline} \label{yjlmylm}
Y^{(N_1,N_2)a}_{j_2 \ell_2 m_2} Y^{(N_2,N_3)}_{j_1 m_1} 
= \sqrt{N_2} \sqrt{(2 j_1+1)(2 \ell_2+1)(2 j_2+1)}
  (-1)^{2 j_1-\frac{N_1-N_3}{2}+N_2}
\\* \times
  \sum_{j_3,\ell_3} (-1)^{j_3} \sqrt{2 \ell_3+1}
  \cg{j_1}{m_1}{j_2}{m_2}{j_3}{m_1+m_2}
  \sixj{j_1}{j_3}{j_2}{1}{\ell_2}{\ell_3}
  \sixj{\ell_2}{j_1}{\ell_3}{\frac{N_3-1}{2}}{\frac{N_1-1}{2}}{\frac{N_2-1}{2}}
  Y^{(N_1,N_3)a}_{j_3,\ell_3,m_1+m_2}.
\end{multline}
Therefore, for square matrices,
\begin{multline} \label{comylmyjlm}
\com{Y_{\ell_1 m_1}}{Y^a_{j_2 \ell_2 m_2}}
= \sqrt{N} \sqrt{(2 \ell_1+1)(2 \ell_2+1)(2 j_2+1)}
  (-1)^{N+1}
\\* \times
  \sum_{j_3,\ell_3} (-1)^{j_3} \left[1-(-1)^{\ell_1+\ell_2+\ell_3}\right]
  \sqrt{2 \ell_3+1}
  \cg{\ell_1}{m_1}{j_2}{m_2}{j_3}{m_1+m_2}
\\* \times
  \sixj{\ell_1}{j_3}{j_2}{1}{\ell_2}{\ell_3}
  \sixj{\ell_1}{\ell_2}{\ell_3}{%
     \frac{N-1}{2}}{\frac{N-1}{2}}{\frac{N-1}{2}}
  Y^{a}_{j_3,\ell_3,m_1+m_2};
\end{multline}
thus only odd $\ell_1+\ell_2+\ell_3$ contribute to the commutator, but
there is no (obvious) restriction on $j_3$.
We should note that one of the 6-$j$ symbols in these expressions
is obviously that from eq.~\eqref{prody} and the presence of the
other is reminiscent of the classical formula~\cite{mathews} and could
presumably be deduced from recursion relations.

\subsubsection{Spinor Spherical Harmonics}

The spinor spherical harmonics are also written in parallel to the
ordinary sphere.  That is,
\begin{align} \label{fsylm}
\boldsymbol{S}^{(N_1,N_2)}_{j+\frac{1}{2},j,m} &=
   \begin{pmatrix} \sqrt{\frac{j+m+\frac{1}{2}}{2j+1}}
         Y^{(N_1,N_2)}_{j,m-\frac{1}{2}} \\
      \sqrt{\frac{j-m+\frac{1}{2}}{2j+1}}
         Y^{(N_1,N_2)}_{j,m+\frac{1}{2}} \end{pmatrix},
\qquad 0\leq j \leq N-1, -j-\frac{1}{2} \leq m \leq
      j+\frac{1}{2} \\
\boldsymbol{S}^{(N_1,N_2)}_{j-\frac{1}{2},j,m} &=
   \begin{pmatrix} -\sqrt{\frac{j-m+\frac{1}{2}}{2j+1}}
         Y^{(N_1,N_2)}_{j,m-\frac{1}{2}} \\
      \sqrt{\frac{j+m+\frac{1}{2}}{2j+1}}
         Y^{(N_1,N_2)}_{j,m+\frac{1}{2}} \end{pmatrix},
\qquad 0\leq j \leq N-1, -j+\frac{1}{2} \leq m \leq
      j-\frac{1}{2}.
\end{align}
The restriction on $j$ is obvious and the restriction on $m$
follows from the usual restriction on $m$, along with an analysis of
the coefficients.  We count \hbox{$[N_1 N_2{+}\min(N_1,N_2)]$}~%
$\boldsymbol{S}^{(N_1,N_2)}_{j+\frac{1}{2},j,m}$\,'s and 
\hbox{$[N_1 N_2{-}\min(N_1,N_2)]$}~%
$\boldsymbol{S}^{(N_1,N_2)}_{j-\frac{1}{2},j,m}$\,'s, for a total of $2
N_1 N_2$
spinor spherical harmonics, as expected.

Some useful identities are
\begin{gather} \label{Sjlmmajorana}
\begin{align} 
\sigma^2 \boldsymbol{S}^{(N_1,N_2)*}_{j+\frac{1}{2},j,m} &=
(-1)^{m-\frac{N_1-N_2}{2}} \boldsymbol{S}^{(N_2,N_1)}_{j+\frac{1}{2},j,-m}, &
\sigma^2 \boldsymbol{S}^{(N_1,N_2)*}_{j-\frac{1}{2},j,m} &=
(-1)^{m+1-\frac{N_1-N_2}{2}}
    \boldsymbol{S}^{(N_2,N_1)}_{j-\frac{1}{2},j,-m}, \end{align} \\
 \label{JonSjlm} \begin{align} 
\sigma^a \com{J^a}{\boldsymbol{S}_{j+\frac{1}{2},j,m}}
  &= j \boldsymbol{S}_{j+\frac{1}{2},j,m}, &
\sigma^a \com{J^a}{\boldsymbol{S}_{j-\frac{1}{2},j,m}}
  &= -(j+1) \boldsymbol{S}_{j-\frac{1}{2},j,m},
\end{align} \\
\com{J^a}{\com{J^a}{\boldsymbol{S}_{j\ell m}}} 
   = \ell(\ell+1)\boldsymbol{S}_{j\ell m}, \\
\Tr \boldsymbol{S}_{j \ell m}^{(N_1,N_2)*}
    \boldsymbol{S}^{(N_1,N_2)}_{j' \ell' m'}
   = \sqrt{N_1 N_2} \delta_{jj'} \delta_{\ell \ell'} \delta_{m m'},
\label{endSjlmprops}
\end{gather}
where $\sigma^a$ are the Pauli matrices,
and the complex conjugation includes Hermitian conjugation of
the U$(N)$ matrices.
In particular, we see that the spinor spherical harmonics are linearly
independent and therefore form a basis.


\begin{thebibliography}{99}

\bibitem{met} R. R. Metsaev,
\ct{Type IIB Green-Schwarz superstring in plane wave Ramond-Ramond
background}
\npb{625}{2002}{70--96};
\phepth{0112044}.

\bibitem{pen} M. Blau, J. Figueroa-O'Farrill, C.~Hull and
G.~Papadopoulos,
\ct{Penrose limits and maximal supersymmetry}
\cqg{19}{2002}{L87--L95};
\phepth{0201081}.

\bibitem{bmn} D. Berenstein, J.~Maldacena and H.~Nastase,
\ct{Strings in flat space and pp waves from ${\mathcal{N}}=4$ Super
Yang Mills}
\jhep{04}{2002}{013};
\phepth{0202021}.

\bibitem{bfss}
T.~Banks, W.~Fischler, S.~H.~Shenker and L.~Susskind,
\ct{M theory as a matrix model: A conjecture},
\citeprd{55}{1997}{5112--5128};
\phepth{9610043}.

\bibitem{dsv1} K. Dasgupta, M.~M.~Sheikh-Jabbarri and M.~Van Raamsdonk,
\ct{Matrix Perturbation Theory For M-theory On a PP-Wave}
\jhep{05}{2002}{056}; \phepth{0205185}.

\bibitem{dsv2} K. Dasgupta, M.~M.~Sheikh-Jabbarri and M.~Van Raamsdonk,
\ct{Protected Multiplets of M-Theory on a Plane Wave}
\jhep{09}{2002}{021}; \phepth{0207050}.

\bibitem{kp} N. Kim and J. Plefka,
\ct{On the Spectrum of PP-Wave Matrix Theory}
\npb{643}{2002}{31--48};
\phepth{0207034}.

\bibitem{kp2} N. Kim and J.-H.~Park,
\ct{Superalgebra for M-theory on a pp-wave}
\citeprd{66}{2002}{106007};
\phepth{0207061}.

\bibitem{jm1} J. Michelson,
\ct{(Twisted) Toroidal Compactification of pp-Waves}
\citeprd{66}{2002}{066002};
\phepth{0203140}

\bibitem{lm} L. Motl,
\ct{Proposals on Nonperturbative Superstring Interactions}
HEP-UK-0003, \hepth{9701025}.

\bibitem{bs} T. Banks and N. Seiberg,
\ct{Strings from Matrices}
\npb{497}{1997}{41--55};
\phepth{9702187}.

\bibitem{dvv} R. Dijkgraaf, E. Verlinde and H.~Verlinde,
\ct{Matrix String Theory}
\npb{500}{1997}{43--61};
\phepth{9703030}.

\bibitem{gb} G. Bonelli,
\ct{Matrix Strings in pp-wave backgrounds from deformed Super
Yang-Mills Theory}
\jhep{08}{2002}{022};
\phepth{0205213}.

\bibitem{sy} K. Sugiyama and K. Yoshida,
\ct{Type IIA String and Matrix String on PP-wave}
\npb{644}{2002}{128--150}; \phepth{0208029}.

\bibitem{hs} S. Hyun and H. Shin,
\ct{$N$=(4,4) Type IIA String Theory on PP-Wave Background}
\jhep{10}{2002}{070}; \phepth{0208074}.

\bibitem{hs2} S. Hyun and H. Shin,
\ct{Solvable $N=(4,4)$ Type IIA String Theory in Plane-Wave Background
and D-Branes}
\npb{654}{2003}{114--134};
\phepth{0210158}.

\bibitem{bfhp} M. Blau, J. Figueroa-O'Farrill, C. Hull and
G. Papadopoulos,
\ct{A New Maximally Supersymmetric Background of IIB Superstring Theory}
\jhep{01}{2002}{047};
\phepth{0110242}.

\bibitem{gopakumar}
R.~Gopakumar,
\ct{String interactions in PP-waves}
\prl{89}{2002}{171601};
\phepth{0205174}.

\bibitem{constable}
N.~R.~Constable, D.~Z.~Freedman, M.~Headrick, S.~Minwalla, L.~Motl,
A.~Postnikov and W.~Skiba,
\ct{PP-wave string interactions from perturbative Yang-Mills theory}
\jhep{07}{2002}{017};
\phepth{0205089}.

\bibitem{myers} R. C. Myers,
\ct{Dielectric-Branes}
\jhep{12}{1999}{022};
\phepth{9910053}.

\bibitem{tv} W. Taylor and M. Van Raamsdonk,
\ct{Multiple D$p$-branes in Weak Background Fields}
\npb{573}{2000}{703--734};
\phepth{9910052}.

\bibitem{giant} J. McGreevy, L. Susskind and N.~Toumbas,
\ct{Invasion of the Giant Gravitons from Anti-de Sitter Space}
\jhep{06}{2000}{008};
\phepth{0003075}.

\bibitem{microgiant}
S.~R.~Das, S.~P.~Trivedi and S.~Vaidya,
\ct{Magnetic moments of branes and giant gravitons}
\jhep{10}{2000}{037};
\phepth{0008203}.

\bibitem{bjy}
D.~Brecher, B.~Janssen and Y.~Lozano,
\ct{Dielectric Fundamental Strings in Matrix String Theory}
\npb{634}{2002}{23--50};
\phepth{0112180}.

\bibitem{jy1}
B.~Janssen and Y.~Lozano,
\ct{On the Dielectric Effect for Gravitational Waves}
\npb{643}{2002}{399--430};
\phepth{0205254}.

\bibitem{jy2}
B.~Janssen and Y.~Lozano,
\ct{A Microscopical Description of Giant Gravitons}
\npb{658}{2003}{281--299};
\phepth{0207199}.

\bibitem{brs}
E.~A.~Bergshoeff, M.~de~Roo and A.~Sevrin,
\ct{On the supersymmetric non-abelian Born-Infeld action}
\forp{49}{2001}{433--440} and
\npps{102}{2001}{50--55};
\phepth{0011264}.

\bibitem{djm} S. R. Das, A. Jevicki and S.~D.~Mathur,
\ct{Vibration modes of giant gravitons}
\citeprd{63}{2001}{024013};
\phepth{0009019}.

\bibitem{giantholo1} V.~Balasubramanian, M.~Berkooz, A.~Naqvi and
M.~Strassler,
\ct{Giant Gravitons in Conformal Field Theory}
\jhep{04}{2002}{034};
\phepth{0107119}.

\bibitem{giantholo2} S.~Corley, A.~Jevicki and S.~Ramgoolam,
\ct{Exact Correlators of Giant Gravitons from dual $N$=4 SYM}
\atmp{5}{2002}{809--839};
\phepth{0111222}.

\bibitem{ppgiantspectrum1} V. Balasubramanian, M.-x. Huang, T.~S.~Levi and
A.~Naqvi,
\ct{Open Strings from N=4 Super Yang-Mills}
\jhep{08}{2002}{037};
\phepth{0204196}.

\bibitem{ppgiantspectrum2} H. Takayanagi and T. Takayanagi,
\ct{Notes on Giant Gravitons on PP-waves}
\jhep{12}{2002}{018};
\phepth{0209160}.

\bibitem{wtl} W. Taylor,
\ct{Lectures on D-branes, Gauge Theory and M(atrices)}
in 
\bt{1997 Summer School in High Energy Physics and Cosmology}
edited by E.~Gava {\em et.\ al.\/},
(World Scientific, Singapore,
1998),  p.~192--271;
\phepth{9801182}.

\bibitem{db}
D. Bak,
\ct{Supersymmetric Branes in the Matrix Model of PP Wave Background}
\citeprd{67}{2003}{045017};
\phepth{0204033}.

\bibitem{bbn1} G. Bonelli, L. Bonora and F.~Nesti,
\ct{Matrix String Theory, 2D SYM Instantons and affine Toda systems}
\plb{435}{1998}{303--311};
\phepth{9805071}.

\bibitem{jl} J. D. Lykken, \ct{Introduction to Supersymmetry} in
\bt{Field Strings and Duality, TASI 96}
edited by C. Efthimiou and B. Green,
(World Scientific, Singapore, 1997), p.~85--153.

\bibitem{jm2} J. Michelson, \ct{A pp-wave with 26 Supercharges}
\cqg{19}{2002}{5935--5949}; \phepth{0206204}.

\bibitem{hms} J. A. Harvey, G. Moore and A. Strominger,
\ct{Reducing $S$-duality to $T$-duality}
\citeprd{52}{1995}{7161--7167};
\phepth{9501022}.

\bibitem{ks}
J.~B.~Kogut and L.~Susskind,
\ct{How quark confinement solves the $\eta\rightarrow 3 \pi$ problem}
\citeprd{11}{1975}{3594--3610}.

\bibitem{gs} G. Grignani and G. W. Semenoff,
\ct{Thermodynamic Partition Function of Matrix Superstrings}
\npb{561}{1999}{243--272};
\phepth{9903246}.

\bibitem{gspp}
G. Grignani, M. Orselli, G.~W.~Semenoff and D. Trancanelli,
\ct{The superstring Hagedorn temperature in a pp-wave background}
\jhep{06}{2003}{006};
\phepth{0301186}.

\bibitem{hpy}
S. Hyun, J.-D. Park and S.-H. Yi,
\ct{Thermodynamic behavior of IIA string theory on a pp-wave}
SNU-TP~03-009,
\hepth{0304239}.

\bibitem{s1} Y. Sugawara,
\ct{Thermal Amplitudes in DLCQ Superstrings on PP-Waves}
\npb{650}{2003}{75--113};
\phepth{0209145}.

\bibitem{s2} Y. Sugawara,
\ct{Thermal Partition Function of Superstring on Compactified PP-Wave}
UT-03-01,
\hepth{0301035}.

\bibitem{gss}
B. R. Greene, K.~Schalm and G.~Shiu,
\ct{On the Hagedorn Behaviour of PP-wave Strings and N=4 SYM Theory at
Finite R-Charge Density}
\npb{652}{2003}{105--126};
\phepth{0208163}.

\bibitem{pv}
L. A. Pando Zayas and D. Vaman
\ct{Strings in RR Plane Wave Background at Finite Temperature}
\citeprd{67}{2003}{106006};
\phepth{0208066}.

\bibitem{hos}
Y. Hosotani,
\ct{Dynamics of Non-integrable Phases and Gauge Symmetry Breaking}
\citeap{190}{1989}{233--253}.

\bibitem{ah} A. Hashimoto,
\ct{Perturbative Dynamics of Fractional Strings on Multiply Wound
D-strings}
\ijmpa{13}{1998}{903--914};
\phepth{9610250}.

\bibitem{tw1} T. Wynter,
\ct{Gauge fields and interactions in matrix string theory}
\plb{415}{1997}{349--357};
\phepth{9709029}.

\bibitem{tw2} T. Wynter,
\ct{Anomalies and large $N$ limits in matrix string theory}
\plb{439}{1998}{37--45};
\phepth{9806173}.

\bibitem{bjsv} M.~Bershadsky, A.~Johansen, V.~Sadov and~C.~Vafa,
\ct{Topological Reduction of 4-D SYM to 2-D Sigma Models}
\npb{448}{1995}{166--186};
\phepth{9501096}

\bibitem{sc} S. Coleman,
\ct{More about the Massive Schwinger Model}
\citeap{101}{1976}{239--267}.

\bibitem{dms2}
S. R. Das, J. Michelson and A.~D.~Shapere,
to appear.

\bibitem{ed} A. R. Edmonds, \bt{Angular Momentum in Quantum Mechanics}
(Princeton University Press, Princeton, 1974).

\bibitem{mathews} J. Mathews, \bt{Tensor Spherical Harmonics} (California
Institute of Technology, Pasadena, 1981).

\bibitem{hill} E. L. Hill, \ct{The Theory of Vector Spherical
Harmonics}
\ajp{22}{1954}{211--214}.

\bibitem{merz} E. Merzbacher, \bt{Quantum Mechanics, Second Edition}
(John Wiley \& Sons, New York, 1970), p.~389--393.

\end{thebibliography}
\end{document}